\documentclass[aps,prd,nofootinbib,twocolumn,superscriptaddress,preprintnumbers,groupedaddress,showpacs]{revtex4}

\usepackage{amssymb}
\usepackage{amsmath}
\usepackage{epsfig}
\usepackage{hyperref}
\usepackage{comment}
\usepackage{color}
\usepackage{cleveref}
\usepackage{multirow}
\usepackage{capt-of}
\usepackage[utf8]{inputenc}

\usepackage{graphicx}
\usepackage{gensymb}

\makeatletter
\def\simgt{\mathrel{\lower2.5pt\vbox{\lineskip=0pt\baselineskip=0pt
           \hbox{$>$}\hbox{$\sim$}}}}
\def\simlt{\mathrel{\lower2.5pt\vbox{\lineskip=0pt\baselineskip=0pt
           \hbox{$<$}\hbox{$\sim$}}}}
\makeatother

\newcommand{\be}{\begin{equation}}
\newcommand{\ee}{\end{equation}}
\newcommand{\bea}{\begin{eqnarray}}
\newcommand{\eea}{\end{eqnarray}}

\newcommand{\es}[2] {\begin{equation} \label{#1} \begin{split} #2 \end{split} \end{equation}}

\begin{document}

\title{The High-Energy Tail of the Galactic Center Gamma-Ray Excess}

\author{Tim Linden}
\affiliation{University of Chicago, Kavli Institute for Cosmological Physics, Chicago, IL}
\affiliation{Center for Cosmology and AstroParticle Physics (CCAPP) and Department of Physics, The Ohio State University Columbus, OH}
\author{Nicholas L. Rodd}
\author{Benjamin R. Safdi}
\author{Tracy R. Slatyer}
\affiliation{Center for Theoretical Physics, Massachusetts Institute of Technology, Cambridge, MA}

\begin{abstract}
Observations by the {\it Fermi}-LAT have uncovered a bright, spherically symmetric excess surrounding the center of the Milky Way galaxy.  The spectrum of the $\gamma$-ray excess peaks sharply at an energy $\sim 2$~GeV, exhibiting a hard spectrum at lower energies, and falls off quickly above an energy $\sim 5$~GeV. The spectrum of the excess above $\sim 10$~GeV is potentially an important discriminator between different physical models for its origin. We focus our study on observations of the $\gamma$-ray excess at energies exceeding $10$~GeV, finding: (1) a statistically significant excess remains in the energy range $9.5-47.5$~GeV, which is not degenerate with known diffuse emission templates such as the {\it Fermi} Bubbles, (2) the radial profile of the excess at high energies remains relatively consistent with data near the spectral peak (3) the data above $\sim 5$~GeV prefer a slightly greater ellipticity with a major axis oriented perpendicular to the Galactic plane. Using the recently developed non-Poissonian template fit, we find mild evidence for a point-source origin for the high-energy excess, although given the statistical and systematic uncertainties we show that a smooth origin of the high-energy emission cannot be ruled out. We discuss the implication of these findings for pulsar and dark matter models of the $\gamma$-ray excess. Finally we provide a number of updated measurements of the $\gamma$-ray excess, utilizing novel diffuse templates and the Pass 8 dataset.
\end{abstract}


\maketitle

\section{Introduction}
\label{sec:int}

Observations by the {\it Fermi} Large Area Telescope ({\it Fermi}-LAT) on board the \emph{Fermi Gamma-Ray Space Telescope} have ushered in a new high-precision era of $\gamma$-ray astronomy. Among its most important findings is a significant and unexpected $\gamma$-ray excess surrounding the dynamical center of the Milky Way galaxy (Galactic Center or GC)~\citep{Goodenough:2009gk, Hooper:2010mq, Hooper:2011ti, Abazajian:2012pn, Gordon:2013vta, Hooper:2013rwa, Abazajian:2014fta, Daylan:2014rsa, Calore:2014xka,  Zhou:2014lva, TheFermi-LAT:2015kwa}. Analyses utilizing standard diffuse emission models have agreed on several robust properties of the $\gamma$-ray excess: (1) it features a spectral peak at an energy of approximately $2$~GeV, (2) it is extended to at least $10^\circ$ from the GC, (3) the emission is centered on the GC and exhibits approximate spherical symmetry. 

While dark matter explanations for the $\gamma$-ray excess (hereafter GCE) have remained extremely popular (for example, see~\citep{Alves:2014yha, Berlin:2014tja, Agrawal:2014una, Liem:2016xpm} for effective and simplified models,~\citep{Ipek:2014gua, Cheung:2014lqa, Gherghetta:2015ysa, Bertone:2015tza} for some UV-complete models, or~\citep{Hooper:2012cw, Ko:2014gha, Martin:2014sxa,Abdullah:2014lla,Freytsis:2014sua,Berlin:2014pya,Cline:2014dwa,Liu:2014cma,Cline:2015qha,Elor:2015tva} for ``dark sector'' models), several astrophysical models have also been posited to explain the key observational properties of the excess, including: (1) populations of either young~\citep{O'Leary:2015gfa, O'Leary:2016osi} or recycled ~\citep{Abazajian:2010zy, Abazajian:2012pn, Yuan:2014rca, Petrovic:2014xra, Lee:2015fea, Bartels:2015aea, Brandt:2015ula} pulsars densely clustered around the dynamical center of the galaxy, (2) outbursts of either leptonic~\citep{Petrovic:2014uda, Cholis:2015dea}  or hadronic~\citep{Carlson:2014cwa} origin originating from Sgr A*, (3) new diffuse emission models including enhanced cosmic-ray injection rates in the central Milky Way~\citep{Gaggero:2015nsa, Carlson:2015ona, Carlson:2016iis}. 

Studies of the GCE have long appreciated that a robust determination of the low-energy spectrum could provide a powerful discriminant between astrophysical and dark matter models for the origin of the excess~\citep{Abazajian:2010zy, Hooper:2013nhl, Cholis:2014lta}. However, low-energy observations of the GCE have been plagued by observational uncertainties stemming from the relatively wide point spread function (PSF) of the {\it Fermi}-LAT instrument at energies below $\sim 1$~GeV, coupled with the large number of $\gamma$-ray point sources (PSs) and bright structured diffuse emission observed in this region of space. 

Less appreciated is the potential for high-energy observations of the GCE to differentiate between astrophysical and dark matter explanations for the excess, and moreover, to differentiate between specific models within these broad categories. For example, the firm detection of a continuation of the excess to high energies, with a consistent morphology, would provide new information on the mass and annihilation mechanism required for a DM interpretation of the excess~\citep{Agrawal:2014oha,Calore:2014nla,Elor:2015tva,Elor:2015bho}, and essentially rule out models of light dark matter as an explanation (such as the $\sim 10$~GeV DM considered in dark photon~\citep{Liu:2014cma} and leptonic-annihilation~\citep{Abazajian:2014hsa} models). In the context of pulsar interpretations of the GCE, a high-energy tail can be naturally accommodated through the inverse-Compton scattering (ICS) of starlight by high-energy electrons accelerated in the pulsar magnetosphere~\citep{O'Leary:2016osi}.

An obvious limitation is that {\it Fermi}-LAT observations suffer from low statistics at high $\gamma$-ray energies; for example, at the time of the original analysis of the GCE~\citep{Goodenough:2009gk}, less than 1000 photons with an energy exceeding $10$~GeV had been observed in the $3^\circ$ surrounding the GC. The results of analyses that cover a broad energy range will accordingly be statistically dominated by the lowest-energy photons in that range, so studying the high-energy regime requires a dedicated analysis. Furthermore, it is possible that the high-energy data might contain photons in this region that are not accounted for by the modeled diffuse backgrounds, but are unrelated to the excess seen at lower energies. This is particularly true if the diffuse background models have been tuned to fit the data over a broad energy range (since such fits are dominated by the more numerous low-energy photons). Such photons might be mistakenly attributed to a continuation of the excess in a template fitting approach, simply because the GCE component is more localized at the GC than the other background templates, and so can be increased without severely impairing the fit elsewhere. For this reason, while previous studies have found some evidence of GCE-correlated emission at high energies, it has not been possible to firmly rule out models for the origin of the GCE that predict no such emission.

In this paper we address the question of whether there is a photon excess at high energy that \emph{independently} prefers the peaked and highly symmetric morphology observed in the $1-3$~GeV range. In Sec.~\ref{sec:models} we describe the analysis framework we employ to study the $\gamma$-ray excess in the GC and Inner Galaxy (IG), using both Poissonian and non-Poissonian template fitting. The latter method (based on~\citep{Lee:2015fea,Malyshev:2011zi,Lee:2014mza,Zechlin:2015wdz}) can be used to account for populations of unresolved PSs. In Sec.~\ref{sec:SpPr} we present our results:\footnote{We use both frequentist and Bayesian statistics in this work. Our frequentist results will be quoted in terms of a test statistic or TS, determined from the improvement in $-2 \Delta \ln \mathcal{L}$, where $\mathcal{L}$ is the Poissonian likelihood. We will use TS to denote the improvement in the quality of fit when adding the GCE over a background only hypothesis, whilst $\Delta$TS will instead indicate the decrease in fit quality when we vary a parameter from its best fit value (or the change in fit quality when we vary a parameter away from a special value). The Bayesian results will be quoted in the form of $2 \ln \left[ {\rm Bayes~factor} \right]$.} we demonstrate that the $\gamma$-ray excess is a statistically significant (TS $>9$) feature at energies up to $\sim 50$~GeV, with a profile slope that is largely energy-independent. However, we find some statistical evidence that the high-energy data prefer a GCE template with an axis ratio elongated perpendicular to the Galactic plane. We also demonstrate that there is not enough statistical power above $10$~GeV to determine whether the GCE in this range is smooth or comprised of PSs. In Sec.~\ref{sec:discussion} we briefly discuss the implications of these results for the origin of the GCE, focusing on dark matter and pulsar models. Specifically we emphasize that ICS from pulsars near the GC (see, for example,~\citep{O'Leary:2015gfa, O'Leary:2016osi}) may be able to explain the high-energy tail of the GCE. Finally our conclusions are presented in Sec.~\ref{sec:conclusion}.

In the Appendices we present several supplemental studies. In App.~\ref{app:Bkg} we evaluate the robustness of the GCE to changes in the modeling of both diffuse and point-like astrophysical backgrounds. Appendix~\ref{app:SelCr} studies the impact of reverting from Pass 8 data to the older Pass 7 Reprocessed data, and of varying our photon quality cuts. In App.~\ref{app:BkgROI} we examine the effect of changing the ROI that we examine, and varying the masking of PSs. Appendix~\ref{app:gc_spec_appendix} directly compares the GC and IG results, while App.~\ref{app:NPTFCheck} provides several cross checks on the analysis of the PS contributions to the excess. In App.~\ref{app:15G} we investigate an apparent downturn in the GCE spectrum at $11.9-18.9$~GeV, and demonstrate that it cannot be robustly established as a physical feature of the excess. Finally, in App.~\ref{app:Binning} we discuss the energy binning; the possible effects of the energy dispersion; and cumulative results for the morphology of the excess, describing the best-fit morphology for all energies above a threshold $E_\mathrm{min}$.

\section{Analysis Framework}
\label{sec:models}

As in~\citep{Daylan:2014rsa} we perform two independent analyses, carried out in the separate yet partially overlapping spatial regions referred to as the ``Galactic Center'' (GC) and ``Inner Galaxy'' (IG), which use somewhat different analysis frameworks. The primary difference is that the IG analysis extends out to larger latitude and longitude than the GC analysis but masks the Galactic plane. In both cases we construct a pixel-based Poisson likelihood, fitting the data to a linear combination of spatial templates, as has been done previously for studies of the GCE~\citep{Hooper:2013rwa,Daylan:2014rsa,Calore:2014xka,Lee:2015fea, Carlson:2016iis} and other features in {\it Fermi} data (e.g.~\citep{Dobler:2009xz,Su:2010qj}). The exact templates we use differ somewhat between the two analyses, due to the different regions of interest (ROIs), as we will discuss in the following subsections: in particular, the GC analysis requires a more careful treatment of known PSs, while the IG analysis employs an additional template for the structures known as the {\it Fermi} Bubbles~\citep{Su:2010qj}. In addition, in the IG region, we apply the non-Poissonian template fitting (NPTF) method outlined and utilized in~\citep{Lee:2015fea} (following earlier work in~\citep{Malyshev:2011zi,Lee:2014mza}), to study the question of whether the excess is comprised of PSs too faint to meet the statistical criteria for inclusion in official {\it Fermi} point source catalogs.

In all analyses we use Pass 8 data collected between August 4, 2008 and June 3, 2015. Note that this includes data taken during the period when the {\it Fermi} satellite modified its scan strategy to increase the exposure of the GC. The \emph{Fermi} public data are subdivided in several different ways: into quartiles based on PSF and thereby angular resolution, quartiles based on energy dispersion, and front-converting vs back-converting events (labeled according to where the photon first produces an $e^+ e^-$ pair in the detector layers). Front-converting and back-converting events each constitute roughly half the dataset; front-converting events tend to have superior angular resolution, so the front-converting events are similar but not identical to the top two quartiles by PSF. Events are also categorized into nested classes according to whether they pass quality cuts intended to reduce cosmic-ray contamination; the relevant event quality classes for this work are ``Source'', ``Clean'', ``Ultraclean'' and ``UltracleanVeto'' (UCV), corresponding to increasingly stringent cuts with successively lower acceptances.

Unlike previous work that has focused on the peak of the excess where there are abundant statistics, the number of photons is limited in the high-energy tail. As such we have chosen to sacrifice angular resolution for enhanced statistics and accordingly generally use all (front- and back-converting) Source class events. We denote this selection by ``All Source''. The exception to this is the NPTF analysis, where we seek a compromise between these factors by using only the top three quartiles of Source data, ranked by PSF, since the worst quartile markedly degrades the angular resolution.\footnote{The average 68\% containment radius of the Source class PSF for all four quartiles at $1$ ($10$)~GeV is $0.98^{\circ}$ ($0.22^{\circ}$); restricting to just the top three quartiles improves this to $0.68^{\circ}$ ($0.12^{\circ}$).} We have checked that the main conclusions of our analysis are robust against variations in the choice of dataset; see App.~\ref{app:SelCr}. In particular, we show results using only the top quartile of events by angular resolution (i.e. the events with the smallest PSF / best angular resolution), which we denote ``BestPSF''. For example, ``UCV BestPSF'' refers to the events in the top quartile by angular resolution that have also passed all the cuts necessary to be classified as UCV; this is our highest-quality (but lowest-statistics) sample of photons.

In all analyses we divide the data into thirty equally log-spaced energy bins between $0.3$ and $300$~GeV; we use only data between $377$~MeV and $47.5$~GeV, dropping the lowest bin and all bins centered above $50$~GeV.\footnote{We drop the lowest energy bin because with All Source class events the PSF is large enough that our PS mask in the IG analysis covers most of the ROI.} We employ the recommended data quality cuts: zenith angle $<90^{\circ}$, \texttt{DATA\_QUAL} $>$ 0, \texttt{LAT\_CONFIG}=1.

We examine contributions to the $\gamma$-ray flux stemming from five emission components: (1) bright $\gamma$-ray PSs of either Galactic or extragalactic origin, (2) diffuse extragalactic emission that is expected be isotropic over the sky, (3) diffuse Galactic $\gamma$-ray emission, stemming from a combination of $\pi^0$-decay emission from the hadronic interaction of cosmic-ray protons with interstellar gas, bremsstrahlung emission from the interaction of cosmic-ray electrons with the same interstellar gas, and ICS of this electron population off the interstellar radiation field and CMB, (4) $\gamma$-ray emission stemming from the recently discovered {\it Fermi} Bubbles~\cite{Su:2010qj} -- large structures extending perpendicular from the galactic disk, and (5) a GCE template. 

Regardless of the origin of the GCE, previous studies have found its spatial morphology to be well described by the line-of-sight integral over the square of a generalized Navarro-Frenk-White (NFW) halo profile~\citep{Navarro:1995iw,Navarro:1996gj}, so we adopt this profile for our GCE template. The generalized NFW density profile is given by
\begin{equation}
\rho(r) = \rho_0 \frac{(r/r_s)^{-\gamma}}{(1+r/r_s)^{3-\gamma}}\,,
\label{NFW}
\end{equation}
and following previous studies we choose a scale radius of $r_s = 20$ kpc and $\rho_0 = 0.4$~GeV/cm$^3$, while leaving $\gamma$ as a free parameter. If the GCE originates from dark matter, then such a distribution could arise naturally from dark matter annihilation around the GC. Nonetheless we remain agnostic on the issue of the excess' origin, and note that since the GCE is quite localized toward the GC, the NFW profile essentially functions as a power law, with the flux per unit volume scaling at small radii as $r^{-2\gamma}$. In analyses where $\gamma$ is not varied and not otherwise specified, we take $\gamma=1.14$ in IG and $1.05$ for the GC analyses, as these are the respective best-fit values when the fit is performed over the full energy range.\footnote{The IG value is somewhat smaller than that obtained in previous IG analyses~\citep{Daylan:2014rsa,Calore:2014xka}, which found best fit values of $1.18$ and $1.28$ respectively. The difference with~\citep{Daylan:2014rsa} is mainly driven by the combination of a move to data with lower angular resolution and changing to a smaller ROI, whilst the distinction with~\citep{Calore:2014xka} is likely due to this as well as the different method the authors of that work used to extract $\gamma$. We discuss these points in more detail in App.~\ref{app:SelCr}.}

\subsection{Inner Galaxy -- Poissonian Analysis}
\label{sec:IGAnFr}
Our default Region of Interest (ROI) for the IG analysis is defined by $1^{\circ}<|b|<15^{\circ}$ and $|l|<15^{\circ}$. This ROI is smaller than some previous analyses of the IG~\citep{Daylan:2014rsa,Calore:2014xka}, which considered either a $40^{\circ} \times 40^{\circ}$ region or the full sky, in both cases masking the plane (at either $|b|=1^\circ$ or $|b|=2^\circ$). However, it was pointed out in~\citep{Daylan:2014rsa} that in a larger ROI, the amplitudes of the diffuse backgrounds are fit primarily in regions outside the IG. This commonly leads to oversubtraction in the IG (especially along the Galactic plane), which can distort any GCE component extracted from this region. We have found our ROI is sufficiently small to avoid these issues and we provide an additional discussion of this point in App.~\ref{app:BkgROI}.

In this IG analysis we compute the pixel-based Poissonian likelihood for each energy bin independently, allowing the amplitude of each template to float in each bin (thus the total number of free parameters is the number of templates multiplied by the number of energy bins). We include four spatial templates in this process: 1. a uniform isotropic map; 2. a map for the {\it Fermi} Bubbles~\citep{Su:2010qj}; 3. a model for the diffuse background; and 4. a template for the GCE. We choose to normalize the templates so that their coefficients correspond to the average flux within some region. For the isotropic emission this region is the full ROI (but the choice of region is irrelevant in this case), for the Bubbles it is the constant surface density interior of the template, for the diffuse model we use a region within a circle of radius $5^{\circ}$ around the GC, except for $|b|<1^{\circ}$, and for the GCE an annulus between $4.9^{\circ}$ and $5.1^{\circ}$.

Our default background diffuse model for the IG analysis is the {\it Fermi} Collaboration \texttt{p6v11} Galactic diffuse model. As in~\citep{Daylan:2014rsa}, this choice is motivated by the fact that this model does not include a spectrally and spatially fixed component for the {\it Fermi} Bubbles (as is the case with the \texttt{p7v6} model), which allows us to fit the Bubbles independently of the normalization for the diffuse model. We cannot use the most recent Pass 8 Galactic diffuse model, \texttt{p8v6}, since it is explicitly unsuitable for studies of extended excesses, by construction\footnote{http://fermi.gsfc.nasa.gov/ssc/data/analysis/LAT\_caveats.html} -- the model includes a component that is obtained by re-adding the spatially filtered residuals between the data and the model, so in general any extended excesses will already be included in this ``diffuse background''.  We do show results for the \texttt{p7v6} and \texttt{p8v6} models in App.~\ref{app:Bkg} and provide additional detail on the various diffuse models used; as expected (by construction), the GCE in the IG analysis is strongly suppressed with the \texttt{p8v6} model.

In addition to our default choice for the diffuse model, we have cross checked our results with sixteen additional diffuse models in App.~\ref{app:Bkg}. In the main text, for the IG analysis we will also make use of the best performing \texttt{GALPROP} model identified in~\citep{Calore:2014xka} -- referred to there as Model F, a convention we follow.\footnote{In~\citep{Calore:2014xka} Model F was taken from~\citep{Ackermann:2012pya} where it was referred to as $^S{\rm L}^Z6^R20^T100000^C5$.} Additionally in the IG and GC analyses we will use Model A from~\citep{Calore:2014xka}, which was used as the reference model in that work; it performs significantly better than Model F if we remove a mask of the plane and move towards the GC.\footnote{Specifically in the GC Model A provides a better fit to the data than Model F by $\Delta {\rm TS} = 4906$.} The main gamma-ray production processes described by these diffuse background models are $\pi^0$ decay, ICS and bremsstrahlung. In \texttt{p6v11}, all three of these contributions are summed, so their contribution in any pixel is a function of only a single coefficient. Conversely, for Model F and A, these components can be fitted independently. Given that both the $\pi^0$ decay and bremsstrahlung maps trace the interstellar gas, we follow~\citep{Calore:2014xka} and choose to combine these templates, while still floating the combined template independently of the ICS component. In this way Model F and A can give us additional insight into the behaviour of the background over the models provided by the {\it Fermi} Collaboration, and we will exploit this in App.~\ref{app:15G}.

We smooth the diffuse model template using the {\it Fermi} Science Tools routine \texttt{gtsrcmaps}. As the remaining three templates are considerably less bright in our ROI, we simply perform Gaussian smoothing before comparing them to the data. In App.~\ref{app:Bkg} we confirm that the use of Gaussian smoothing for the diffuse model has minimal impact on our results. We include PSs as a fixed contribution in our template fit and further mask the 300 brightest and most variable sources in the 3FGL catalogue~\citep{Acero:2015hja}, where the size of the mask is determined by the 95\% containment radius.

\subsection{Inner Galaxy -- Non-Poissonian Template Fitting}
\label{sec:NPTF_intro}

Much of the previous subsection carries over to the NPTF analysis in the IG. We keep the four templates previously discussed (or five templates in the case of Model F or A and other \texttt{GALPROP}-based models, where the ICS template is floated separately),\footnote{To facilitate comparison with the previous work of~\citep{Lee:2015fea}, for the NPTF analysis we use a GCE template constructed from an NFW with $\gamma=$1.25, which is the value used in that reference. We find changing this to 1.14 has a much smaller impact than the other sources of uncertainties in our analysis.} which are taken to have Poissonian statistics, but also add two new templates with \emph{non}-Poissonian statistics. These templates correspond to populations of PSs, with (a) a thin-disk doubly exponential distribution, with the density of sources proportional to $e^{-|z|/0.3\mathrm{kpc}} e^{-r/5\mathrm{kpc}}$, or (b) a centrally peaked spherical distribution, with the density of sources tracing a generalized NFW profile squared, so that the average flux per pixel follows the inferred distribution of flux for the GCE. We refer to these templates respectively as the ``disk PS'' and ``GCE PS''  templates.\footnote{In~\citep{Lee:2015fea} isotropically distributed PSs were also considered. In the present analysis we use a smaller ROI to that work within which we find an isotropic NPTF template to be poorly constrained. As such we have chosen to exclude it.} 

The source count function for each template, i.e. the number of sources producing a given number of counts $S$ in a pixel $p$, is parameterized as a broken power law:
\es{SourceCount}{
\frac{dN_p}{dS} = A_p \begin{cases} (S/S_b)^{-n_1}, & S > S_b \\  (S/S_b)^{-n_2}, & S < S_b  \end{cases}\,.
}
Here $A_p$ is the pixel-dependent normalization factor that accounts for the spatial dependence of the source distribution.  More specifically, $A_p$ is taken to follow the disk (GCE) template for the disk PS (GCE PS) model.  The two indices $n_1$, $n_2$ and the break $S_b$, with units of counts, are assumed to be constant between pixels.\footnote{Note that the assumption that $S_b$ does not vary between pixels is a good approximation since the exposure map does not vary significantly over our small region~\cite{Lee:2015fea}, specifically changing by less than 4\% of the mean.}  These three parameters, along with the overall normalization for $A_p$, are treated as independent model parameters for each template. Thus, our default model ${\cal M}$ has 12 model parameters $\theta$, in each energy bin: 4 normalization factors for the Poissonian templates (if we instead use a \texttt{GALPROP}-based diffuse model, one additional parameter is added for the independent ICS component), plus $2\times4 = 8$ for the source count function parameters for the 2 PS templates.  We will also consider a simplified model that does not include the GCE PS template but is otherwise the same.

We follow the procedure outlined in the literature~\citep{Malyshev:2011zi,Lee:2014mza,Lee:2015fea,Zechlin:2015wdz} to calculate the photon-count probability distribution in each pixel for the combined template model as a function of the model parameters $\theta$.  Given these distributions, we may evaluate the likelihood function~\cite{Lee:2015fea}:
\es{LLNPTF}{
P(d | \theta,{\cal M} ) = \prod_p P^{(p)}_{n_{p}} (\theta) \,,
}
where the data set $d$ consists of $n_p$ counts in each pixel $p$.
We use Bayesian methods (implemented with \texttt{MultiNest}\footnote{Specifically we run with 400 live points, disable both importance nested sampling and constant efficiency mode, and the sampling efficiency is set for model-evidence evaluation.}~\citep{Feroz:2008xx,Buchner:2014nha}) to compute the posterior distribution $P(\theta | d, {\cal M})$ for the model parameters. Our prior ranges for the model parameters are shown in Table~\ref{table:PNPTF},\footnote{The choice of 2.05 and 1.95 as boundaries for the prior ranges of $n_1$ and $n_2$ respectively is chosen for numerical stability of the code. The origin of the instability is that the total flux associated with an non-Poissonian template diverges if $n_1 = 2$ or $n_2 = 2$. Regardless we confirmed in all cases the preferred index was converged away from these boundaries.} where $A_p = A \tilde{A}_p$ and $\tilde{A}_p$ is the template with baseline normalization. For the Poissonian templates these baseline normalizations are set by summing over energy the best-fit values determined in the IG analysis of the previous section. The GCE PS template inherits the same baseline normalization as the GCE Poissonian template, while the disk PS is normalized such that the mean number of photons per pixel is one over the full sky.\footnote{The disk PS normalization is arbitrary, but the prior range is sufficiently large that the posterior is well converged.} Note that all priors are flat on a linear scale except for the normalizations of the Poissonian GCE and GCE PS templates, which are flat on a logarithmic scale. The prior ranges are sufficiently large such that all model parameters are well converged within the prior ranges.
\begin{table}[h]
\begin{center}
\begin{tabular}{| c || c |}
\hline
Parameter & Prior Range \\ \hline \hline
$A_{\rm diff}$ & $[0,2]$ \\ \hline
$\log_{10} A_{\rm GCE}$ & $[-6,6]$ \\ \hline
$A_{\rm iso}$ & $[-2,2]$ \\ \hline
$A_{\rm bub}$ & $[0,2]$ \\ \hline
$\log_{10} A_{\rm PS}$ & $[-6,1]$ \\ \hline
$S_b$ & $[0.05,600]$ \\ \hline
$n_1$ & $[2.05,30]$ \\ \hline
$n_2$ & $[-2,1.95]$ \\ \hline
\end{tabular}
\end{center}
\caption{Prior ranges used for the non-Poissonian template fitting analyses. For all fits, the parameters were confirmed to be well converged within these ranges. The bottom four parameters are used to describe an NPTF template, and the same ranges were used for the GCE PS and disk PS templates. The only exception to this was a cross check performed in App.~\ref{app:NPTFCheck} where we fixed $n_2$ to be $-1.5$ for the GCE PS, and $-1.4$ for the disk PS. The results of that check are shown on the left of Fig.~\ref{fig:BayesCheck}.}
\label{table:PNPTF}
\end{table}

For all of our NPTF analyses, we will perform two template fits, one with a model that includes the GCE PS template and one that does not.  Each of these fits returns a Bayesian evidence $p(d | {\cal M})$; the ratio of the Bayesian evidence between the two models is known as the Bayes factor.  We will use the convention that the Bayesian evidence for the model with GCE PS's is in the numerator, so that a Bayes factor greater than unity indicates evidence for spherical PSs.  All of our results will be quoted in terms of $2 \ln \left[ {\rm Bayes~factor} \right]$. 

As shown in~\citep{Lee:2015fea}, the disk PS template can largely describe the identified gamma-ray PSs.  As such, we do not mask any PSs in this analysis. Further, we use slightly less data than in the IG analysis -- only the top 3 PSF quartiles of source data -- in order to improve the PSF, which is helpful in looking for sources too faint to be included in existing {\it Fermi} point source catalogs.

In~\cite{Lee:2015fea} a significant Bayes factor was found in preference of a model with a GCE PS template in the energy range $\sim 2-12$~GeV.  Here, we are interested in determining whether the preference for spherical PSs persists at higher energies.  Our approach is to work in single large energy bins, but to repeat the analysis progressively increasing the lower boundary of the bin. Specifically, while keeping the maximum energy fixed at $50$~GeV, we move the lower energy bound by 10 log spaced steps between $\sim 2$ and $\sim 15$~GeV. In this way, we can determine which energies dominate the statistical preference for the GCE PS template.\footnote{A more thorough inclusion of energy dependence directly into the NPTF is the subject of future work~\cite{edephighlatpaper,edepigpaper,nptfpaper}.}

In order to facilitate the interpretation of the NPTF results, we also generate a large number of simulated data sets and analyze these using the same NPTF framework.  Our simulated data are based on the best-fit parameter values, as extracted from the posterior distribution, from the NPTF analyses on the real data.  In particular, we have two sets of simulated data; the first includes spherical PSs, and uses the best-fit values from the NPTF that also includes this template, while the second does not.  In order to accurately convolve the simulated data with the {\it Fermi} instrument response function, which is energy dependent, we must assume a spectrum for each source component.  The Poissonian spectra are assumed to follow the spectra extracted from an energy-dependent Poissonian template fit on the real data.  We assume  that the GCE PS template has the spectrum extracted by the Poissonian GCE template and consider a variation on this in App.~\ref{app:NPTFCheck}.  For the disk template we use a more data-driven method. By moving the minimum of the lowest energy bin we determine how the integrated flux associated with the disk PS template varies with that lower energy. Specifically, we assume the average spectrum of this population is a power law so that we can use the variation in the integrated flux to constrain the parameters. We then use this derived spectrum when creating the simulated data.

\subsection{Galactic Center}
\label{subsec:gc}

We define a GC analysis to cover the dense region \{$|l|,|b|$\}$<$7.5$^\circ$, where the fractional intensity of the GCE component is maximized. In this ROI, the emission from bright $\gamma$-ray PSs cannot be masked without significantly diminishing the ROI. Thus, in this analysis, all bright PSs are modeled and the flux of each is allowed to float independently in each energy bin. We also allow the $\gamma$-ray intensity from the \texttt{p7v6} {\it Fermi}-LAT Galactic diffuse emission model and an isotropic background model to float freely. A template for the {\it Fermi} Bubbles is already included in the \texttt{p7v6} diffuse emission model, and thus no additional template is added. While in the IG analysis we prefer not to use \texttt{p7v6} because of this fixed Bubbles template, we expect the impact of this template to be far less important in the GC. Thus, given that \texttt{p7v6} has superior resolution and modeling of the Galactic plane we use this as our default diffuse template in the GC analysis. We also utilize an alternative diffuse emission model, using the results from~\citep{Calore:2014xka}, and in this case we add a {\it Fermi} Bubbles template from~\citep{Su:2010qj}.

We note that the choice to allow our background components to float freely in each energy bin differs from previously published models of the GC ROI (e.g.~\citep{Daylan:2014rsa}), where the background components (besides the GCE) were fit over the full energy range assuming simple spectral parameters (however, see the recent results of~\citep{Carlson:2016iis}). While~\citep{Daylan:2014rsa} found this approximation to not significantly affect the characteristics of the GCE component near the spectral peak, it is imperative for analyses of the high-energy tail that we do not constrain the normalization of emission components to be fixed by low-energy data.

In order to compute the spectrum, intensity, and statistical preference for the GCE template in each energy bin, we performed a binned likelihood analysis using the {\it Fermi}-LAT tools. Due to the focus of our analysis on the high-energy regime, where the photon flux is greatly lessened and the angular resolution of the {\it Fermi}-LAT is good, we utilize all events passing through the {\it Fermi}-LAT instrument, placing no constraints on front/back conversion or PSF class. We first utilize {\tt gtbin}, dividing the {\it Fermi}-LAT data into $150 \times 150$ angular bins of size $0.1^\circ$, and convolve each input template with the {\it Fermi}-LAT PSF using {\tt gtsrcmaps}. We then utilize the {\it Fermi}-LAT python tools to run {\tt MINUIT}~\cite{James:1975dr} and calculate the normalization of each $\gamma$-ray emission template, before using {\tt gtmodel} to calculate the expected source counts from our normalized model. Finally, we calculate the fit of our model to the {\it Fermi}-LAT data in each energy bin.

\section{Properties of the High-Energy Tail}
\label{sec:SpPr}
In this section, we describe the results of the various analyses described above.

\subsection{The High-Energy Spectrum}
\label{subsec:spectrum}

\begin{figure*}[t!]
\centering
\includegraphics[scale=0.45]{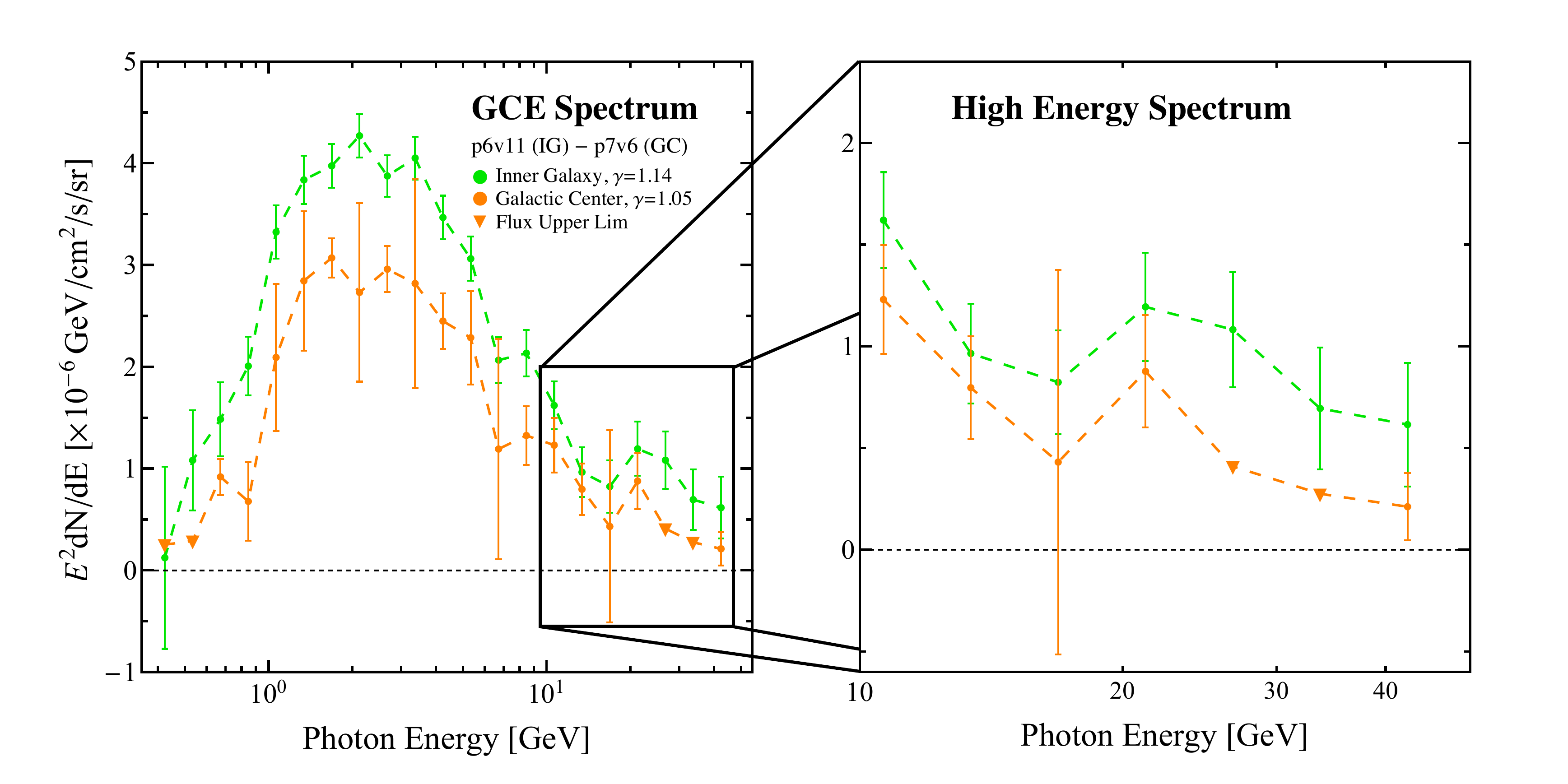}
\caption{\footnotesize{The spectrum of the GCE extracted in the IG (green) and GC (orange) regions over the full energy range (left) and $10-50$~GeV (right). This spectrum was determined using the analysis framework outlined in Sec.~\ref{sec:IGAnFr} for the IG and Sec.~\ref{subsec:gc} for the GC. We show the spectrum for the best fit generalized NFW profile, which corresponds to $\gamma=1.14$ for the IG and $1.05$ for the GC. A similar figure showing the spectra for identical $\gamma$ values or identical diffuse emission models is shown in Fig.~\ref{fig:gcSpec_compare}. In both cases the flux is normalized to its value at $5^{\circ}$ from the plane. In the GC analysis, no flux is observed in several bins, and so we instead show the 90\% upper limit. Note the IG and GC analyses differ in their ROI, diffuse modeling and treatment of the point sources. See text for details.}}
\label{fig:BaseSpec}
\end{figure*}

Using the default choices outlined above for the IG, we extract a spectrum for the GCE template shown in Fig.~\ref{fig:BaseSpec}. On the left hand side of this figure we show the spectrum over the energy range $0.377-47.5$~GeV, while on the right we focus on the range $9.5-47.5$~GeV -- the high-energy range, which we will scrutinize in the following sections.

We see that the GCE template does favor a non-zero coefficient at energies above $10$~GeV in the IG analysis, with a falling spectrum in $E^2 dN/dE$ out to energies above $40$~GeV. The formal significance of the excess above 10~GeV in the IG is ${\rm TS} \sim 127$. There appears to be some evidence for structure in the spectrum, albeit not at high significance. However, as we will discuss in App.~\ref{app:15G}, the apparent ``dip'' at $\sim 15$~GeV may well be an artifact of background mis-subtraction.

Although, as already emphasized, the spectrum alone is not enough to conclude that the GCE extends to higher energies. We must also show that this spectral feature is robust against reasonable changes to the diffuse emission model.  This point is outlined in detail in App.~\ref{app:Bkg}, where we show that a very similar spectrum is obtained for many different diffuse background models, with the only substantial variation stemming from models that have large scale residuals added, which make them poorly suited for studying the GCE.

\begin{figure*}[t!]
\centering
\includegraphics[scale=0.45]{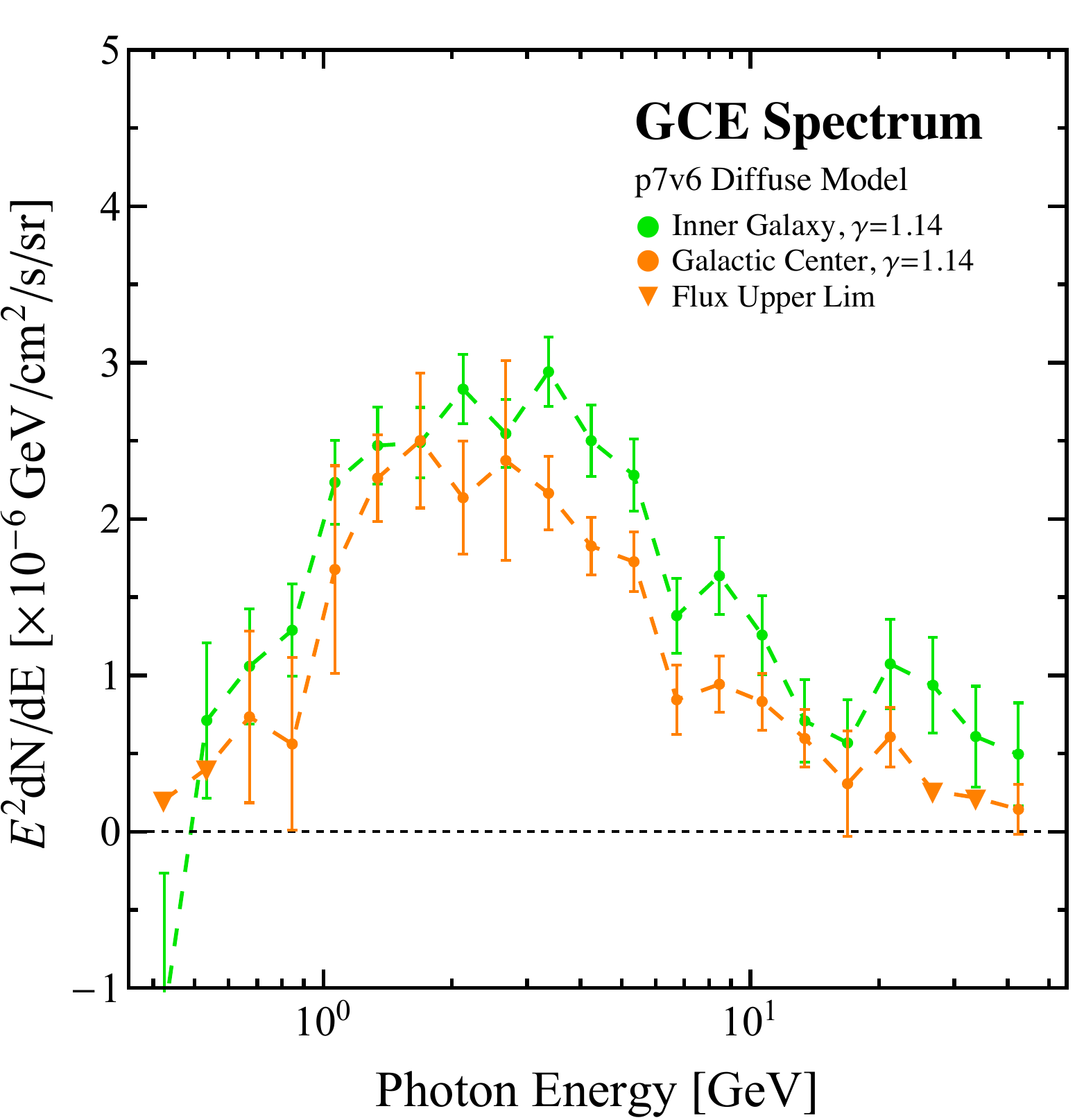} \hspace{0.15in}
\includegraphics[scale=0.45]{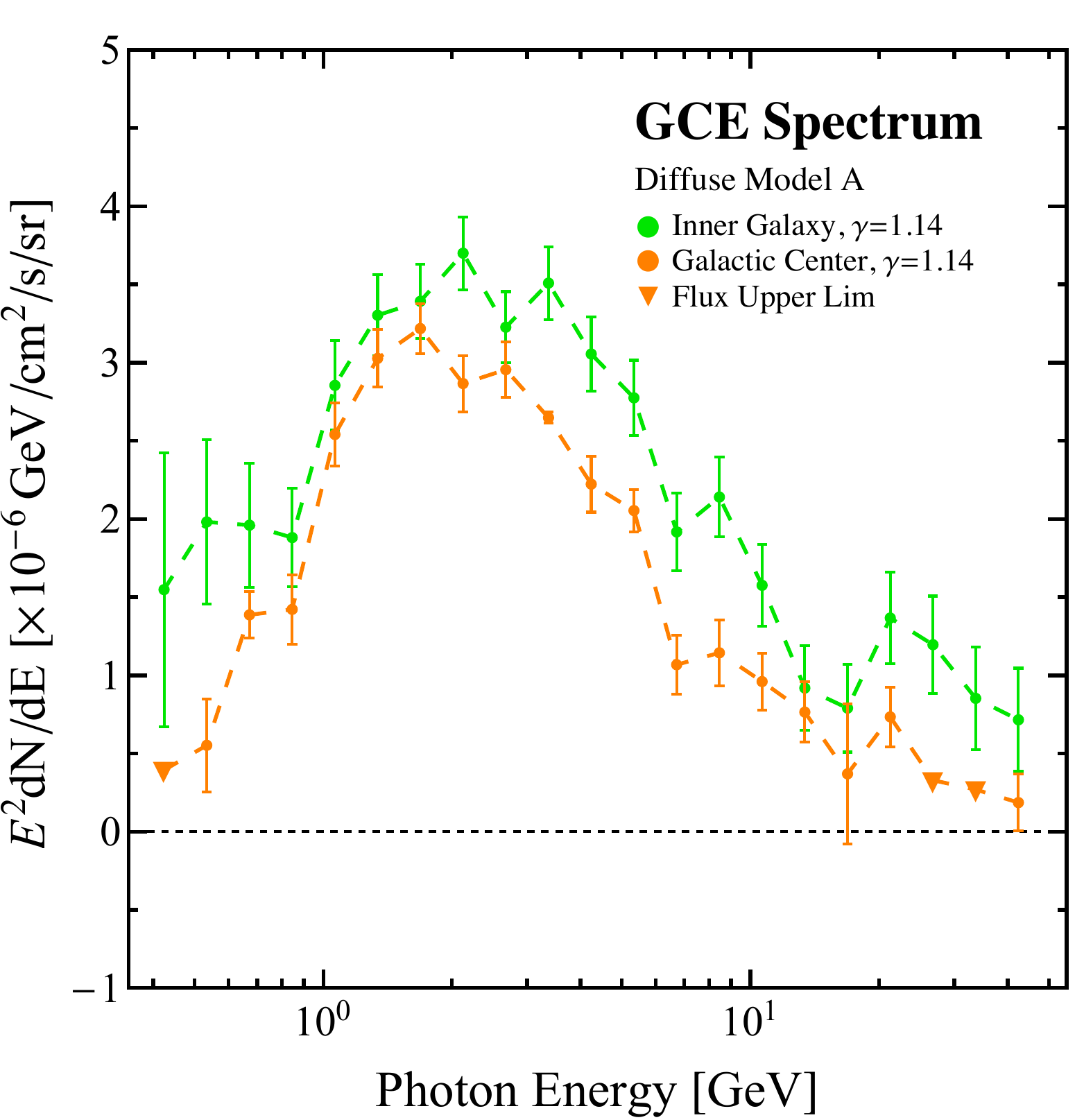}
\caption{Same as the left panel of Fig.~\ref{fig:BaseSpec}, except $\gamma = 1.14$ for both the IG and GC analyses.  Also, in this case both the IG and GC analyses use the same diffuse mode: {\tt p7v6} in the left panel and \texttt{GALPROP} Model A in the right panel.  Using the same diffuse model in both regions alleviates some of the tension in the overall normalization computed between the two analyses.  However, fixing $\gamma = 1.14$ in the GC analysis has an almost comparable effect in magnitude on the normalization of the spectrum as changing diffuse model. We further explore this in App.~\ref{app:gc_spec_appendix}.}
\label{fig:SpecComp}
\end{figure*}

Figure~\ref{fig:BaseSpec} also shows the spectrum of the GCE in the GC analysis. Two results are immediately apparent: (1) the GC analysis prefers an overall normalization of the GCE that is smaller than the IG by $\approx 30\%$, (2) the spectral features of the GCE in each case are very similar. We note that there are several systematic differences between the IG and GC analyses that could contribute to the offset normalization of the GCE between each study, including: (1) a variable radial profile of the GCE, (2) the change in diffuse emission models ({\tt p6v11} in the IG analysis to {\tt p7v6} in the GC analysis), and (3) the treatment of point sources near the Galactic Center in the GC analysis. 

To further illustrate the differences between the IG and GC results, in Fig.~\ref{fig:SpecComp} we show the spectrum for the GCE computed in both the IG and GC regions, but this time using the same diffuse background and the same radial-profile parameter $\gamma$ for each analysis.  In particular, the left (right) panel uses the {\tt p7v6} model (\texttt{GALPROP} Model A) in both regions, with $\gamma$ fixed to $1.14$.  In both cases, we see that using the same diffuse model in the IG and GC regions alleviates some of the tension between the spectrum computed in the two analyses.  However, we may also see---comparing the GC result in Fig.~\ref{fig:BaseSpec} with that in the left panel of Fig.~\ref{fig:SpecComp}---that changing $\gamma$ from $1.05$ to $1.14$ in the GC itself causes a systematic decrease in the normalization at a level $\sim$20\%.  Thus, there are a variety of factors that may contribute to the offset between the spectrum computed between the two regions.  This is further explored in App.~\ref{app:gc_spec_appendix}.

Focusing instead on the spectral characteristics of the excess in both ROIs, we find broad qualitative agreement. Namely, we find a statistically significant excess that extends above an energy of $10$~GeV. However, in the GC analysis we do not find any statistical preference for GCE emission in the two energy bins spanning $23-38$~GeV. With that said, the GC upper limits are not inconsistent with the values determined by the IG analysis, once the smaller normalization of the GC template is taken into account. We note that unlike in the case of the IG, we utilize a value of $\gamma=1.05$ in the GC, which provides the best fit to the data in this ROI. We find that this choice has a negligible effect on the spectral properties of the excess, as we demonstrate in Fig.~\ref{fig:gcSpec_compare}.

\subsection{The High-Energy NPTF Analysis}
\label{sec:NPTF}

\begin{figure*}[t!]
\centering
\begin{tabular}{c}
\includegraphics[scale=0.5]{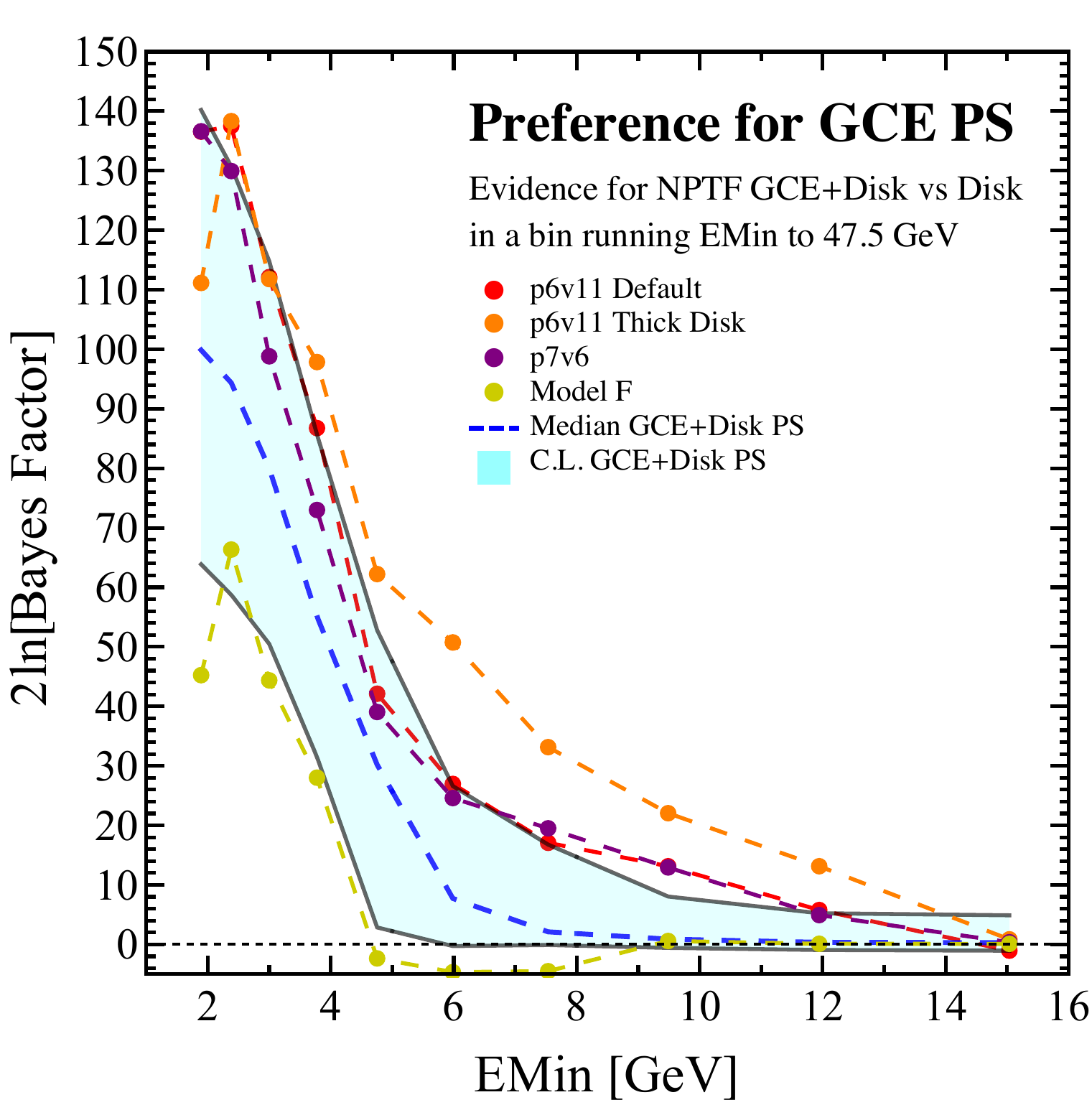} \hspace{0.15in}
\includegraphics[scale=0.5]{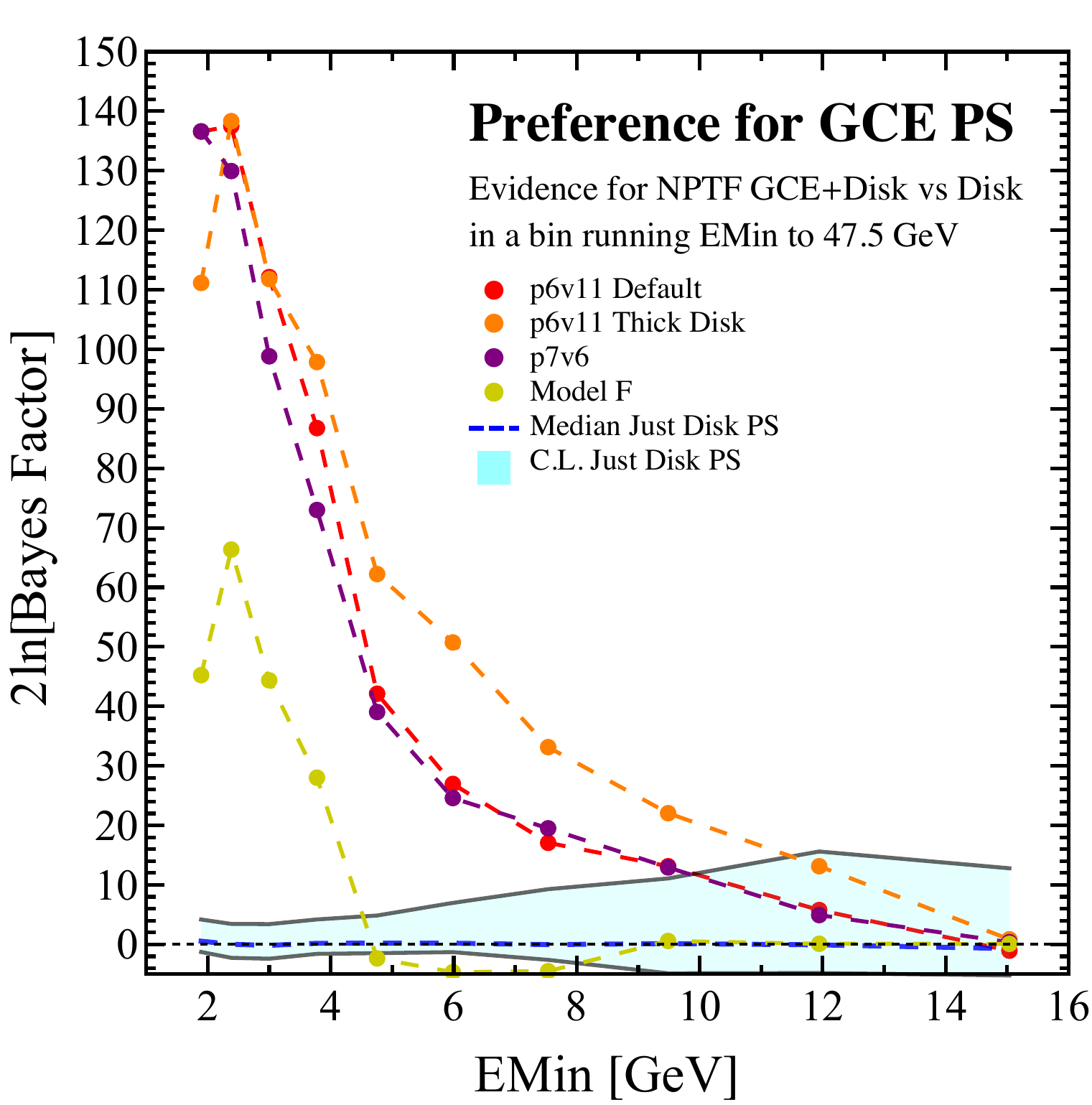}
\end{tabular}
\caption{\footnotesize{Preference for adding GCE correlated PSs to our default fit as a function of the minimum energy bin, whilst the maximum is kept at $47.5$~GeV. The result for our default analysis is shown in red, whilst varying the disk PS to use a thicker disk template is shown in orange, and changing the diffuse model from \texttt{p6v11} to \texttt{p7v6} or Model F is shown in purple and khaki respectively. In addition, in blue we shown the median and 90\% confidence limits from analysing a large ensemble of mock data maps assuming both disk and GCE correlated PSs (left) or just a disk PS template (right). Mock data sets were created using the \texttt{p6v11} diffuse model. See text for details.}}
\label{fig:Bayes}
\end{figure*}

In Fig.~\ref{fig:Bayes} we show the results of our default NPTF analysis of the IG in red on both the left and right panels. We find evidence in favor of the model with a spherical PS population up to $\sim 15$~GeV, with a moderately large Bayes factor; e.g., $2\ln\left[{\rm Bayes~factor}\right]\sim 13$ for the bin with $E_\mathrm{min} \sim 10$~GeV.

To assess the true significance of these results, we take two approaches. First, to correctly interpret the statistical significance of such a detection, we create a large ensemble of simulated data maps based on the best-fit model, and then we repeat our analysis on the simulated data. As described in Sec.~\ref{sec:NPTF_intro}, we create two types of simulated data to contrast differing hypotheses.  Our first simulated data set is based on the best-fit values from the NPTF on the real data that includes four Poissonian templates -- isotropic, Bubbles, diffuse and a smooth GCE -- as well as disk PSs.  In this scenario, the GCE is fully accounted for by the smooth GCE template.  Our second set of simulated data is based on the NPTF that also includes the GCE PS template; in this case, the GCE is produced by a population of spherical PSs.

The second approach we take for assessing the significance is to repeat the analysis with three different background models in order to estimate the systematic uncertainty associated with the choice of background model. Specifically, we 1. replace our thin-disk non-Poissonian template with a thicker disk of scale height 1 kpc rather than 0.3; 2. we replace the \texttt{p6v11} diffuse model with \texttt{p7v6}; and 3. we again replace the diffuse model with Model F, a \texttt{GALPROP} model. Additional background-model variations and variations on the simulated-data tests, as well as the best fit source-count functions for each case considered, are shown in App.~\ref{app:NPTFCheck}.

The full results of these tests are shown in Fig.~\ref{fig:Bayes}. On the left, we show simulated data generated assuming both disk and GCE correlated PSs, whilst on the right, only disk-distributed PSs are included in the Monte Carlo. In both cases we show the 90\% confidence limits in blue constructed from multiple Monte Carlo simulations. On the one hand, we find that given the best-fit model with GCE PSs, the expected $2\ln\left[{\rm Bayes~factor}\right]$ for the model containing the GCE PS template becomes less significant ($\lesssim 10$) for energy bins with minimum energies above $\sim 10$~GeV. The Bayes factors extracted from the real data are somewhat high compared to expectations from the mock data, consistently across energy bins, but generally lie within the 90\% confidence band of expected Bayes factors. Furthermore, when we construct simulated data with \emph{no} GCE PS contribution, the Bayes factors we find become consistent with the simulated-data prediction, within the 90\% confidence band, above $\sim 10$~GeV. This suggests that above $10$~GeV it is not possible to significantly distinguish a model where the GCE is comprised entirely of PSs from one with no GCE PSs with this method and data set. Secondly, the fact that our results consistently overshoot the simulated-data prediction (albeit at low significance) should be cause for some caution. One possible interpretation of this result is that in the real data we do not have a perfect diffuse model as we do in the case of the simulated data. As discussed in~\citep{Lee:2015fea}, NPTF templates can help alleviate imperfect diffuse modeling, which may partly explain why the data often has a high Bayes factor with respect to the simulated data. The relation between this overshoot and background mismodeling in the real data is further supported by the fact that when we repeated this analysis using the top PSF quartile of UCV data we found greater consistency between data and Monte Carlo. This dataset has an improved angular resolution making results more robust to background mismodeling, but this comes at the cost of statistics which is why we have chosen not to use it for this analysis.

Finally, when we perform the analysis with different diffuse models or disk templates, the Bayes factors that we find vary substantially, at the same level as the width of the band from the simulated-data studies.  In particular, using Model F we find no significant detection of GCE PSs at energies above $\sim 6$~GeV.  However, at low energies we always find a preference for GCE PSs. This again emphasizes the interplay between the modeling of the diffuse background and the preference for a PS template. Note that in Model F there is an additional degree of freedom in that the ICS is floated separately from the $\pi^0$ and bremsstrahlung components. It may be that this ICS template improves modeling around the GC where the GCE is bright, thereby reducing the impact a GCE non-Poissonian template can have.

Accordingly, we can neither robustly favor nor disfavor the PS interpretation of any extension of the GCE above $\sim 5$~GeV.

\subsection{Spatial Morphology}

Modeling the gamma-ray sky is a challenging and open problem. Despite great progress being made with data provided by the {\it Fermi} Gamma-Ray Telescope, current diffuse $\gamma$-ray models are still a long way from describing the data to the level of Poisson noise. As such, the difference between $\gamma$-ray data and best-fit models will inevitably contain spatial residuals. This issue is particularly acute around the GC where the modeling of the $\gamma$-ray sky is most challenging.

\begin{figure*}[t!]
\centering
\includegraphics[scale=0.65]{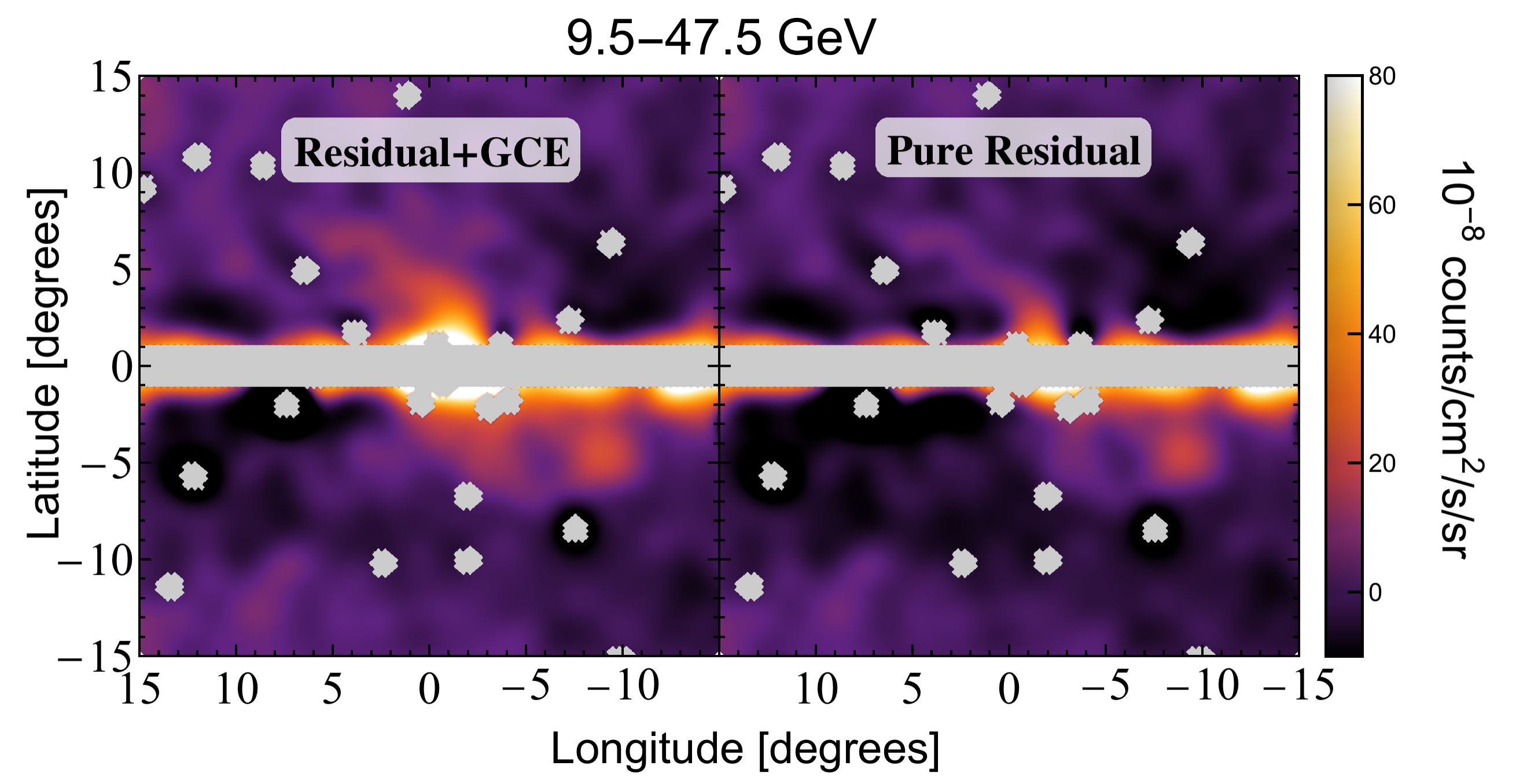}
\caption{\footnotesize{Spatial residual in our default ROI for the IG subtracting all templates except (left) or including (right) the GCE template. The data have been smoothed to $2^{\circ}$, but the PS masks have not -- these are the masks corresponding to the lowest energy where the PSF is largest, here $9.5$~GeV. A number of regions of over and under subtraction are evident in both maps, symptomatic of imperfect background models. Masks are shown in gray. See text for details.}}
\label{fig:FullResidual}
\end{figure*}

\begin{figure*}[t!]
\centering
\includegraphics[scale=0.45]{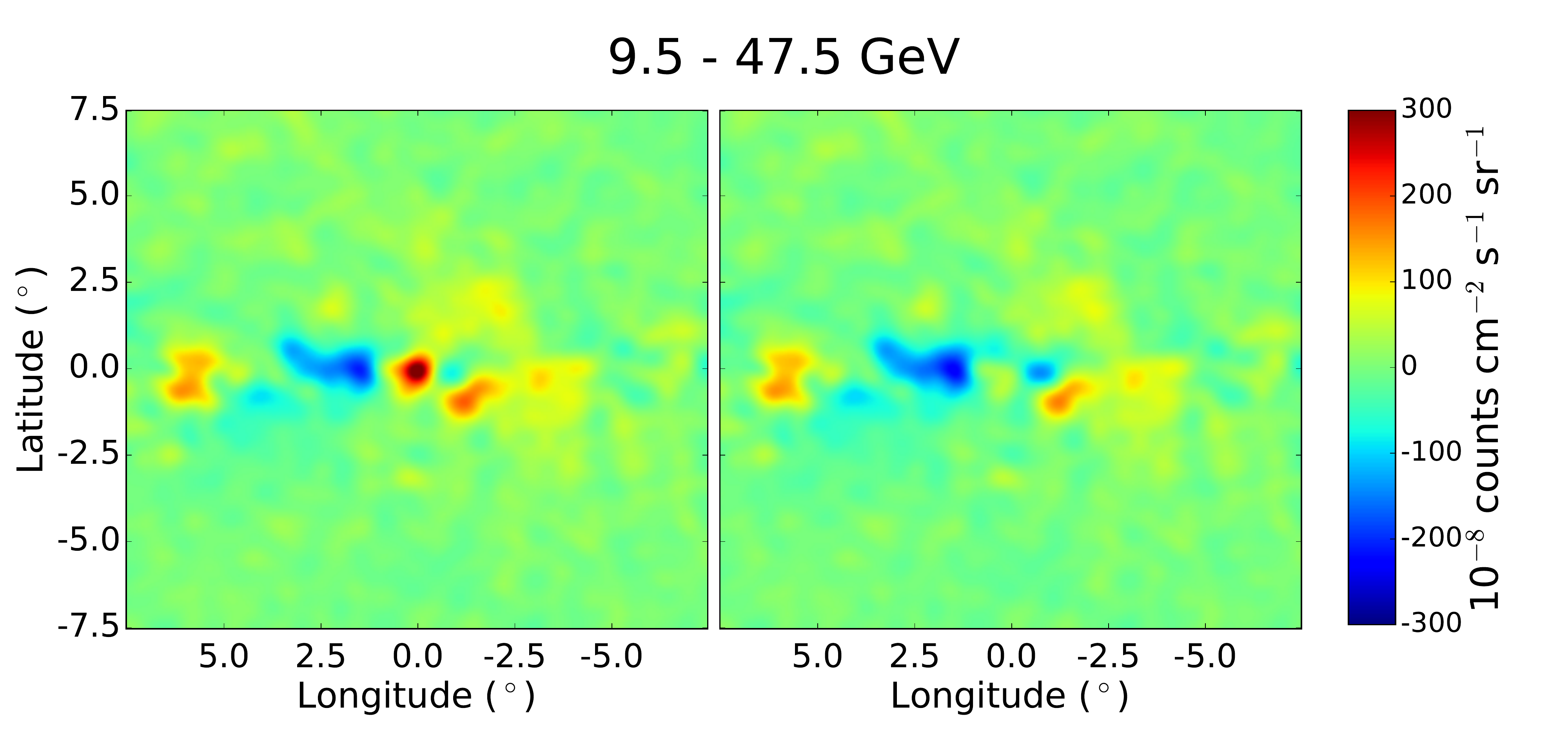}
\caption{\footnotesize{Same as Fig.~\ref{fig:FullResidual} for the GC analysis, and with all data and models smoothed to $0.25^\circ$. The left figure shows the residual when the GCE template is not included, while the right shows the residual after the GCE template is included in the model prediction. }}
\label{fig:GCResidual}
\end{figure*}

If the GCE does extend above $10$~GeV, then it is likely to have an intensity comparable to these spatial residuals. It is easy to imagine a situation where the model underestimates the total $\gamma$-ray emission near the GC, producing to a residual that could be absorbed by a GCE in a template fit. For this reason, the coefficient extracted for our GCE template in Fig.~\ref{fig:BaseSpec} may not be a reliable indicator of the true intensity of the GCE. To highlight this issue, in Fig.~\ref{fig:FullResidual} and \ref{fig:GCResidual} we show the spatial residuals in our ROI before and after the subtraction of the GCE template used to obtain the spectra in Fig.~\ref{fig:BaseSpec}. The maps have been smoothed to $2^{\circ}$ for the IG and $0.25^{\circ}$ for the GC. Although clear emission associated with the excess is evident in the left hand side of both figures, there are a number of regions of over and under subtraction in the ROI, which could also affect the Galactic Center.

To ameliorate this concern, we examine the spatial morphology of the high-energy emission. Among the most striking features of the GCE near its spectral peak, are the simplicity and consistency of its spatial morphology. As pointed out in~\citep{Daylan:2014rsa}, its radial distribution is well described by the square of a generalized NFW profile (projected along the line of sight), it is approximately spherically symmetric and not elongated along the plane of the Milky Way, and it appears very well-centered on the dynamical center of the galaxy at Sgr A*. Furthermore, the first two properties have been shown to be robust against the inclusion of systematic uncertainties~\citep{Calore:2014xka}. While background mismodeling can lead to spurious emission near the GC, it is unlikely that such emission would mimic the peculiar spatial properties exhibited by the GCE.

Focusing on the high-energy emission, we will examine the three basic spatial properties characteristic of the GCE,  investigating the consistency of the radial variation, sphericity, and (in the GC analysis only) the preferred emission center, compared to the GCE component near the $1-3$~GeV spectral peak. We consider the morphology independently in each of the high-energy bins between $10$ and $50$~GeV. In order to help mitigate issues associated with limited statistics we combined the 6 highest energy bins into pairs, so that our four high-energy bins are $[9.5,11.9]$, $[11.9,18.9]$, $[18.9,30.0]$, and $[30.0,47.5]$~GeV. This is the binning we use for our statistical analyses,\footnote{To clarify, when we speak of combining bins, we are summing TS values across bins not redoing the fit in a larger bin. Thus TS values quoted here (as done in Table~\ref{table:TSP7P8} within App.~\ref{app:SelCr} - c.f. Table~\ref{table:TSAllBins} for the values before combining the bins) for the presence of the excess in these combined bins can loosely be interpreted as following a $\chi^2$ distribution with two degrees of freedom.} while for spectral plots we maintain the log-spaced binning. More details on this choice and results from using equally log spaced bins are given in App.~\ref{app:Binning}. Generally if we consider the global, rather than bin-by-bin features of the excess, the spatial properties are driven by the morphological preferences near the spectral peak, around $\sim 1-2$~GeV. This point is explored further in App.~\ref{app:Binning} where we show cumulative results (for all photons above some threshold energy) rather than showing each energy bin individually.

The results for each spatial property are shown below, but we summarize the basic details here. In the IG, the first and third high-energy bins demonstrates similar properties to the low-energy GCE, although with greater significance in the first bin. The second bin at around $15$~GeV is noticeably more statistically limited. Taken together this might indicate a non-trivial spectral variation of the GCE, but we believe this ``dip'' is more likely to be due to issues with the background model impacting this bin, a point we explore in App.~\ref{app:15G}. Finally the fourth bin does not appear to share similar properties to the GCE at lower energies, but the results are highly statistically limited and so it seems at present no definite conclusions can be reached about the extent of the GCE above $30$~GeV. In the Galactic Center analysis we find that the radial slope of the NFW profile is consistent with our best-fitting global value at the $\approx 1.5 \sigma$ level in every high-energy bin. The ellipticity in energy bins above $18.9$~GeV show some evidence ($\Delta {\rm TS} = 4.58$) for an ellipticity that is more strongly stretched perpendicular to the Galactic plane than our global analysis. Investigating the centering of the NFW emission profile, we find that the $\gamma$-ray emission is sourced to within $0.2^\circ$ of Sgr A* in all energy bins.

\subsubsection{Radial Variation}

\begin{figure*}[t!]
\centering
\includegraphics[scale=0.55]{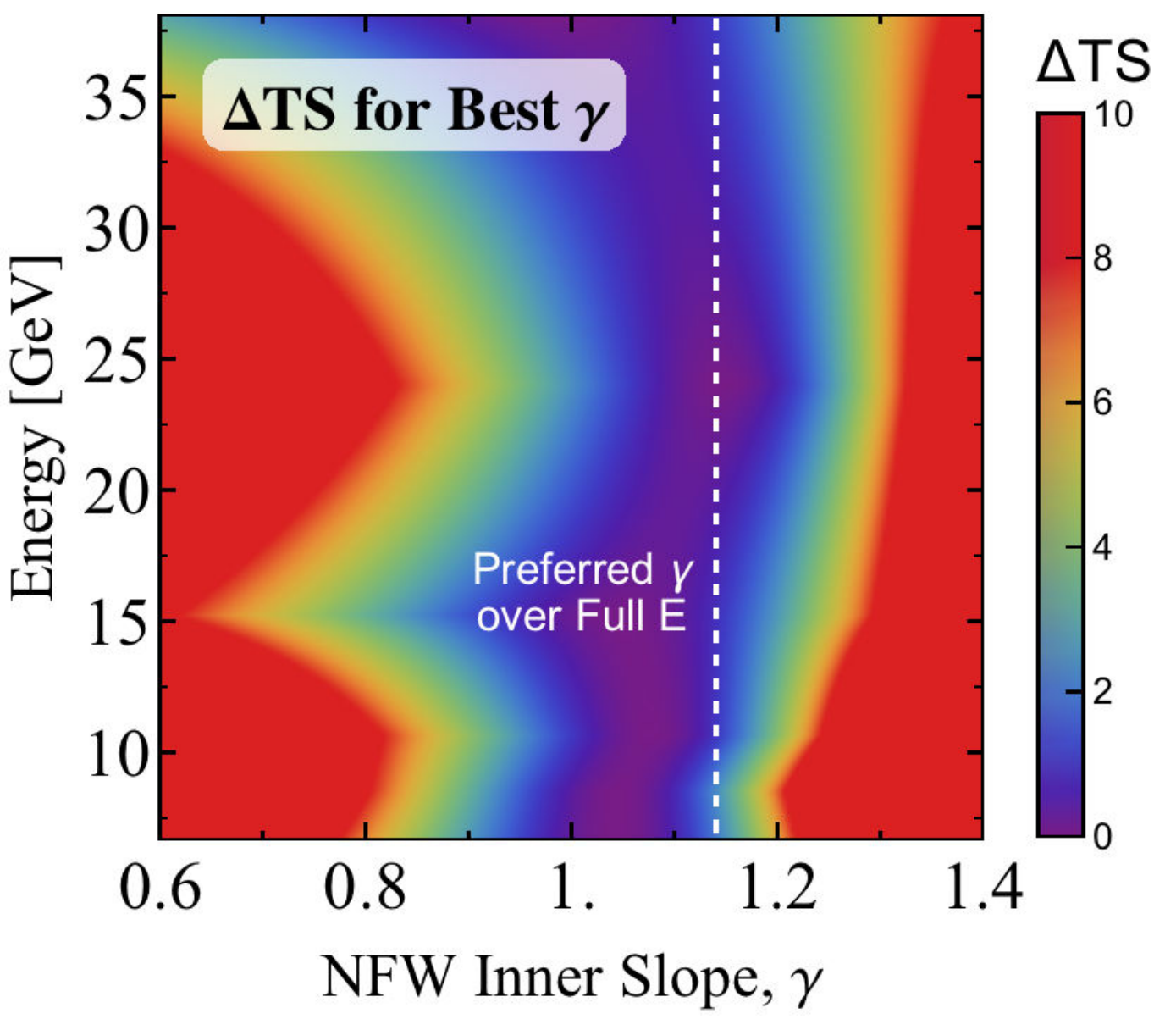} \hspace{0.15in}
\includegraphics[scale=0.55]{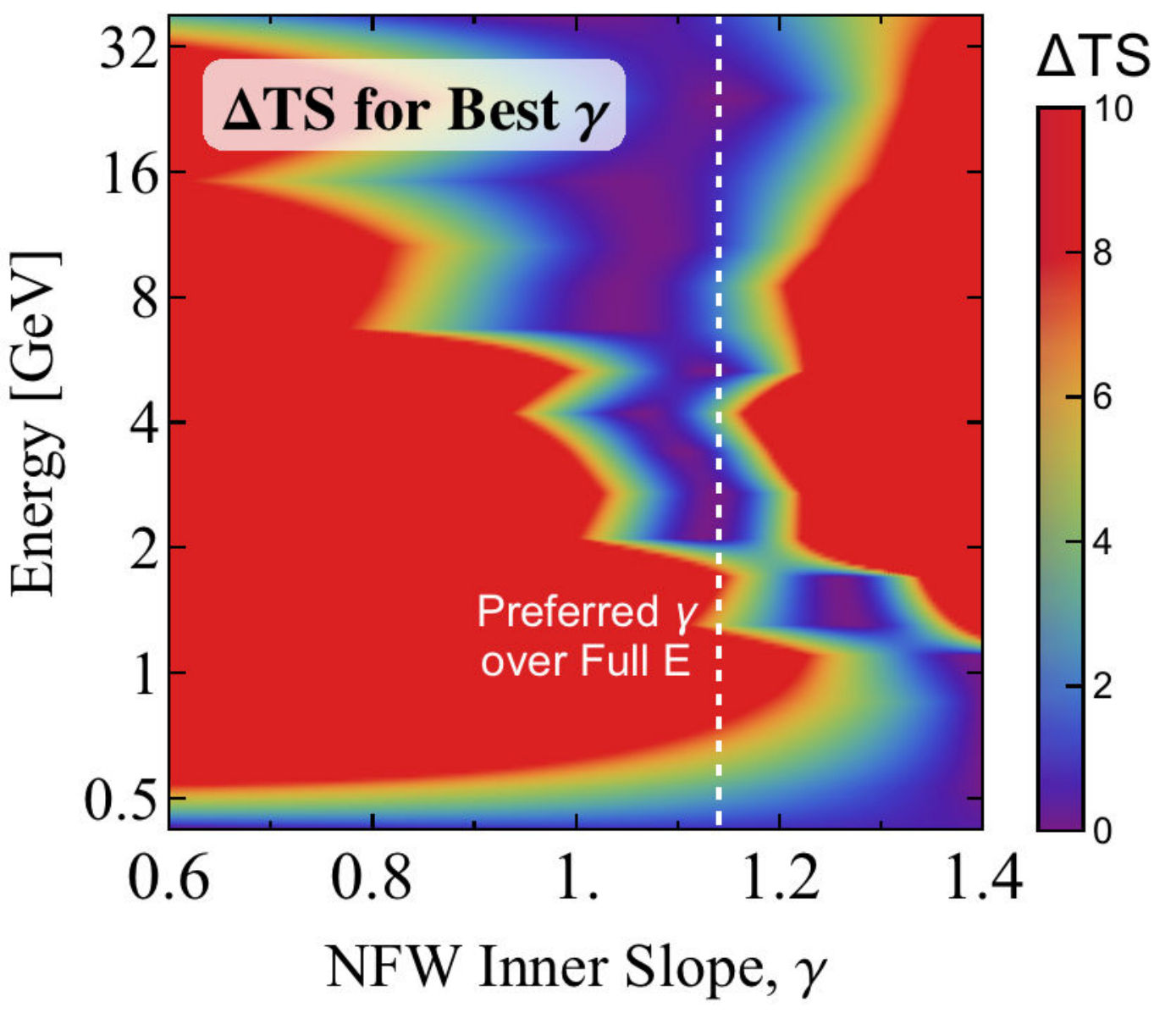}
\caption{\footnotesize{We show the IG preferred variation in $\Delta$TS with energy and inner slope, $\gamma$, of a generalized NFW profile from which we form our GCE template. On the left panel we zoom in to higher energies, and find that the statistical power to discriminate between different values of $\gamma$ is reduced, leading to an opening up of the preferred value. Also in the bin around $15$~GeV we notice an increased preference for lower $\gamma$ values.}}
\label{fig:BestGamma}
\end{figure*}

First we consider the radial variation of the GCE template at high energies by repeating our template analysis for various choices of the inner slope $\gamma$. For each model, we calculate the fit to the data, and thus eventually determining the best fit value of $\gamma$. In Fig.~\ref{fig:BestGamma} we show our best fit results for each individual energy bin in the IG analysis. We first note the greatly reduced statistics at high energies, which decreases the sensitivity of our analysis to the value of $\gamma$. However, in all energy bins above $\sim 2$ GeV, the preferred value of $\gamma$ appears to be statistically consistent (to within $\Delta {\rm TS} \sim 2$) with the globally preferred value that is dominated by emission at low energies. By combining the likelihoods from all $\gamma$-ray energies above $9.5$~GeV, we find that the best fit value of $\gamma$ is $1.08$, and the globally preferred value of $1.14$ differs at the level of only $\Delta {\rm TS}=1.78$.

\begin{figure*}[t!]
\centering
\includegraphics[scale=0.22]{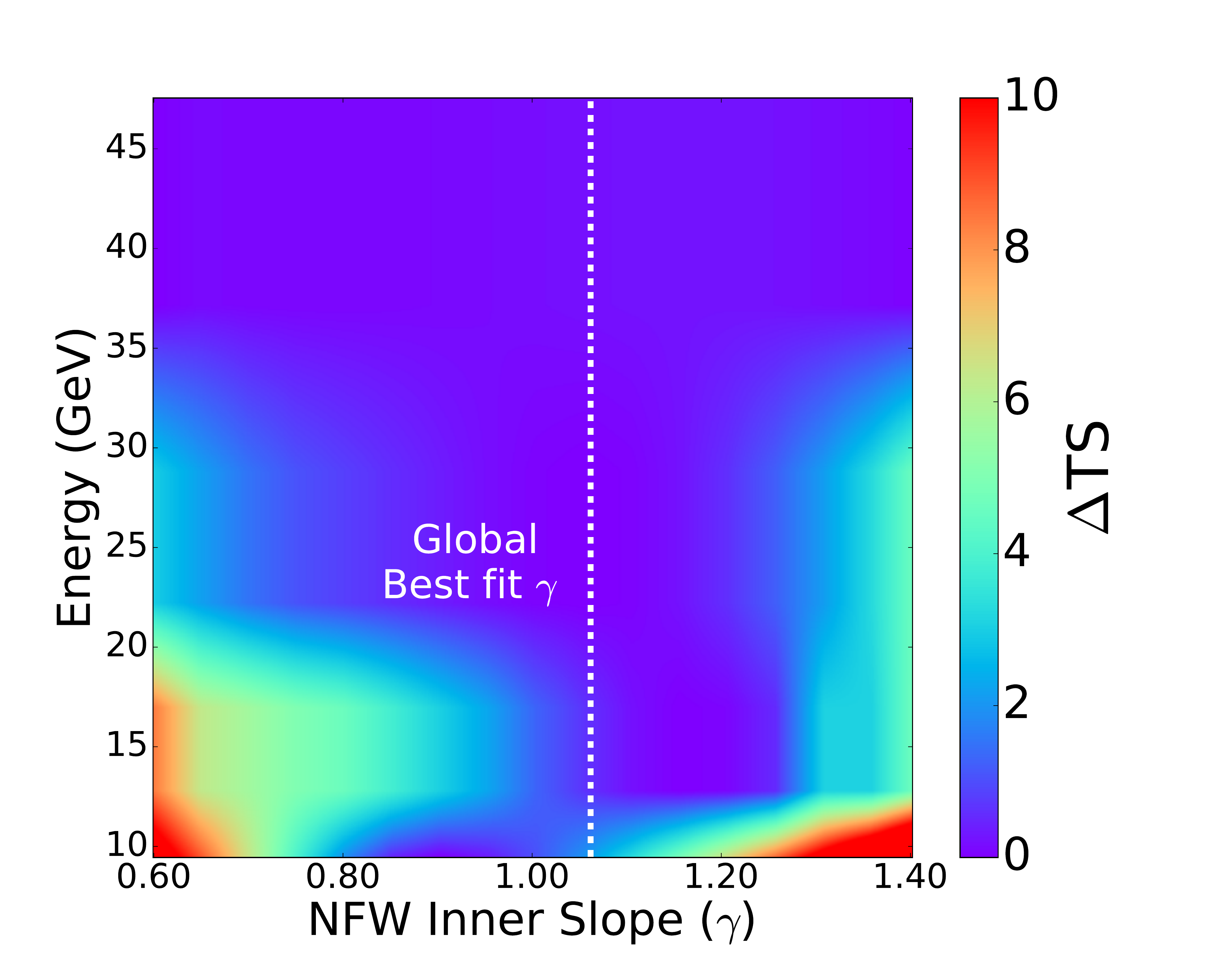} \hspace{0.15in}
\includegraphics[scale=0.22]{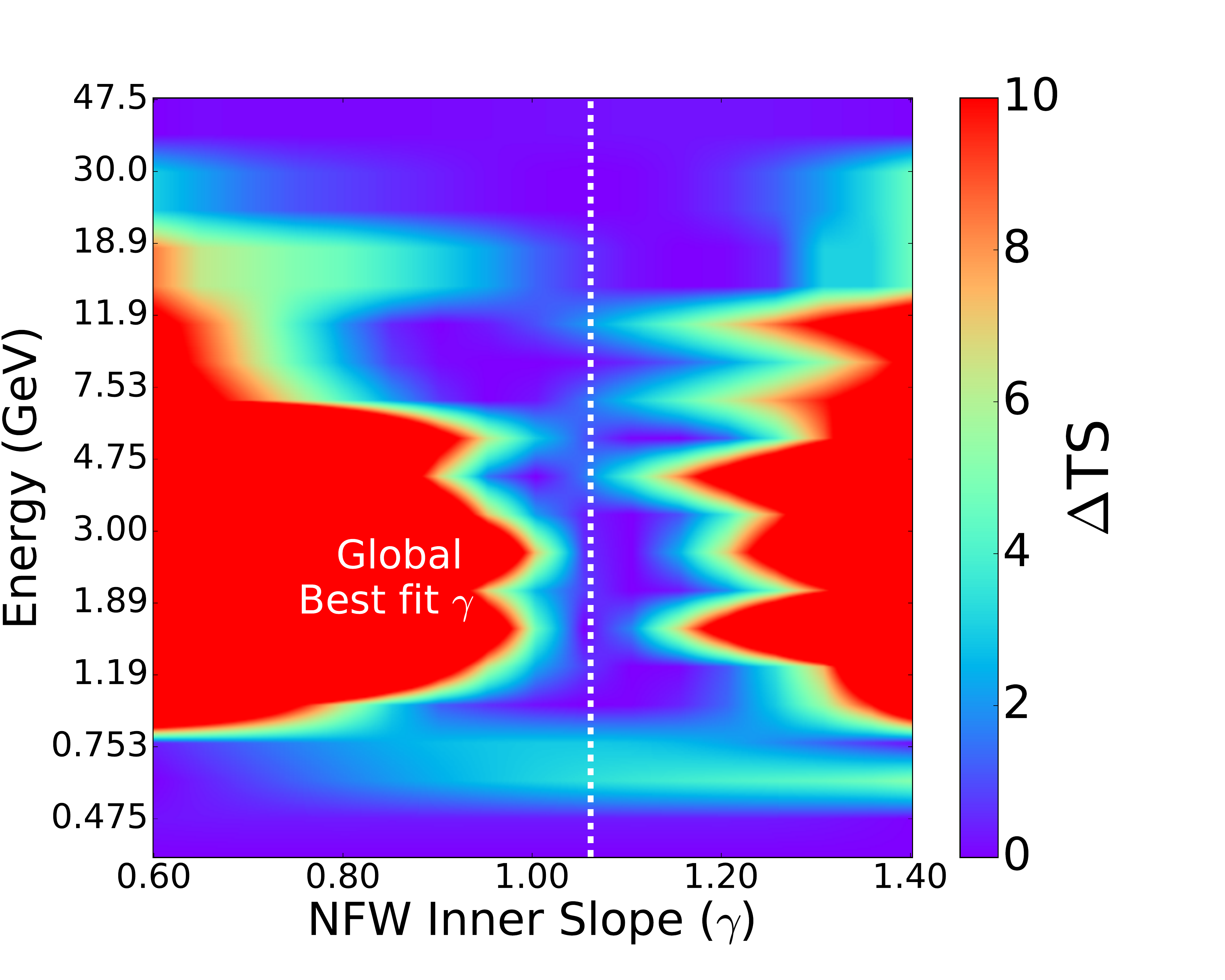}
\caption{\footnotesize{Same as Fig.~\ref{fig:BestGamma} for an analysis of the GC ROI. We find results consistent with the IG analysis, though we note that the relatively weak statistical evidence for a GCE component in the energy range 23-38~GeV makes our constraint on the value of $\gamma$ weak at high energies. The left panel is identical to the right, but shown in only the high energy regime.}}
\label{fig:gc_morphology_default}
\end{figure*}

In Fig.~\ref{fig:gc_morphology_default} we show our best fit values for the inner profile slope in an analysis of the GC ROI. We find our results to be qualitatively similar to those from the IG analysis, despite several quantitative differences. First, the globally preferred value for the GCE component in the GC analysis is $\gamma=1.05$, rather than $\gamma=1.14$ as in the IG analysis, though we note that a global value $\gamma=1.14$ reduces the fit by $\Delta {\rm TS} =24$ (which is small compared to the preference for the excess as a whole). This discrepancy is reasonable, given that the analyses probe different ROIs, and there is no theoretical reason (even in dark matter models) to believe that the GCE component is a constant power-law in regions very close to the GC. In determining whether the high-energy portion of the GC excess differs from the low-energy data, we compare our results to the GC reference value of $1.05$. We note that the two energy bins spanning the range $23-37$~GeV in this analysis are split and combined with higher and lower energy bins (as in the IG analysis). This means that there is a statistically significant detection of the GCE component in every energy bin shown in all morphological plots in the GC analysis.

As in the case of the IG analysis, we find that the results of our analysis are generally consistent with the default value of $\gamma=1.05$, with the exception of the highest energy bin (above $47.5$~GeV) where a value of $\gamma=0.75$ is preferred and the global value is disfavored at $\Delta {\rm TS}=3.71$. However, after stacking all energy bins above an energy of $9.5$~GeV, we find that the best fit value of $\gamma=0.91$, though a value of $\gamma=1.05$ is disfavored at only $\Delta {\rm TS}=1.80$.

\subsubsection{Ellipticity}
\begin{figure*}[t!]
\centering
\begin{tabular}{c}
\includegraphics[scale=0.55]{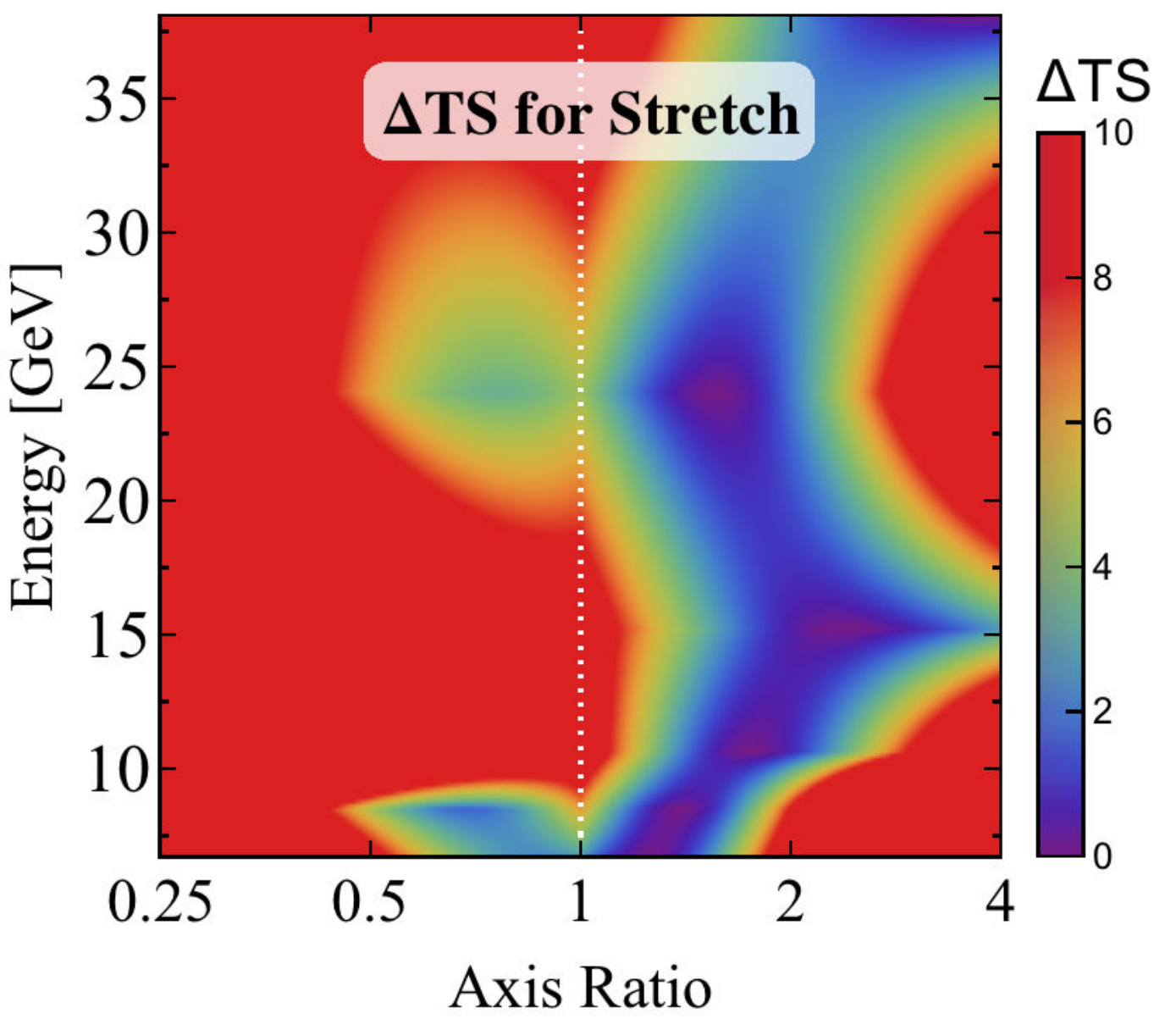} \hspace{0.15in}
\includegraphics[scale=0.55]{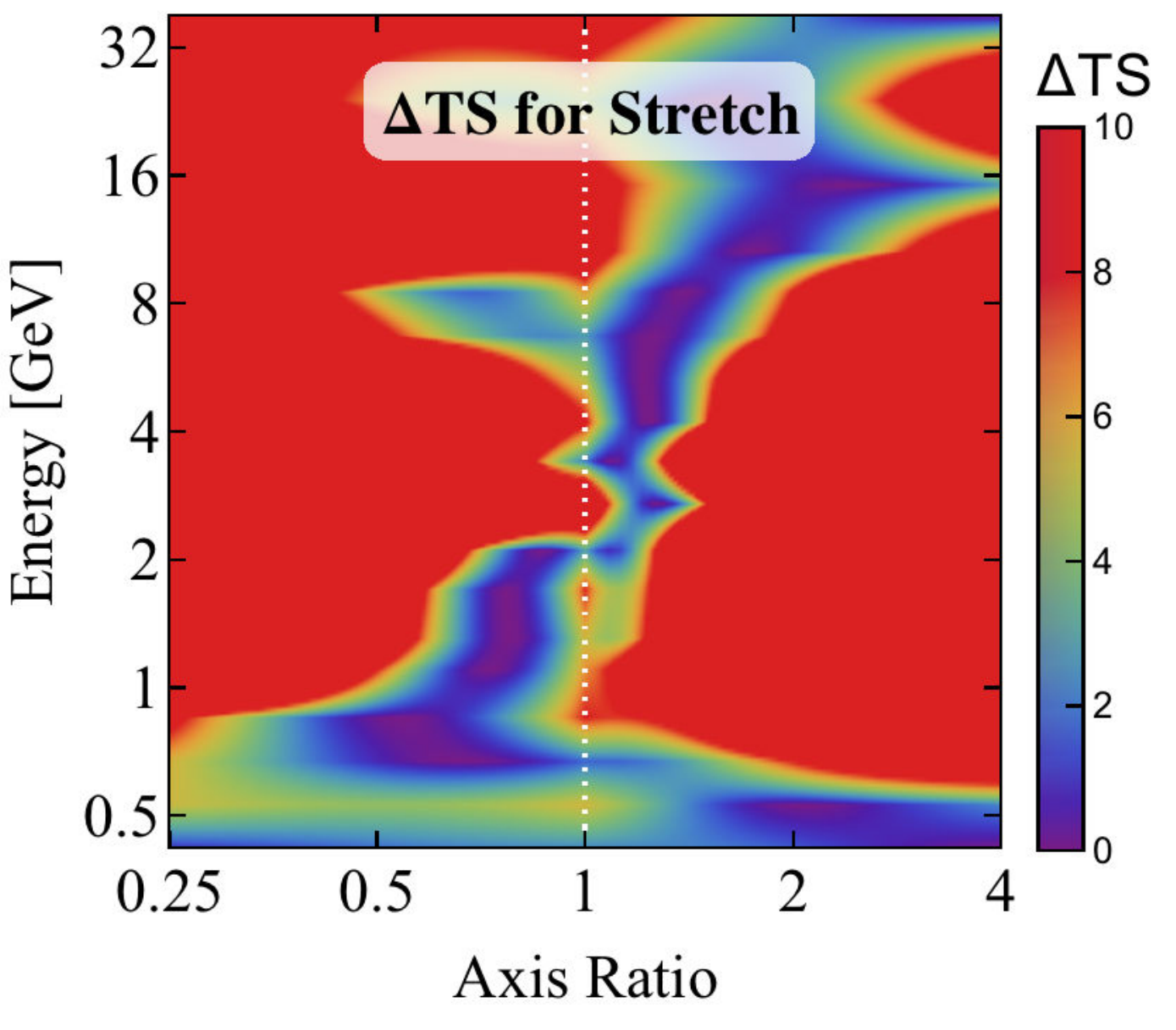}
\end{tabular}
\caption{\footnotesize{Preferred axis ratio as a function of energy in the IG. The axis ratio is defined such that values greater than 1 correspond to a stretch perpendicular to the plane, whilst value less than 1 indicate a stretch along the plane. The left panel is identical to the right, but shown in only the high energy regime.}}
\label{fig:IGsphericity1}
\end{figure*}

\begin{figure*}[t!]
\centering
\begin{tabular}{c}
\includegraphics[scale=0.55]{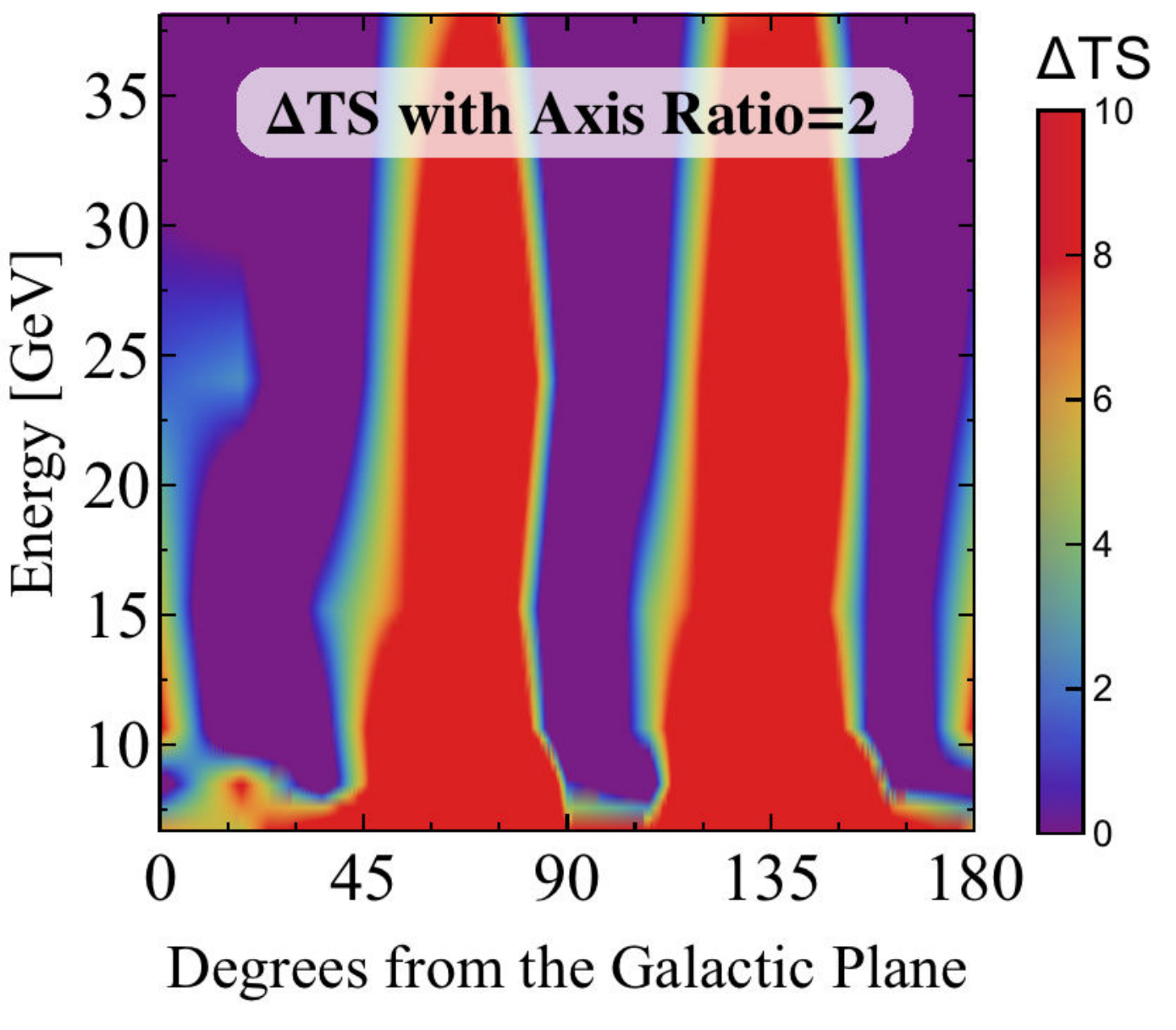} \hspace{0.15in}
\includegraphics[scale=0.55]{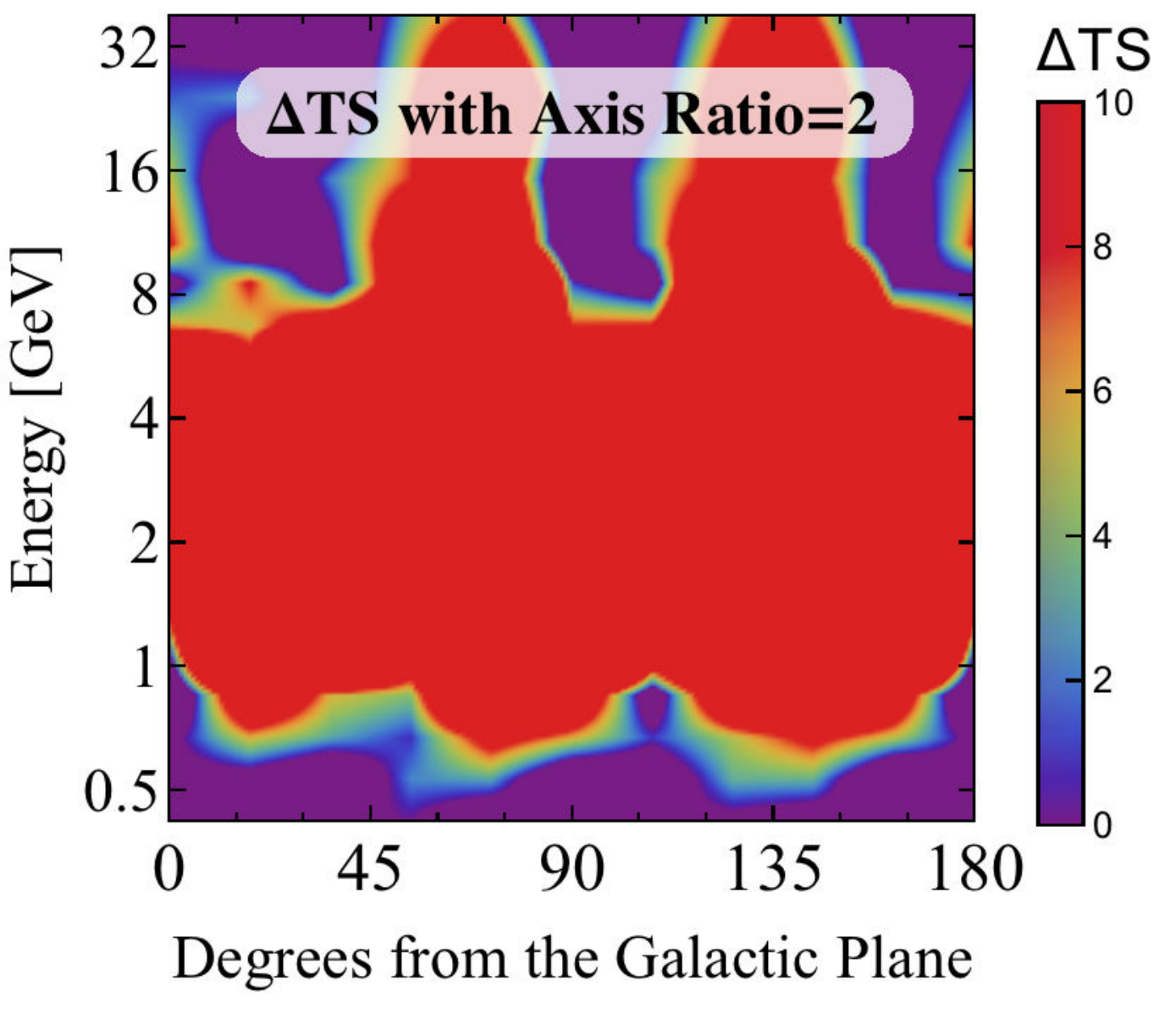}
\end{tabular}
\caption{\footnotesize{For a fixed axis ratio of 2 we show the IG preference for a stretch along various axes for each of the energy bins, as compared to the quality of fit for no stretch at all. The left panel is identical to the right, but shown in only the high energy regime.}}
\label{fig:IGsphericity2}
\end{figure*}

To determine the spherical symmetry of the $\gamma$-ray excess, we repeat our analysis with elliptical versions of the GCE template and calculate the change in the quality of fit to the $\gamma$-ray data with respect to our default, spherically symmetric, GCE model. In our analyses of ellipticity, we constrain our results to GCE templates with $\gamma=1.0$, and examine changes in three relevant parameters: the axis ratio of the major to minor axes, the angle of the major axis with respect to the Galactic plane, and the energy range of our analysis. We note that while $\gamma=1.0$ is not statistically the best fit, this choice has very little effect on the best fit value of the eccentricity distribution, which is relatively independent of $\gamma$. In Fig.~\ref{fig:IGsphericity1} and ~\ref{fig:IGsphericity2} we show two cross-sections of this three-dimensional space in the IG. In the first figure, we show the preference for ellipticity along and perpendicular to the Galactic plane for each energy bin in our analysis. Compared to the inner profile slope $\gamma$, the change in preferred ellipticity is somewhat more pronounced at high energies. Overall, the data above $9.5$~GeV prefers an axis ratio of $1.9$ which is incompatible with the global value of $1.17$ at $\Delta {\rm TS}=20.17$.

Further in Fig.~\ref{fig:IGsphericity2} we instead choose a fixed axis ratio of 2 and show the variation in the quality of fit with respect to a spherical template as a function of  energy and rotation of the elongation axis. Note that the degrees from Galactic plane is for a clockwise rotation from the positive $l$ axis, such that $90^{\circ}$ rotation turns $+l$ into $+b$. We can see that at high energies there are certain directions along which a stretch is preferred, perpendicular to the plane and along shallow angles relative to the plane. This behavior may be due to oversubtraction issues apparent in Fig.~\ref{fig:FullResidual} -- a magnified version of these plots can be seen in Fig.~\ref{fig:ResBin1and2} and will be discussed in App.~\ref{app:15G}. The angles along which a stretch improves the fit generally moves the GCE template away from these regions of oversubtraction, and so this apparent lack of sphericity may be the result of the apparent ``GCE'' emission being comparable to the spatial residuals. Note one of the angles along which the fit is improved, $\sim 35^{\circ}$, was already identified as giving an improved fit in~\citep{Daylan:2014rsa}.

In the GC analysis, we obtain qualitatively similar results, again with some slight quantitative differences. The best fit axis ratio over the full data-set is $1.21$, indicating a slight eccentricity perpendicular to the Galactic plane. However, we find that a spherically symmetric GCE profile is still consistent with the data, providing a fit that is worse by only $\Delta {\rm TS}=0.68$. Interestingly, the two energy bins above $18.9$~GeV have emission that is moderately inconsistent with our best fit global value. The energy bin spanning $18.9-30$~GeV is best fit with an axis ratio of $2.51$, and is inconsistent with the global best fit value at $\Delta {\rm TS}=3.99$. The energy range $30-47.5$~GeV favors an axis ratio of $3.16$, but due to limited statistics is only inconsistent with the global best fit at $\Delta {\rm TS}=0.64$. Thus, the $\gamma$-ray data above $\sim 18$~GeV is inconsistent with our best fitting global axis ratio at a level $\Delta {\rm TS}=4.58$, and, furthermore, is inconsistent with spherical symmetry at the level $\Delta {\rm TS} =6.63$. 

From these results we can see that the excess at high energies does not favor elongation along the plane, but it is much more difficult to rule out elongation in other directions; in particular, elongation perpendicular to the plane appears to be mildly favored by the IG data and the higher-energy GC data. It is natural to hypothesize that this preference for elongation is due to mismodeling of the {\it Fermi} Bubbles. In light of this possibility we test several different templates for the Bubbles in App.~\ref{app:Bkg}, and we find that this trend persists irrespective of the Bubbles templates considered. However, the behavior of the true Bubbles may not be adequately captured by the possibilities we have tested.

\begin{figure*}[t!]
\centering
\includegraphics[scale=0.22]{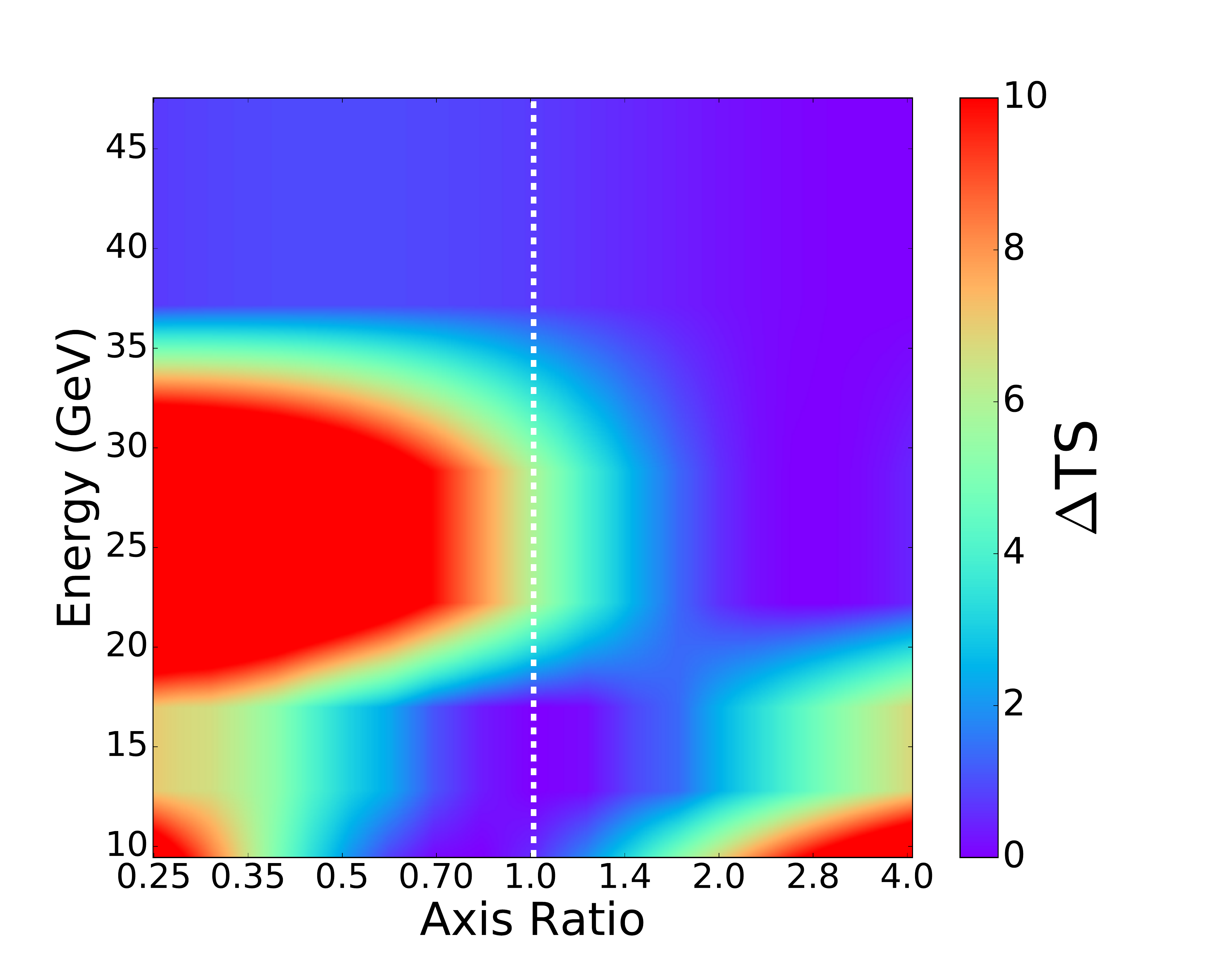}
\hspace{0.05in}
\includegraphics[scale=0.22]{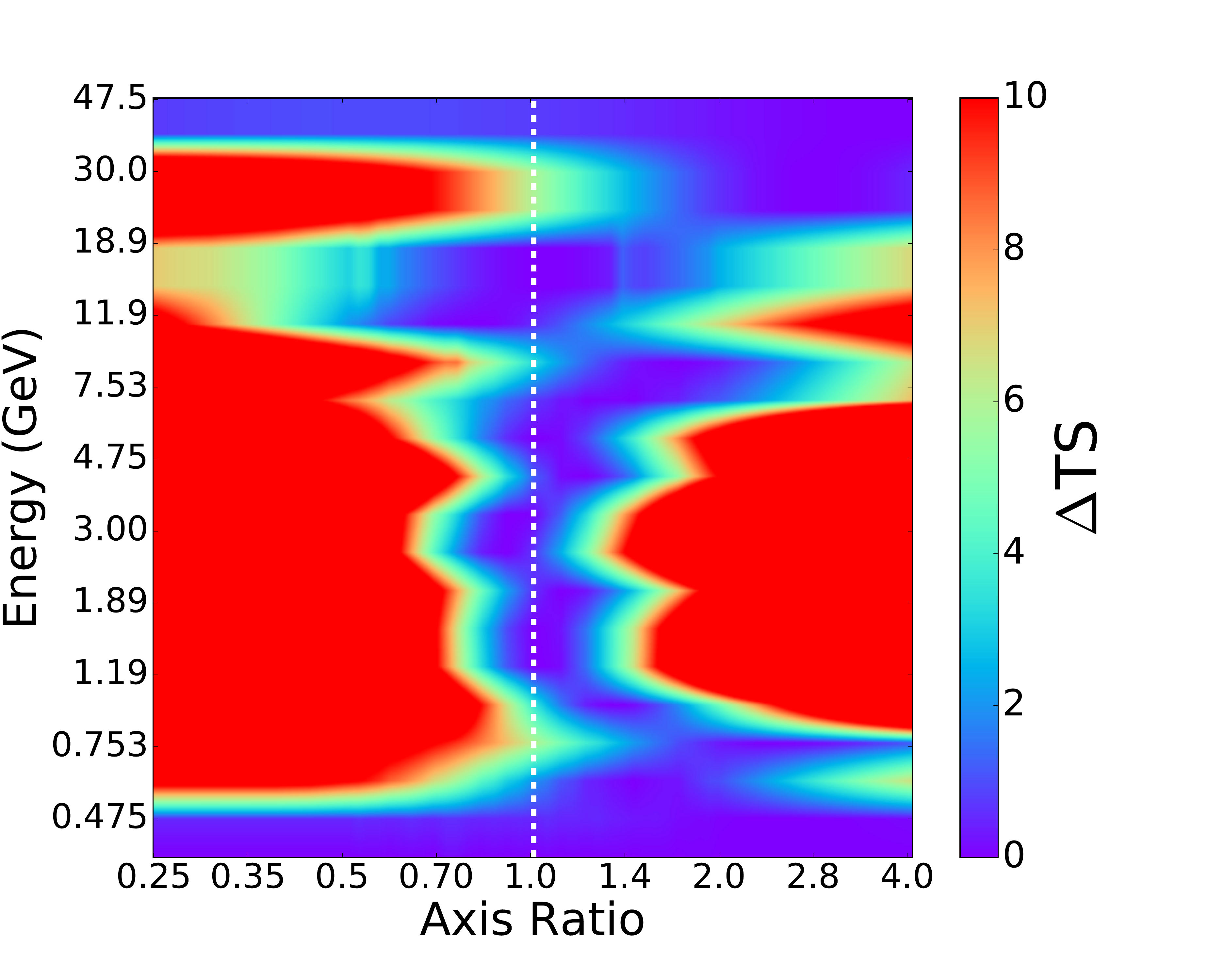} 
\caption{\footnotesize{Same as Fig.~\ref{fig:IGsphericity1} for an analysis of the GC ROI. We again find results consistent with the IG analysis, including a preference for a GCE profile stretched perpendicularly to the Galactic plane at high $\gamma$-ray energies. The left panel is identical to the right, but shown in only the high energy regime.}}
\label{fig:gc_sphericity}
\end{figure*}

\subsubsection{Preferred Center}

\begin{figure*}[t!]
\centering
\includegraphics[scale=0.45]{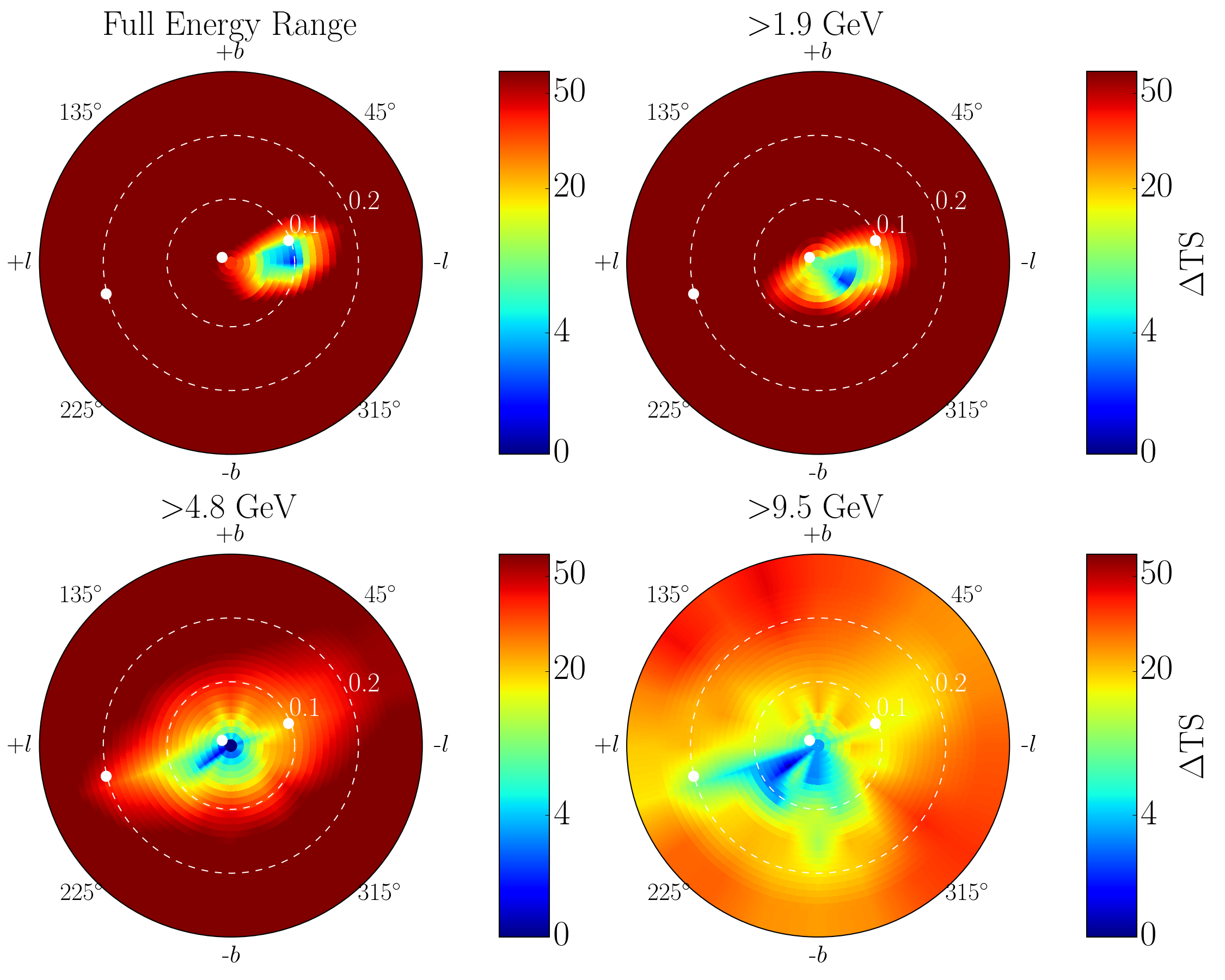} 
\caption{\footnotesize{The best fit position of the GCE template compared to the dynamical center of the Milky Way galaxy, as a function of the \emph{minimum} energy of the GC analysis. A small offset is found, in particular at high energies. However, this result is best understood as demonstrating the degeneracy between the GCE template and the multiple PS degrees of freedom densely clustered around the position of Sgr A*. The three white dots denote the positions of nearby 3FGL point sources (from left to right 3FGL J1746.3-2851c, 3FGL J1745.6-2859c, 3FGL J1745.3-2903c).}}
\label{fig:gc_centering}
\end{figure*}

So far all our results have adopted a GCE template centered on Sgr A*. Here we test this assumption by considering templates centered away from this point. While we attempted this analysis in both the IG and GC ROIs, we found that the IG is unable to constrain the position of the GCE center to a level better than $1^\circ$.\footnote{While our default analysis masks the region with $|b|<1^{\circ}$, we did test the effect of changing our plane mask to $|b|<0.3^{\circ}$; this did not assist in determining a preferred center for the excess.} This is not surprising as the IG analysis masks the GC itself from the analysis. In what follows, we report results for only the GC analysis. 

In Fig.~\ref{fig:gc_centering} we show the best fit position of the GCE compared to the dynamical center of the Milky Way galaxy for different choices of the \emph{minimum} analysis energy in our model (including data from all energy bins above a certain cutoff energy). While we find that the emission is well centered on the position of Sgr A* (to within $0.2^\circ$) at high significance, we find that our full energy analysis prefers a GCE component that is centered on a position approximately $0.1^\circ$ from the Galactic Center, pointed primarily towards negative longitude. This offset is prefered by $\Delta$TS~=~37 compared to an excess template centered on the position of Sgr A*. (We note that we do not quote the 1$\sigma$ statistical error since it is smaller than the bin sized used in our analysis (0.025$^\circ$ radial bins and 20 angular bins) --- thus any error would be based on the interpolation of data between the best fit point and its nearest neighbors.) At higher energies, the residual emission is more coincident with the dynamical center of the Milky Way Galaxy. We note that the GCE component is slightly more offcenter in this analysis than in the previous work of~\citep{Daylan:2014rsa}, where the emission center was confined to within 0.05$^\circ$ of Sgr A*, with a best fit that fell only $0.025^\circ$ away from Sgr A*. We find some evidence that this is due to the inclusion of all source class events (including those that passed through the back of the Fermi-LAT instrument and thus have a bad angular reconstruction). This event selection is well-motivated for investigations into the high-energy excess, but may introduce additional systematic uncertainties very close to the Galactic center and at low $\gamma$-ray energies. In App.~\ref{app:SelCr} we repeat this analysis using only $\gamma$-ray events with the best angular reconstruction, finding that the global fit then prefers a GCE component with a center only $0.05^\circ$ from Sgr A*, and we discuss several explanations for this offset. 

\section{Discussion}
\label{sec:discussion}

As we have shown above, the GCE is best-fit by an emission morphology that is spherically symmetric around the position of Sgr A* and has  an inner profile slope that is slightly adiabatically contracted compared to a standard NFW profile. The most important deviations from this picture are: (1) slight evidence for elongation perpendicular to the Galactic plane in energy bins above 9.5~GeV, and (2) an offset of the GCE profile center from the position of Sgr A* by approximately 0.05$^\circ$ --- 0.1$^\circ$. While the focus of this work is data analysis rather than interpretation, it is worth briefly mentioning the implications of a high-energy GCE for the most frequently suggested $\gamma$-ray emission models.

For dark matter models, it would be quite difficult to explain a change in morphology at high energies, especially for any scenario where the high-energy emission morphology becomes \emph{less} peaked towards the GC. One potential dark matter explanation involves a two-component $\gamma$-ray emission model. For example, the higher-energy emission might stem from prompt photons, while the lower-energy emission from stems from the ICS of electrons produced by DM annihilation~\citep[see e.g.][]{Abazajian:2014hsa}. In this case the electrons would propagate before losing all their energy, and the morphology of the ICS signal would depend on the interstellar radiation field strength and diffusion properties of the medium. However, in general one would expect the profile of the ICS emission to be broader than that of the prompt photons~\citep{Lacroix:2014eea}. Our results hint at the opposite trend, where the high-energy $\gamma$-ray emission prefers a spatial profile that is slightly more extended than emission near the spectral peak.

This morphological preference is quite weak, and one might disregard it. However, one would also generally expect the ICS \emph{spectral} profile to be broader than that of the prompt photon emission. If photons and electrons are produced with similar energies by the annihilation process, then the cooling of the electrons will lead to a steady-state electron population with a broader (less peaked) spectrum than the photons, and furthermore the process of ICS will in general lead to a photon spectrum that is broader and less peaked than the original electron spectrum (since an electron at a fixed energy can scatter photons to a range of different energies). Thus it would be somewhat surprising to obtain a broad, hard prompt photon spectrum combined with a lower-energy peaked excess originating from ICS off a sharply peaked electron spectrum, although individual models may evade this generic argument.

The most natural prediction for DM annihilation is thus that the high-energy excess should share the spatial morphology of the photons from the few-GeV peak of the excess. In the IG analysis, this prediction appears in tension with the preference for elongation perpendicular to the plane at high energies. In the context of a DM origin for the excess, the most natural hypothesis would be that the apparent elongation reflects contamination from mismodeling of one of the other emission components; in particular, as discussed previously and in App.~\ref{app:Bkg}, the shape of the {\it Fermi} Bubbles close to the plane, which is not well-understood. Thus, caution should be used in interpreting the high-energy spectrum of the excess (e.g. Fig.~\ref{fig:BaseSpec}) as originating solely from DM annihilation; obtaining full consistency with the expected spatial distribution likely requires some modification of the background model, and omitting any such modification has the potential to bias the extracted spectrum.

A second possible explanation for the GeV component of the GCE is the emission from a population of $\gamma$-ray pulsars densely clustered in the Galactic bulge. While only a handful of pulsars have currently been observed in the inner kpc of the Galaxy~\citep{Taylor:1993ba}, the population of which is incapable of explaining the $\gamma$-ray excess~\citep{Linden:2015qha}, it is possible that a substantial population of currently undetected pulsars resides in the Galactic center and contributes a significant diffuse $\gamma$-ray flux throughout the inner kpc of the galaxy~\citep{Abazajian:2010zy, Abazajian:2012pn}. Numerous studies have cast doubt on this interpretation by a comparison of the luminosity distribution of $\gamma$-ray pulsars observed in the Galactic plane with the lack of individually detected $\gamma$-ray pulsars near the Galactic center~\citep{Hooper:2013nhl, Cholis:2014noa, Cholis:2014lta, Hooper:2016rap} (see, however, \citep{Petrovic:2014xra, Yuan:2014rca, Brandt:2015ula,  O'Leary:2016osi, O'Leary:2015gfa} for alternative arguments). On the other hand, recent studies of the fluctuations in the $\gamma$-ray data have found significant ``hotspots" consistent with a population of sub-threshold point sources, potentially indicative of a significant pulsar contribution~\citep{Bartels:2015aea, Lee:2014mza}. While significant work remains in assessing the fit of pulsar models to the Galactic center data, it is worth investigating the emission from such a pulsar population at high $\gamma$-ray energies. 

In the case of emission from $\gamma$-ray pulsars, the GCE morphology can be broken down into ``prompt" and ICS components, which may have separate morphologies. Moreover, in the case of $\gamma$-ray pulsars, we expect the ICS emission to be produced at ``higher" energies than the prompt emission --- while {\it Fermi}-LAT observations indicate that pulsars produce the majority of their $\gamma$-ray emission at $\sim 2$~GeV~\citep{TheFermi-LAT:2013ssa}, both models and observations indicate that the e$^+$e$^-$ flux from pulsars may extend to energies $\sim 1$~TeV. Most notably, a hard cosmic-ray injection spectrum ($\alpha$~$\sim$~1.5---1.7, compared to the typical ) is needed for pulsar populations to fit the rising positron fraction observed by PAMELA and AMS-02~\citep{Cholis:2013psa}, although the initial injection spectrum could be softer if local pulsars dominate the positron flux~\citep{Linden:2013mqa}. While many models of the AMS-02 data have concentrated specifically on the e$^+$e$^-$ population from young pulsars, it is not currently known from observations or theoretical arguments whether young or recycled pulsars (or some subpopulation of each) would be most likely to dominate the total e$^+$e$^-$ injection rate.  

Recent work~\citep{O'Leary:2016osi,Yuan:2014yda} has suggested that ICS from the e$^+$e$^-$ pairs sourced by young pulsars could indeed produce a high-energy tail for the GCE. In this paper, we remain agnostic about whether the GCE is powered primarily by young or recycled pulsars, and in both cases employ a cosmic-ray lepton injection model matching fits to the local AMS-02 data~\citep{Cholis:2013psa}, and given by:

\begin{equation}
\frac{dN}{dE} = E^{-1.50} \mathrm{exp}(-E/600~\mathrm{GeV})\,,
\end{equation}

We propagate this injected electron population through the \texttt{GALPROP} cosmic-ray propagation code~\citep{Strong:1998pw} and calculate the resulting ICS spectrum. We choose standard \texttt{GALPROP} parameters throughout our calculation. Because the fraction of the pulsar spindown power that is converted to $\gamma$-rays and e$^+$e$^-$ pairs is uncertain, we allow the relative normalizations of the ``prompt" and ICS pulsar spectra to float arbitrarily in order to produce the best fit to the $\gamma$-ray data. For the prompt spectrum we take the best fit millisecond pulsar model from~\citep{Cholis:2014fja}. The spectrum of this prompt component should be independent of the sky location in our analysis. For the ICS component, the ICS spectrum may shift as a function of sky position, and thus we choose to evaluate the ICS spectrum at a location $5^\circ$ above the GC. This matches the default sky positions chosen throughout the analysis portion of the paper. We note that changes in the morphology of the interstellar radiation field (ISRF), Galactic magnetic field and Galactic diffusion parameters can produce morphological changes in the intensity and spectrum of the ICS signal. Future studies of the high-energy excess could be sensitive to these morphological changes.

\begin{figure}[t!]
\centering
\includegraphics[scale=0.45]{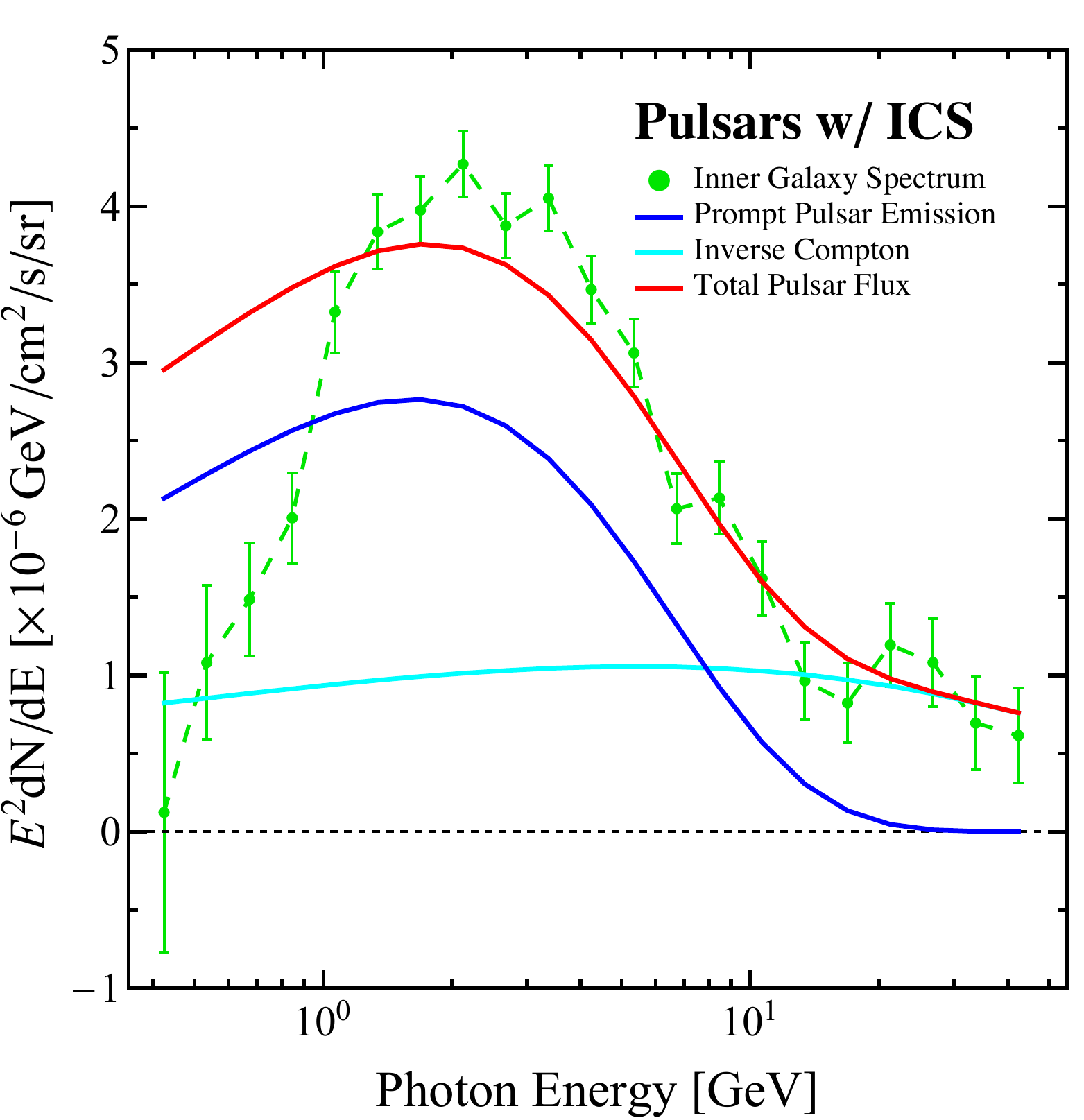}
\caption{\footnotesize{The combined spectrum of a pulsar model that includes both prompt emission from pulsars (with a spectrum following that of~\citep{Cholis:2014fja}), along with a ICS spectrum produced by \texttt{GALPROP} utilizing a cosmic-ray lepton injection model as described in the text. The normalizations of the prompt and ICS components have been adjusted to give the best fit to the data.}}
\label{fig:ics_fit}
\end{figure}

In Fig.~\ref{fig:ics_fit} we show our combined pulsar $\gamma$-ray spectrum compared to the spectra observed in our default IG model. We find that the addition of the ICS template improves the TS of the fit  by $\Delta {\rm TS} = 90.9$. While we note that the this is still not statistically a good fit to the data ($\Delta\chi^2=109.1$ with 19 d.o.f.), we have not marginalized this fit over the multitude of reasonable \texttt{GALPROP} models, and could easily adjust many tunable knobs in order to significantly improve the fit to the $\gamma$-ray data. Finally, we note that the energetics of this component are reasonable, with similar total emission intensities stemming from both the prompt and ICS emission. Since the $\gamma$-ray efficiency of pulsars is typically $\sim 1-5\%$~\citep{TheFermi-LAT:2013ssa}, this would correspond to an electron injection efficiency of $\sim 10-50\%$ with a $10\%$ conversion efficiency of electron energy into ICS. These numbers are reasonable in regions of space with very high ISRF energy density and very low diffusion constants (such as the GC). Thus, we consider this high-energy flux to be a reasonable, and perhaps expected, component in pulsar interpretations of the GCE.

Another potential observable consequence of high-energy ICS from a pulsar population would be a continuation of ``point-source" residuals in the $\gamma$-ray data at high energy. Notably, the electron energy loss time due to the ICS of the ISRF is given by:
\begin{equation}
\tau_{\rm ICS} = 9.8 \times 10^{15} \left(\frac{\rho_{\rm ISRF}}{\rm {eV}/{\rm cm}^3}\right)^{-1} \left(\frac{E}{\rm GeV} \right)^{-1} {\rm s}\,,
\end{equation}
assuming a typical ISRF energy density of $40$~eV~cm$^{-3}$ near the Galactic Center~\cite{Porter:2006tb}, and assuming a typical electron energy of $100$~GeV (to produce a $50$~GeV $\gamma$-ray, using $E_{\gamma, f} = \frac{4}{3}\gamma_{e}^2E_{\gamma, i}$ for $1$~eV starlight), we calculate an average timescale for electron energy losses of $2.45\times 10^{12}$~s. Assuming a $\sim 0.25^\circ$ angular scale corresponding to the grid size of the NPTF template, which corresponds to $\approx 40$~pc at the Galactic Center, we note the diffusion constant must exceed:
\begin{equation}
D_{xx} = \frac{\ell^2}{6\tau} = \frac{(40 {\rm pc})^2}{1.47\times10^{13}~{\rm s}} = 1.1\times10^{27}~{\rm cm}^2~{\rm s}^{-1}\,,
\end{equation}
in order for electrons to diffuse out of a pixel in the NPTF analysis. While this diffusion constant lies approximately a factor of 100 below the nominal diffusion constant in the local Galaxy ($2\times 10^{29}$~cm$^2$~s$^{-1}$ for a $100$~GeV electron), the diffusion parameters of the Galactic Center region are highly unknown.\footnote{However, large changes to the diffusion parameters would also modify the propagation of protons and therefore may measurably impact gas-correlated $\gamma$-ray emission. We thank Ilias Cholis for making this point.} Furthermore, numerous effects may decrease the energy-loss time of electrons in the Galactic Center. For example, if $100$~$\mu$G magnetic fields are present in the region of interest, the electron energy loss time decreases to $3.4 \times 10^{11}$~s, and would decrease further to $6.8\times 10^{10}$~s for $500$~GeV electrons that also contribute to the $50$~GeV $\gamma$-ray signal. Additionally, even if the typical electron propagates farther than $0.25^\circ$, the over-density in ICS emission centered on candidate pulsars may leave a detectable signature on the high-energy $\gamma$-ray sky. Thus, we consider that a solid NPTF detection of point-source emission at high $\gamma$-ray energies to be a highly specific, but not a necessary feature of pulsar contributions to the GCE. However note that if this tail is due to inverse Compton from pulsars, this would be a unique feature only seen in pulsars close to the GC where the ISRF is large. For more nearby pulsars, the lower local values of the ISRF would prevent such a tail from being observed.

\section{Conclusions}
\label{sec:conclusion}

In this paper, we have shown that there is a statistically significant preference for $\gamma$-ray emission steeply peaked toward the GC at energies $>10$~GeV with properties similar to the GCE previously identified at $\mathcal{O}$(GeV) energies. In the Inner Galaxy the formal significance of the excess above 10~GeV is ${\rm TS} \sim 127$. Emission correlated with the GCE template is (statistically) significantly detected up to $\sim 50$~GeV, although above $\sim 30$~GeV its morphology is essentially unconstrained. This component is found with a consistent spectrum in analyses of both the Galactic Center and Inner Galaxy and appears quite robust to changes in the diffuse modeling.

We find mild evidence for an elongation of this high-energy GCE perpendicular to the Galactic plane. When all data above $9.5$~GeV is combined, this high-energy component appears to be centered on the Galactic Center to within $\sim 0.15^\circ$, though a statistically significant offset $\Delta$TS~=~37) is found for an emission profile offset from the GC by $\sim$0.1 degrees. We have demonstrated that for energies above $\sim 5$~GeV the photon statistics, as captured by the non-Poissonian template fit, indicate mild evidence in favor of a point-source explanation of the excess. However when systematic uncertainties are taken into account we cannot reliably distinguish whether this component is diffuse or arises from a population of faint unresolved point sources. With that said, below $\sim 5$~GeV a point-source origin is preferred within the systematic tests we have performed.

While we have focused on data analysis rather than interpretation in this work, we note that if the GeV-scale peak of the excess were due to prompt photon emission from pulsars in the GC region, a high-energy tail could potentially be generated by the ICS of electrons produced by the spindown of those same pulsars. If DM annihilations were responsible for the full excess, the mass of the DM particle must be sufficiently high to produce $\gamma$-rays at energies up to $\sim 50$~GeV -- disfavoring very light models of DM. 

The slight evidence for elongation perpendicular to the Galactic plane also suggests a possible association with or contamination by the {\it Fermi} Bubbles, which may have presently unmodeled features close to the Galactic Center. We have verified that changing the modeling of the Bubbles does not severely impact our results; however, it is possible the true spatial distribution of photons from the Bubbles does not lie anywhere in the space probed by our models. If a mechanism associated with the Bubbles is responsible for the bulk of the high-energy emission we observe, it would need to yield a signal centered on and peaked toward the Galactic Center. It is worth noting that in the context of DM interpretations of the GCE, a contamination of the high-energy $\gamma$-ray emission by the Fermi Bubbles is well-motivated, as it is difficult for DM models to produce a $\gamma$-ray morphology that is more elliptical at higher $\gamma$-ray energies. Future studies which theoretically motivate a morphological model for the Bubbles spectrum near the GC are thus imperative to resolving this possible degeneracy.

\section*{Acknowledgements}

The authors thank JoAnne Hewett for inspiring this study by pointing out the potential importance of high-energy data for interpretation of the excess. We are grateful to Dan Hooper and Manuel Meyer for comments that greatly improved the quality of this manuscript, as well as Meng Su, Ilias Cholis and Eric Carlson for providing several of the diffuse and Bubbles emission templates utilized throughout this work. We further thank Ilias Cholis and Hongwan Liu for a careful reading of the manuscript and very helpful comments. NLR thanks Lina Necib for helpful discussions. TL is supported by the National Aeronautics and Space Administration through Einstein Postdoctoral Fellowship Award Number PF3-140110. NLR is supported by the American Australian Association’s ConocoPhillips Fellowship. BRS is supported by a Pappalardo Fellowship in Physics at MIT. This work was supported by the U.S. Department of Energy under grant Contract Numbers DE$-$SC00012567 and DE$-$SC0013999. This work also made use of computing resources and support provided by the Research Computing Center at the University of Chicago.

\newpage

\appendix

\section{Stability Under Variations to the Background Modeling}
\label{app:Bkg}

In this appendix we discuss the dependence of our results on the choice of diffuse background model and PS model, as well as other more subtle variations in the background modeling. As outlined in the main text we consider seventeen different models outlined in the next section. 

Here and in subsequent appendices, we use variation in the preferred inner slope value $\gamma$ -- determined over the full energy range -- as an indicator of stability. We have crossed checked that when the inner slope is stable, the conclusions of our high-energy analysis are generically unchanged. The full energy range is chosen to maximise statistics, thereby emphasising systematic variations. With this choice, the $1\sigma$ statistical uncertainties on the preferred $\gamma$ values are $\sim 0.01$ or smaller. As this is much smaller than many of the systematic effects considered we usually show values without associated statistical uncertainties.

\subsection{Employing Different Galactic Diffuse Models}

\subsubsection{Inner Galaxy}

As mentioned in the main text, we consider 17 different diffuse models in this work, which can be naturally divided into two sets. The main gamma-ray processes described by these diffuse background models are $\pi^0$ decay, ICS and bremsstrahlung. In one set of diffuse models we combine these physical processes together into a single diffuse template and then let the normalization of this template float independently in each energy bin, while in the other set we combine only the $\pi^0$ and bremsstrahlung emission, letting the ICS float independently. We will describe each of these cases in turn.

The first set includes three official LAT background models provided by the {\it Fermi} team: \texttt{gll\_iem\_v02\_P6\_V11\_DIFFUSE} (\texttt{p6v11}), \texttt{gal\_2yearp7v6\_v0} (\texttt{p7v6}) and the recently released \texttt{gll\_iem\_v06} (\texttt{p8v6}). Note we did not consider the Pass 7 Reprocessed model (\texttt{gll\_iem\_v05\_rev1}), as by construction it includes any large-scale residuals between the underlying physical model and the data, making it inappropriate for studying an extended emission component like the GCE. This issue also exists for the \texttt{p8v6} diffuse model, as discussed in the main text. We examine the \texttt{p8v6} diffuse emission model since it is the only official Pass 8 model available at this time, but caution that the suppression of the GCE is expected in this case and our results are unlikely to have a physical interpretation. The \texttt{p7v6} diffuse model suffers from a similar but less acute problem, as it has had the large scale structures of the {\it Fermi} Bubbles added as a fixed component. We prefer to float the Bubbles independently in our fits, so we employ the \texttt{p6v11} model as our default for the IG analysis (where the Bubbles contribute significantly, unlike in the GC ROI), although in any of the ROIs considered the \texttt{p8v6} and then \texttt{p7v6} models gave better quality fits.

In addition to the official {\it Fermi} models, we also consider 14 of the \texttt{GALPROP} models used in~\citep{Calore:2014xka}. The models we used were referred to in that reference as Model A, and F-R, a naming scheme we follow here. Models F-R were taken from~\citep{Ackermann:2012pya}, where they were given different names,\footnote{For example Model F was referred to as $^S{\rm L}^Z6^R20^T100000^C5$. See~\citep{Calore:2014xka} for details concerning each model.} while Model A was created using \texttt{GALPROP v54}~\citep{Strong:1998fr}.\footnote{We thank Ilias Cholis for providing us with the galdef file for this model, and Eric Carlson for providing the version used in the Galactic Center analysis.} For these models each of the three diffuse components can be fit independently. However, given that both the pion and bremsstrahlung maps trace the gas, as in~\citep{Calore:2014xka} we choose to combine these and float them independently of the ICS component.

\begin{figure*}[t!]
\centering
\begin{tabular}{c}
\includegraphics[scale=0.45]{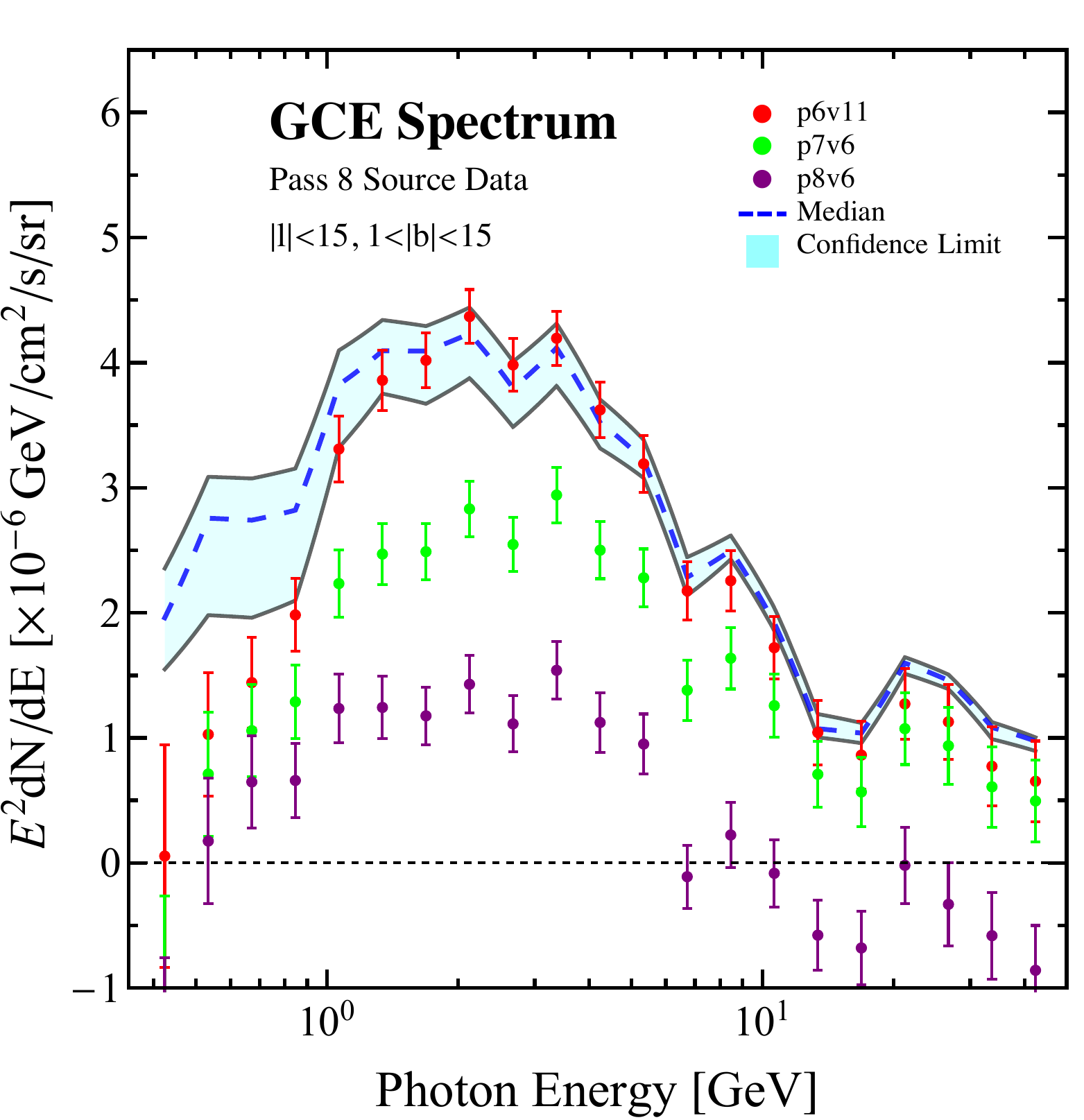} \hspace{0.15in}
\includegraphics[scale=0.45]{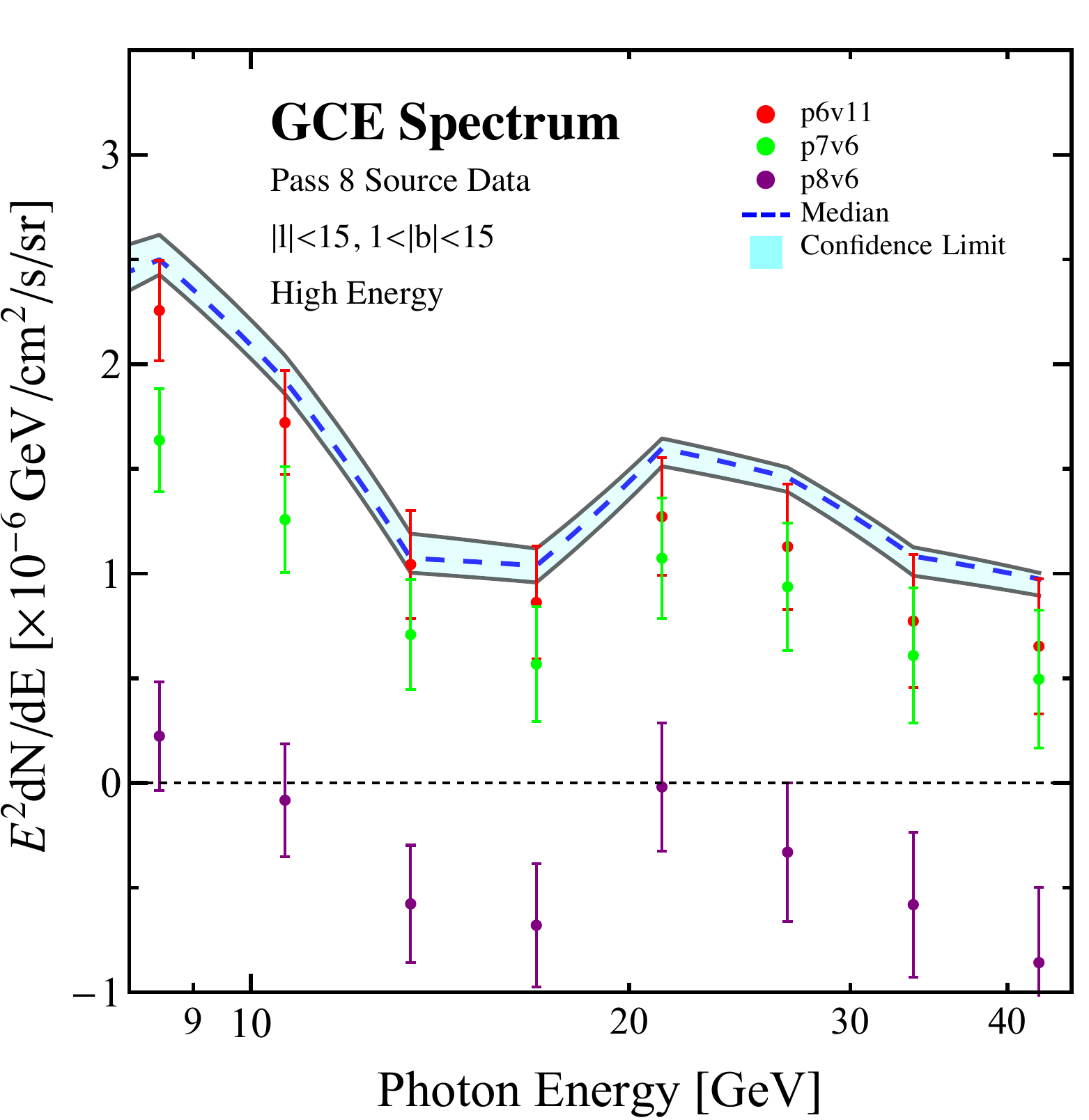}
\end{tabular}
\caption{\footnotesize{Spectrum obtained by a GCE template with $\gamma=1.0$ in our default ROI for the IG, performed for the 17 different diffuse models. We explicitly show the spectrum extracted for the \texttt{p6v11}, \texttt{p7v6} and \texttt{p8v6} models, while we use 14 of the diffuse models taken from~\citep{Calore:2014xka} to form the median and the 16\% and 84\% percentiles as confidence limits.  We show the full energy range (left) as well as the high-energy region (right). See text for details.}}
\label{fig:AllCoeff}
\end{figure*}

Using these background models, and following our default IG analysis procedure, but holding $\gamma=1.0$ fixed for the GCE, we can look at the spectrum obtained over the full and high-energy region, with the result shown in Fig.~\ref{fig:AllCoeff}. In that plot we explicitly show the resulting GCE spectrum for fits using all the {\it Fermi} models, while for Models A and F-R we show the mean and 68\% confidence limits on this spectrum, based on the 16\% and 84\% percentiles. The results using our default model, \texttt{p6v11}, are clearly consistent with expectations from the sample of \texttt{GALPROP} models. The primary reason the other {\it Fermi} diffuse models differ concerns the other components added into these data-driven diffuse models, as discussed above. An important observation is that the full and high-energy spectrum is robust against variations within the space of \texttt{p6v11} and \texttt{GALPROP} models considered, including features like the dip between $12-20$~GeV (to be discussed in App.~\ref{app:15G}) and resurgence at high energies. The only differences are observed at lower energies, which is likely related to the large PS mask applied for those energies in the Source-class data, a point we return to in App.~\ref{app:SelCr}.

To further quantify the impact of varying the diffuse model, we also consider how the preferred value of $\gamma$ varies between the different models. We show the best fit values of $\gamma$ for several models in Table~\ref{table:GamValDiff}. We see that the $\gamma$ values extracted for \texttt{p6v11}, \texttt{p7v6} and Model A are similar, while \texttt{p8v6} is higher and Model F is lower. Note that the values for \texttt{p7v6} and \texttt{p8v6} should be interpreted in light of the issues these models have in analyses of the GCE, given the extra internal components they include. Furthermore the $\Delta$TS of the GCE component as a function of $\gamma$ is relatively flat in \texttt{p8v6}, so the difference between this high value of $\gamma$ and values closer to other models is not particularly significant. As for Model F, note that whilst in Table~\ref{table:GamValDiff} it looks like an outlier, in the context of all the models considered, it is actually representative of the values coming from many \texttt{GALPROP} models. Despite this, the extracted value of $\gamma$ is generically stable to a value of about $\Delta \gamma \sim 0.1$ and this is small enough that within this range the qualitative conclusions of the paper are unchanged. 

In Table~\ref{table:GamValDiff}, we also compare the variation in $\gamma$ in a higher-quality photon dataset (UCV BestPSF), with better angular resolution and cosmic-ray rejection, but lower statistics (we will expand on this comparison in App.~\ref{app:SelCr}). In this case the spread in $\gamma$ between different Galactic diffuse models is somewhat reduced (especially when we exclude the \texttt{p8v6} model, since as discussed above we do not believe the suppression of the excess in this model is physical). We believe this behavior can be largely attributed to the larger point-source mask in All Source data; at low energies this mask can remove a substantial fraction of the ROI, and it preferentially removes regions where the excess is brightest (toward the Galactic Center), which likely renders $\gamma$ more sensitive to small changes in the Galactic diffuse modeling. Most of the photons in the excess are at relatively low $\gamma$-ray energies, so in a global analysis where the spatial morphology is assumed to be energy-independent, the best-fit $\gamma$ will be largely determined by the low-energy data.

\begin{table}[h]
\begin{center}
\begin{tabular}{| c || c | c |}
\hline
& \multicolumn{2}{ |c| }{Preferred $\gamma$} \\ \cline{2-3}
Model & All Source & UCV BestPSF \\ \hline \hline
\texttt{p6v11} & 1.14 & 1.08 \\ \hline
\texttt{p7v6} & 1.16 & 1.15 \\ \hline
\texttt{p8v6} & 1.25 & 1.24 \\ \hline
Model A & 1.13 & 1.15 \\ \hline
Model F & 1.04 & 1.09 \\ \hline
\end{tabular}
\end{center}
\caption{The preferred NFW inner slope value, $\gamma$, for the GCE template in the default IG analysis using 5 different diffuse models and two different Pass 8 datasets. The $1 \sigma$ statistical uncertainties are $\sim 0.01$ or less and are omitted. See text for details.}
\label{table:GamValDiff}
\end{table}

\begin{figure*}[t!]
\centering
\begin{tabular}{c}
\includegraphics[scale=0.55]{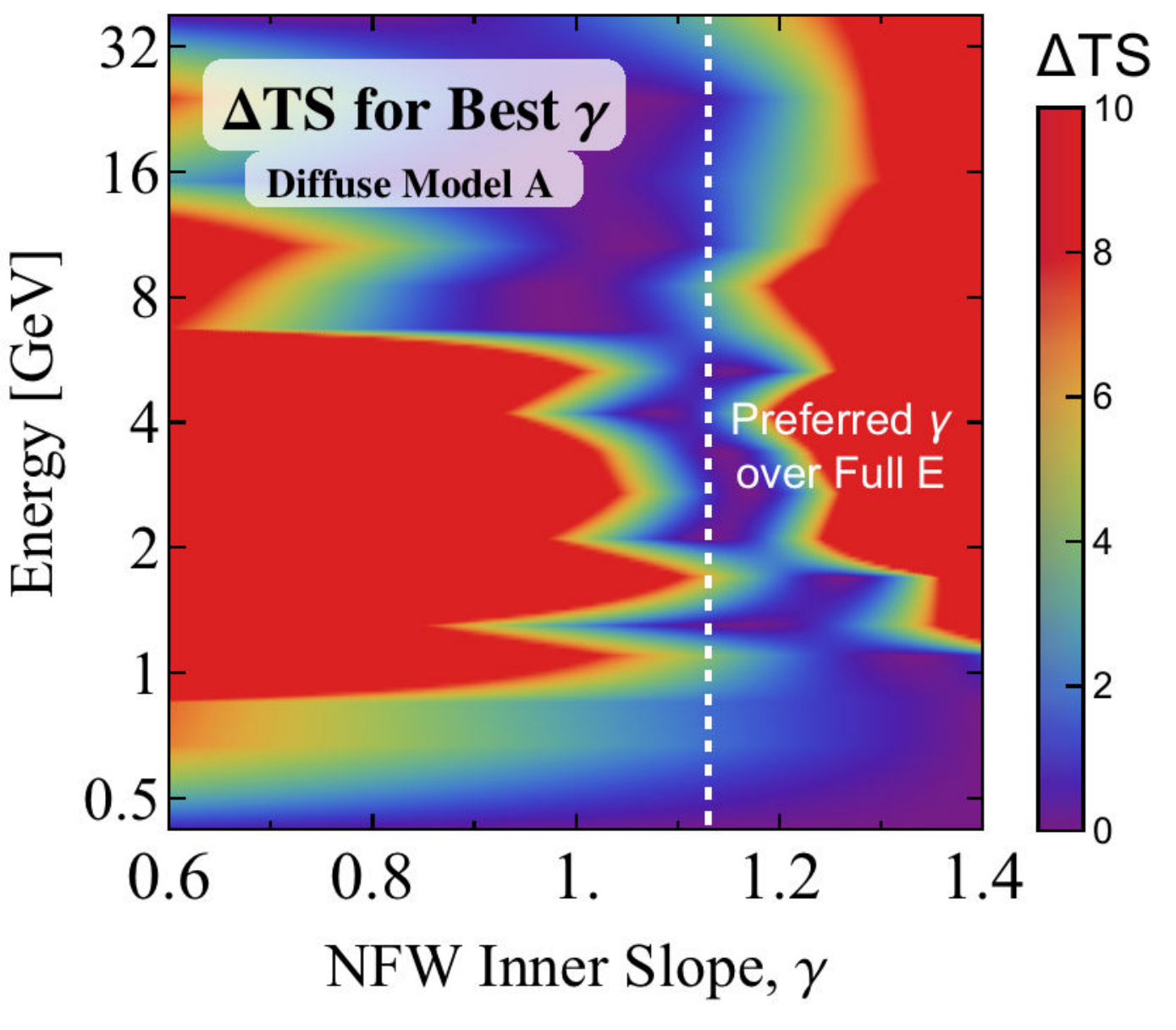} \hspace{0.15in}
\includegraphics[scale=0.55]{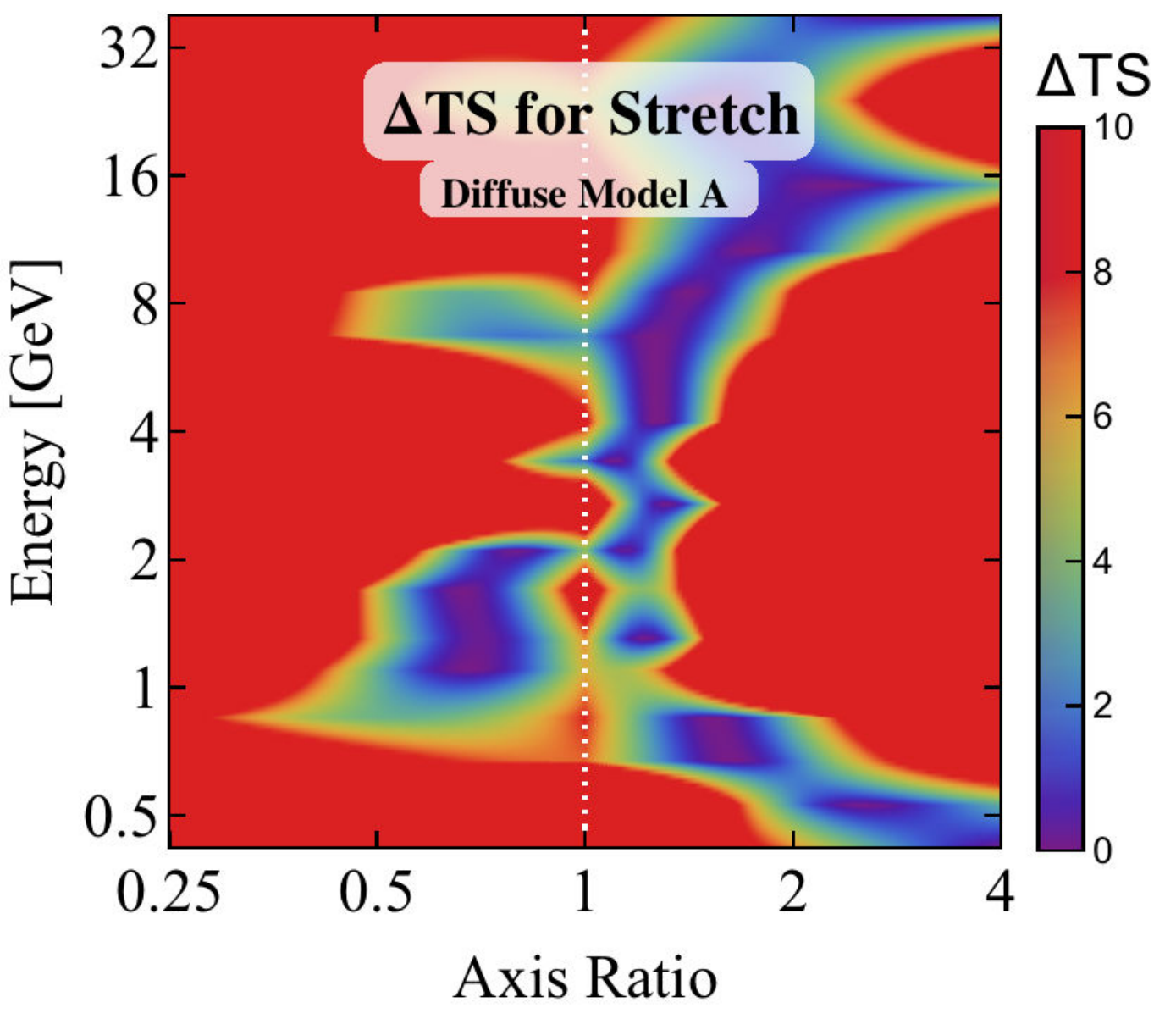}
\end{tabular}
\caption{\footnotesize{The bin-by-bin preferred NFW inner slope $\gamma$ (left) and axis ratio (right), repeating the default IG analysis using the \texttt{GALPROP}-based Model A, rather than \texttt{p6v11}. For the inner slope, the globally preferred value of $1.13$ is close to our default value of $1.14$. Further we see that at high energies the fit again prefers a flatter profile. In terms of the axis ratio, as for our default analysis we see a general preference for a stretch perpendicular to the plane, with a greater stretch being preferred in the high-energy regime.}}
\label{fig:ModAGam}
\end{figure*}

\begin{figure*}[t!]
\centering
\begin{tabular}{c}
\includegraphics[scale=0.55]{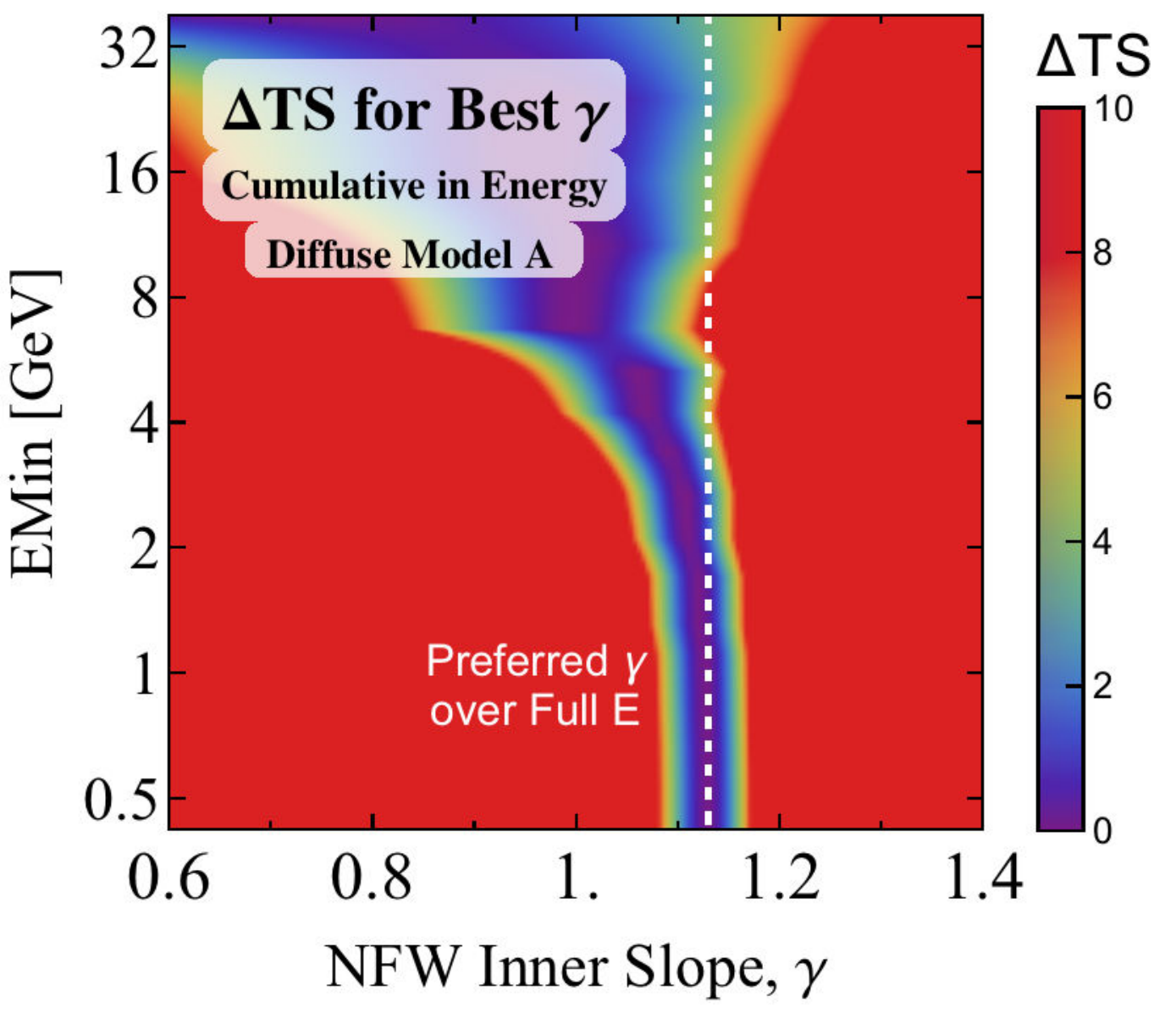} \hspace{0.15in}
\includegraphics[scale=0.55]{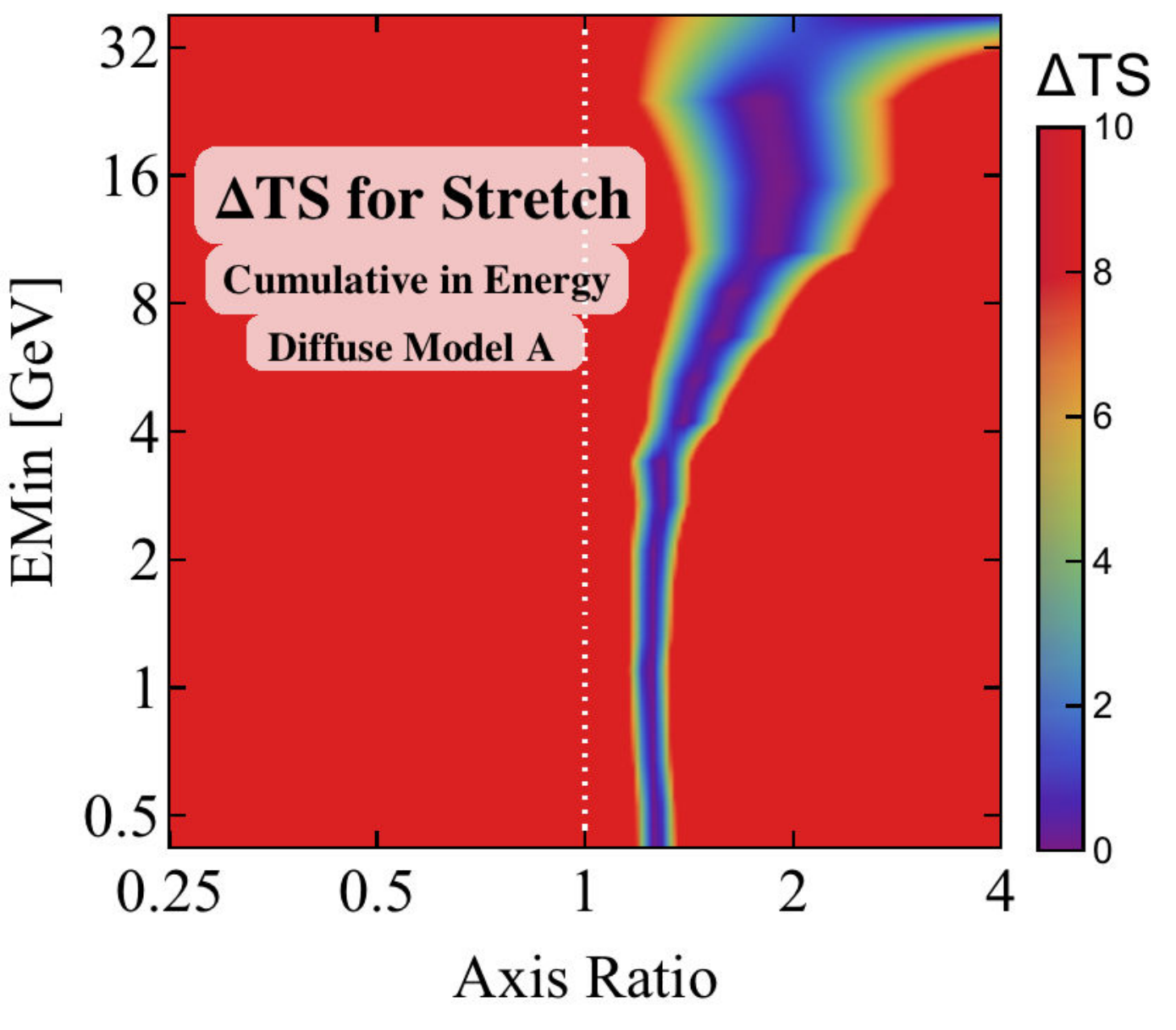}
\end{tabular}
\caption{\footnotesize{As for Fig.~\ref{fig:ModAGam} but showing results cumulative in energy.}}
\label{fig:ModAAxis}
\end{figure*}

Finally, we consider in detail the impact on our radial variation and ellipticity analysis if we replace our default diffuse model \texttt{p6v11} with the \texttt{GALPROP}-based Model A in Fig.~\ref{fig:ModAGam} and \ref{fig:ModAAxis}. For the radial variation, as seen in Table~\ref{table:GamValDiff}, the globally preferred value is $1.13$ -- close to our default value of $1.14$, which is only disfavored at $\Delta {\rm TS} = 0.5$. As in the default analysis, above $10$~GeV, the fit prefers a flatter profile, with a best fit value here of $\gamma=1.01$. The global best fit value is disfavored by $\Delta {\rm TS} = 4.6$. For the ellipticity, over all energies the preferred axis ratio is $1.25$ -- a stretch perpendicular to the plane, as in our default analysis. The value differs from the preferred \texttt{p6v11} value of $1.17$ by $\Delta {\rm TS} = 11.6$. At high energies, the fit prefers a profile stretched even further, with an axis ratio of $1.85$ -- the global value differing by $\Delta {\rm TS} = 15.1$. Again this behavior mirrors our default analysis and we see that whilst the specifics can change, the qualitative features we presented in the main text are unaltered by the move to Model A. 

Note comparing the right of Fig.~\ref{fig:ModAGam} and the right of Fig.~\ref{fig:IGsphericity1} we see that at low energies the preferred stretch is quite different, the fits prefer an axis ratio greater than one and less than one respectively. We emphasize, however, that the fit at low energies, where the spectrum drops off, is unlikely to be related to the GCE. Diffuse mismodeling likely plays a larger role and our results in this regime likely highlight the difference between these diffuse models.

\subsubsection{Galactic Center}

We can perform a similar exercise in the GC ROI. Here, due to the computational complexity of the GC analysis, we examine only the results from Model A in~\citep{Calore:2014xka}. Because Model A does not include an emission component tracing the {\it Fermi} Bubbles (contrary to the default p7v6 diffuse emission model), we add a Bubbles component identical to the default choice in the IG analysis. We will investigate alterations to this Bubbles template in the GC analysis later in this section.

\begin{figure*}[t!]
\centering
\begin{tabular}{c}
\includegraphics[scale=0.22]{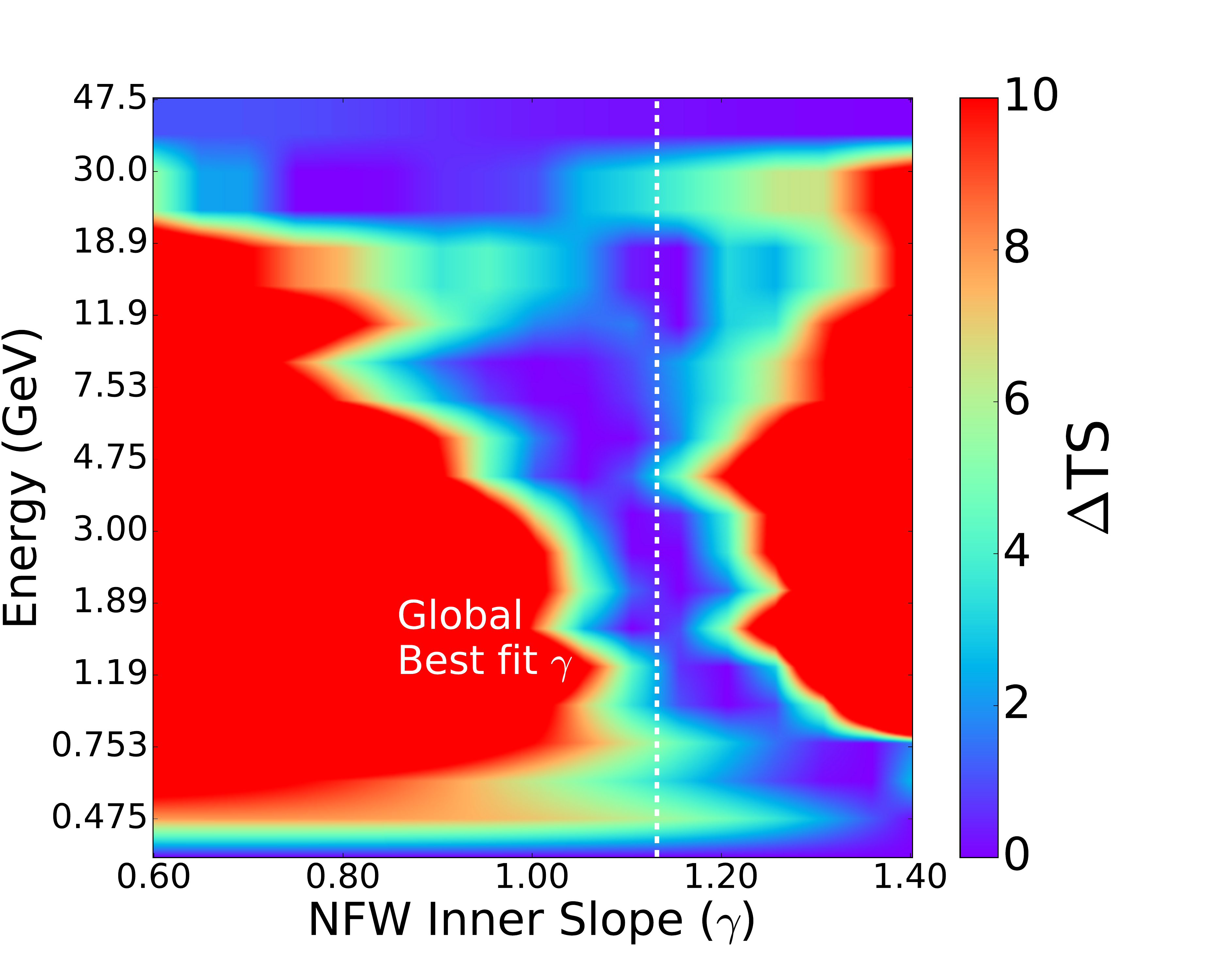} \hspace{0.15in}
\includegraphics[scale=0.22]{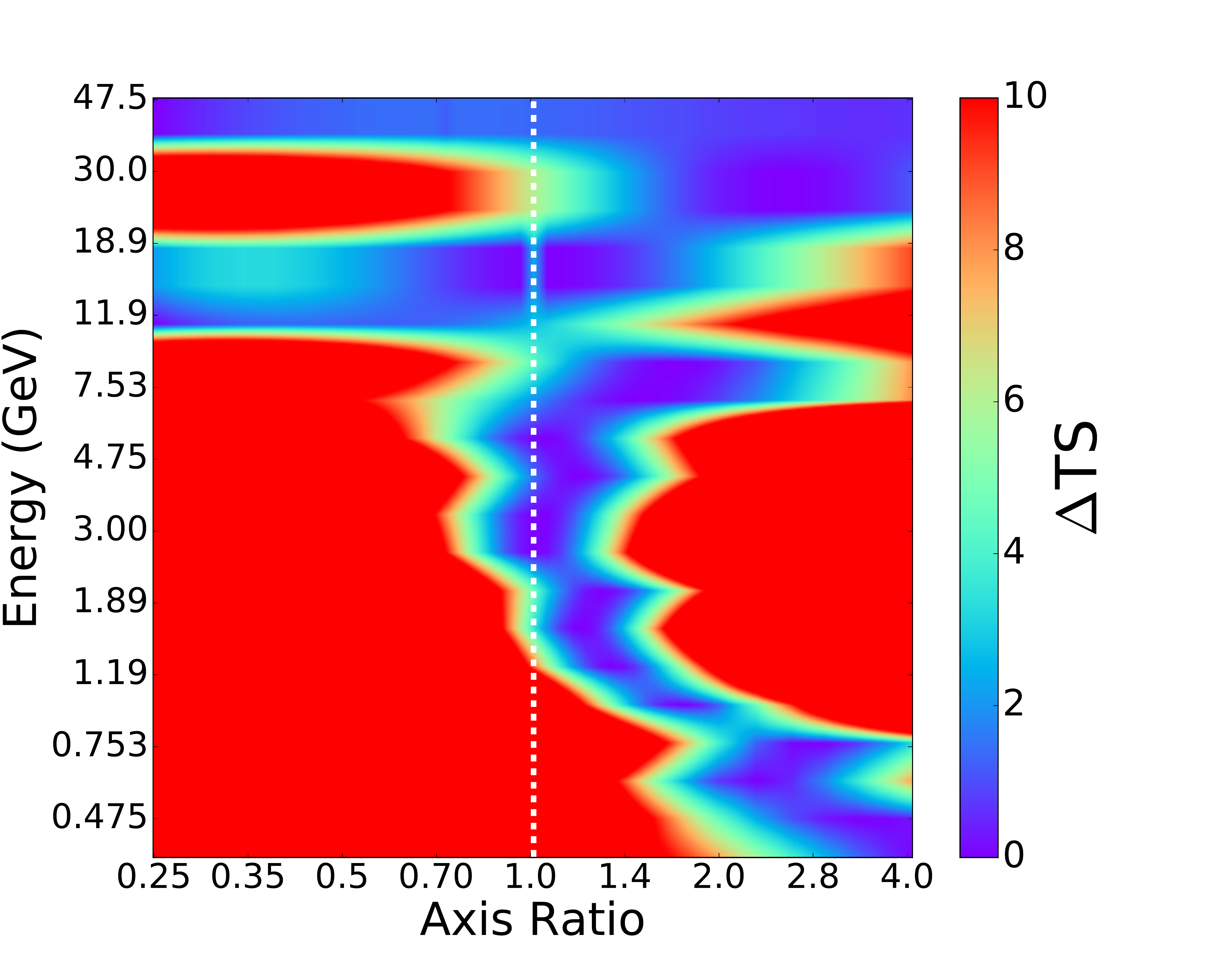}
\end{tabular}
\caption{\footnotesize{Preferred value for $\gamma$ and the axis ratio for a GC analysis that utilizes the diffuse emission Model A~\cite{Calore:2014xka} as opposed to the {\it Fermi}-LAT based {\tt p7v6} diffuse emission model. In this case we also add a model for the {\it Fermi} Bubbles taken from~\citep{Su:2010qj}. We find features that are generically consistent with our default model near the spectral peak of the GCE ($\approx 1-3$~GeV), but which prefer a steeply peaked profile stretched perpendicular to the Galactic plane at low $\gamma$-ray energies. This may be due to a clear oversubtraction of the plane by Model A near the GC, which was not intended to fit the $\gamma$-ray data very close to the Galactic plane.}}
\label{fig:gc_modA_fits}
\end{figure*}

\begin{figure*}[t!]
\centering
\begin{tabular}{c}
\includegraphics[scale=0.22]{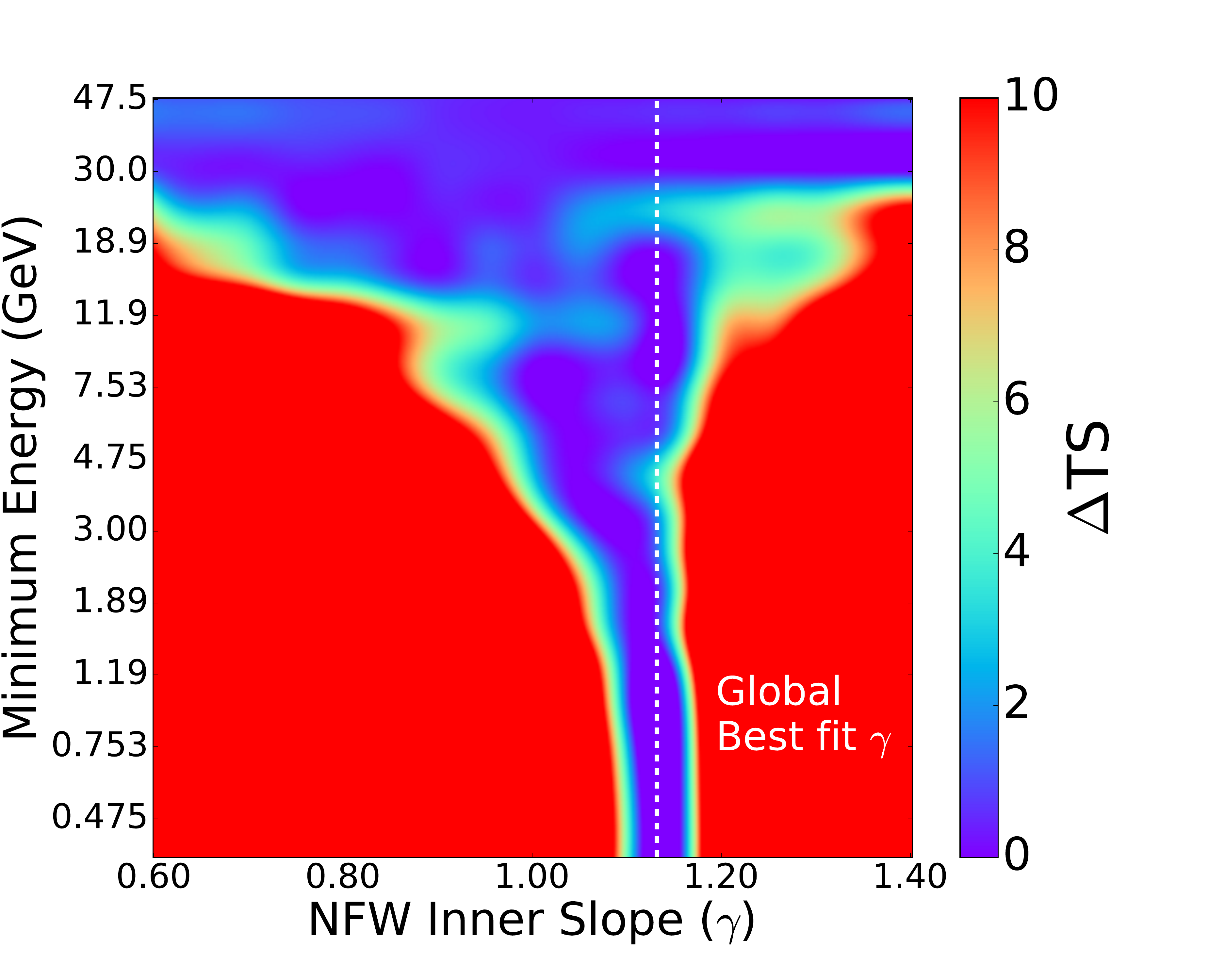} \hspace{0.15in}
\includegraphics[scale=0.22]{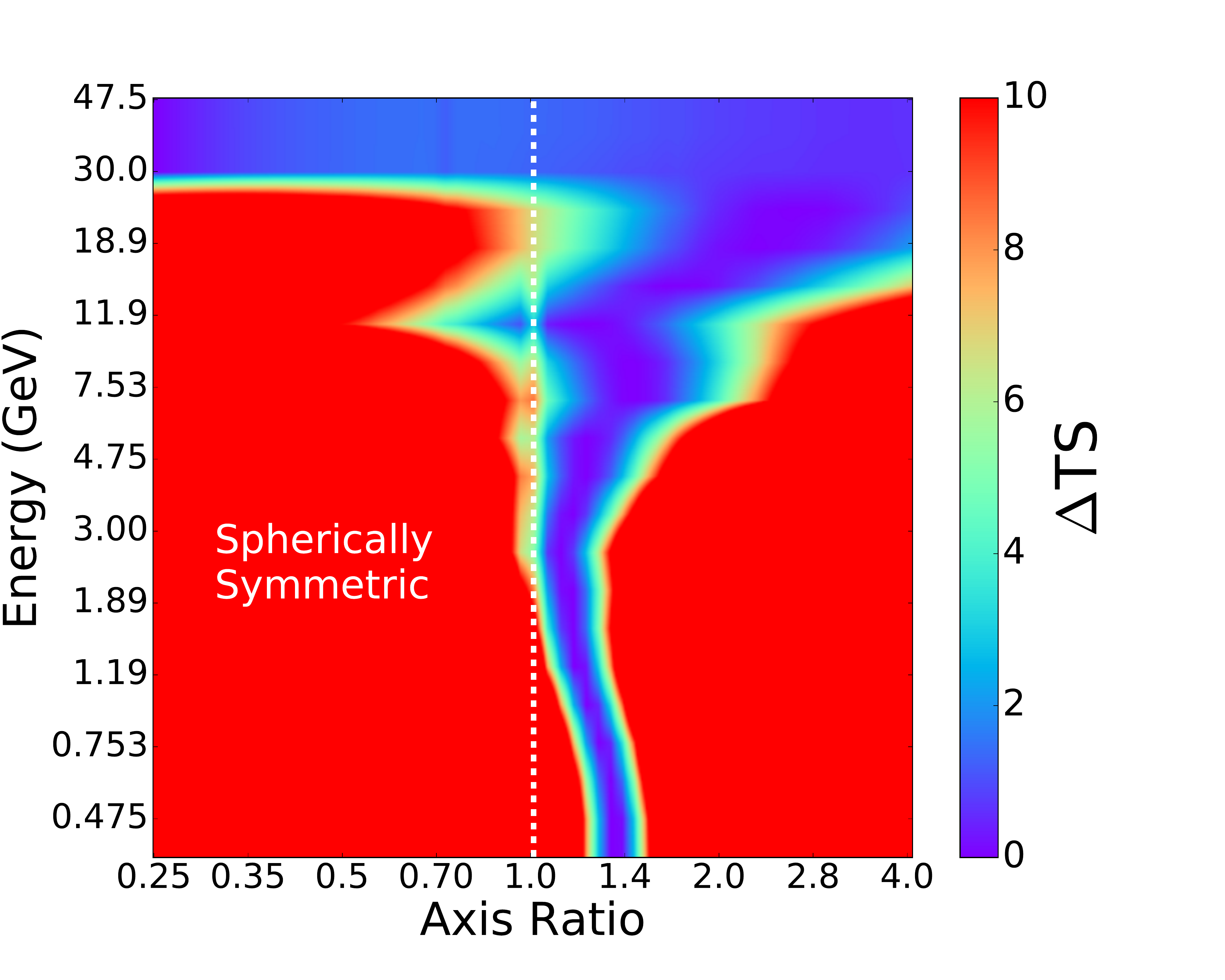}
\end{tabular}
\caption{\footnotesize{Same as Fig.~\ref{fig:gc_modA_fits} plotted as a function of the minimum analysis energy, rather than individually in each energy bin. In this case, we see that the global results are dominated by the emission near the peak of the GCE intensity ($\sim 1-3$~GeV). We then find results that are globally qualitatively similar to those of our default analysis, prefering $\gamma=1.13$ and an axis ratio of $1.32$.}}
\label{fig:gc_modA_fits_mine}
\end{figure*}

In Fig.~\ref{fig:gc_modA_fits} we show the resulting change in the value of $\gamma$ and the ellipticity in our GC analysis. We note that our results are significantly affected, compared to an analysis utilizing the p7v6 diffuse model provided by the Fermi-LAT collaboration. The effect is most significant at the lowest energy bins, however, while bins near the spectral peak tend to prefer values similar to our default analysis. To illustrate this point, in Fig.~\ref{fig:gc_modA_fits_mine} we show the same results plotted as a function of the minimum analysis energy, rather than individually in each energy bin. In this case, we see that the global fit to the data is dominated by the fit near the peak of the GCE emission ($1-3$~GeV). 

The strong divergence from both $\gamma\approx 1.1$ and spherical symmetry at energies below $1$~GeV is intriguing, though its interpretation is unclear. Notably, Model A was not designed to fit the $\gamma$-ray data in regions along the Galactic plane, a region which was masked during the fitting of Model A to the $\gamma$-ray data by~\citep{Calore:2014xka}. Thus, it is possible that the low-energy behavior of this data represents a strong oversubtraction of the Galactic plane. A similar feature was found recently in the analysis of \texttt{GALPROP}-based diffuse emission models of the GC data~\citep{Carlson:2016iis}, where the residual was solved through the introduction of strong Galactic advection. 

\subsection{Adding an Independent ICS template to the Default Diffuse Model}

In~\citep{Calore:2014xka} it was pointed out that the \texttt{p6v11} diffuse model contains a very hard ICS component, which is likely to markedly overestimate the ICS emission at high energies. Given that this is our default diffuse model, in this section we confirm that our results are unchanged if we attempt to alleviate this problem by adding a freely-floating ICS template to absorb the overestimation.

\begin{figure*}[t!]
\centering
\begin{tabular}{c}
\includegraphics[scale=0.45]{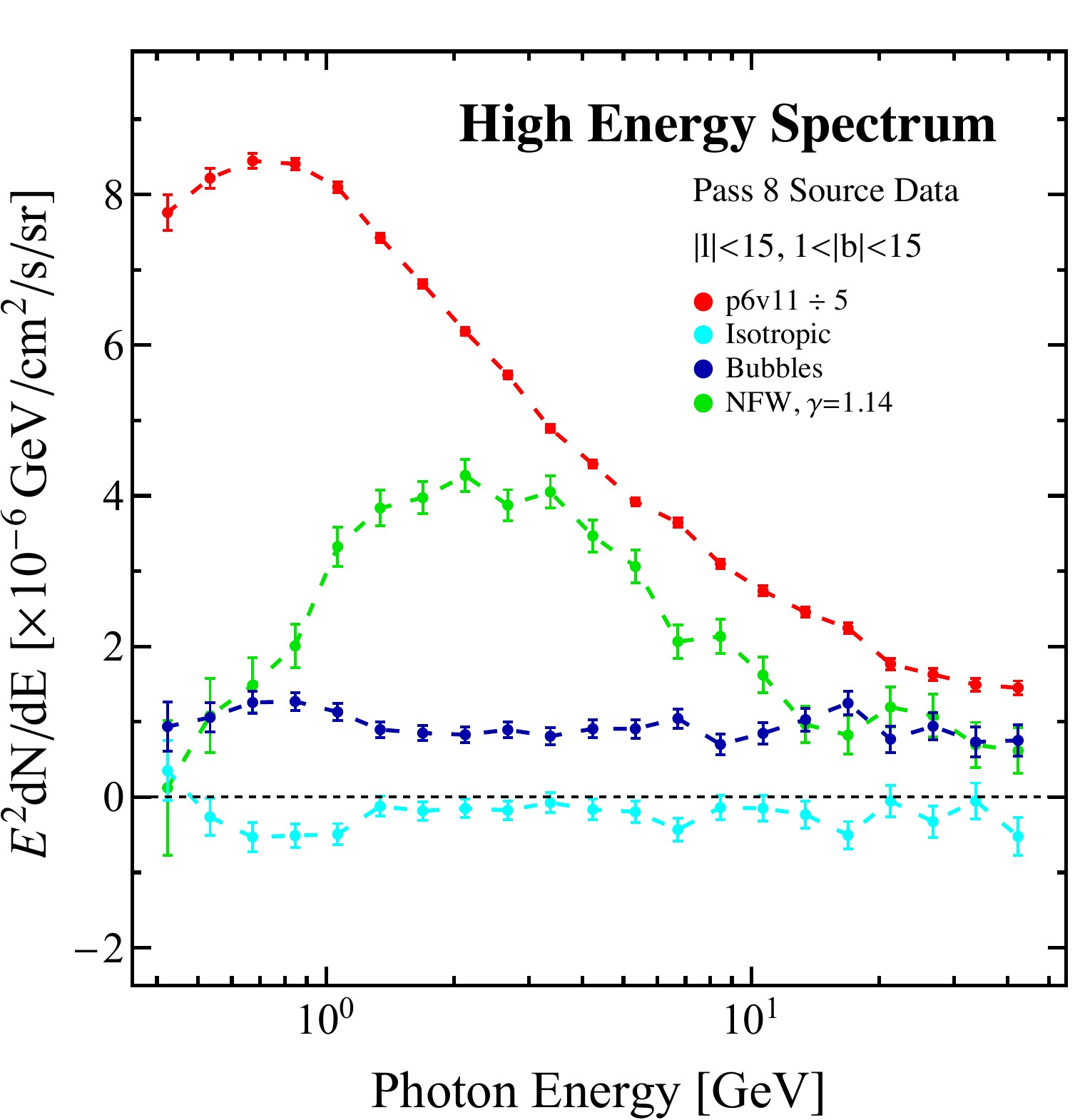} \hspace{0.15in}
\includegraphics[scale=0.45]{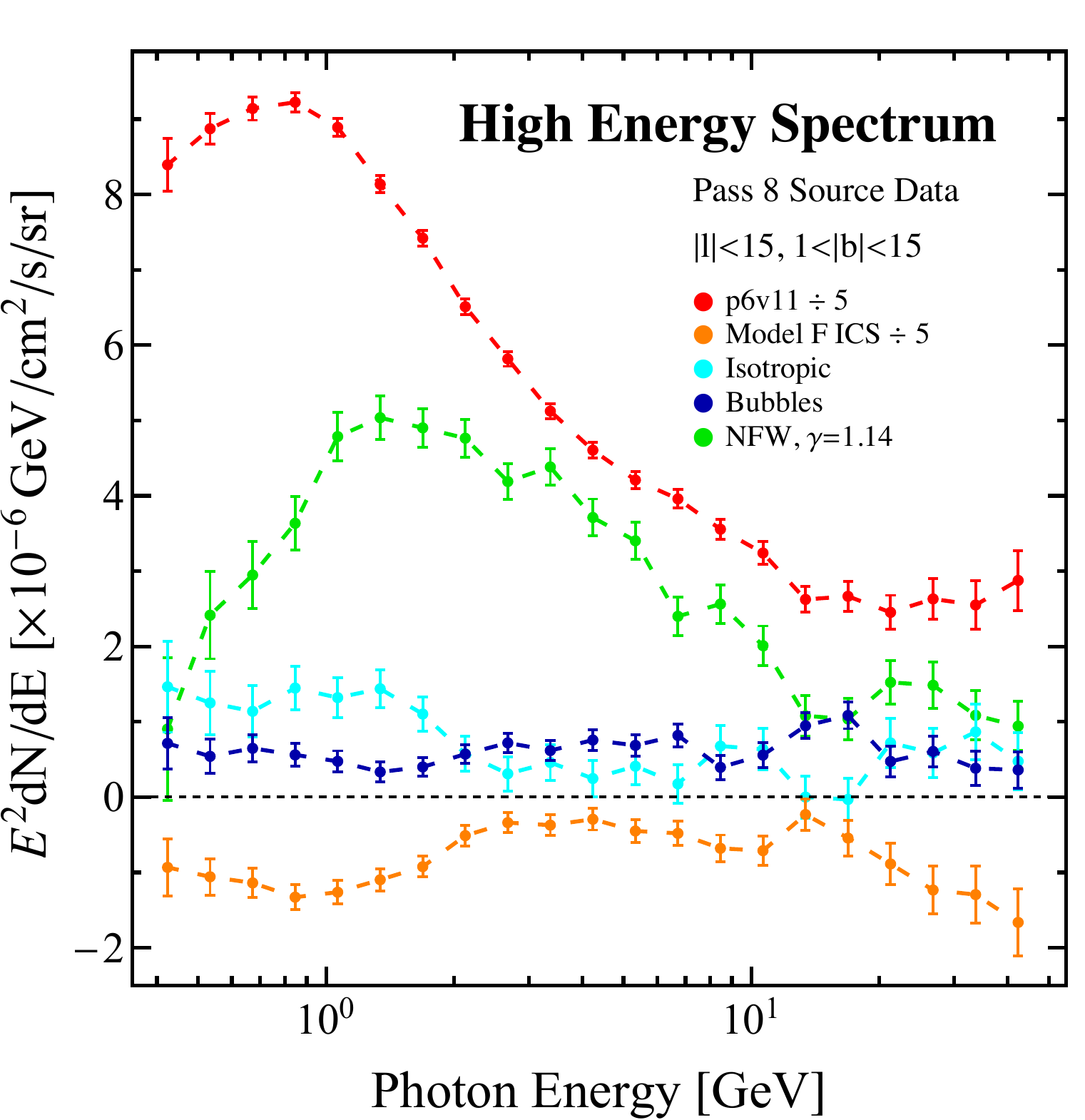}
\end{tabular}
\caption{\footnotesize{Spectrum for our default IG analysis (left) and an identical analysis with an extra template in the form of the Model F ICS contribution (right). This addition is motivated by the observation in~\citep{Calore:2014xka} that the \texttt{p6v11} diffuse model gets the ICS modeling wrong at high energies. Given the large coefficient obtained by the ICS template this appears to be true, but this issue appears to leave the basic features of the GCE unchanged. In both figures the coefficients of the diffuse components are rescaled to facilitate the comparison between templates. See text for details.}}
\label{fig:ADDICS}
\end{figure*}

While the ICS contribution to the \texttt{p6v11} Galactic diffuse model is not provided independently, as a proxy we can use the spatial ICS templates derived from the 14 \texttt{GALPROP} models discussed above. For example in Fig.~\ref{fig:ADDICS} we show our default IG fit on the left, and then on the right an identical fit except for the addition of the ICS model associated with Model F. We tested the ICS components from all 14 models, and the results were all similar to the figure shown. In all cases the ICS template was negative across the whole energy range; the spectrum of the diffuse model flattens out at higher energies; and the isotropic template is now generally positive. The quantitative details of the extracted GCE spectrum change somewhat -- in particular, the amplitude of the high-energy tail is larger -- but its basic features are unaffected, increasing our confidence that our results are not strongly dependent on the (likely mismodeled) hard ICS component in the \texttt{p6v11} model.

\subsection{Changing the Template used to Model the {\it Fermi} Bubbles}

The template describing the {\it Fermi} Bubbles~\citep{Su:2010qj} is somewhat ad hoc, as the physical origin of the Bubbles is not yet well understood. In particular, at low latitudes the spectrum and morphology of the Bubbles are uncertain, and several different templates have been employed in the literature.

\begin{figure*}[t!]
\centering
\begin{tabular}{c}
\includegraphics[scale=0.5]{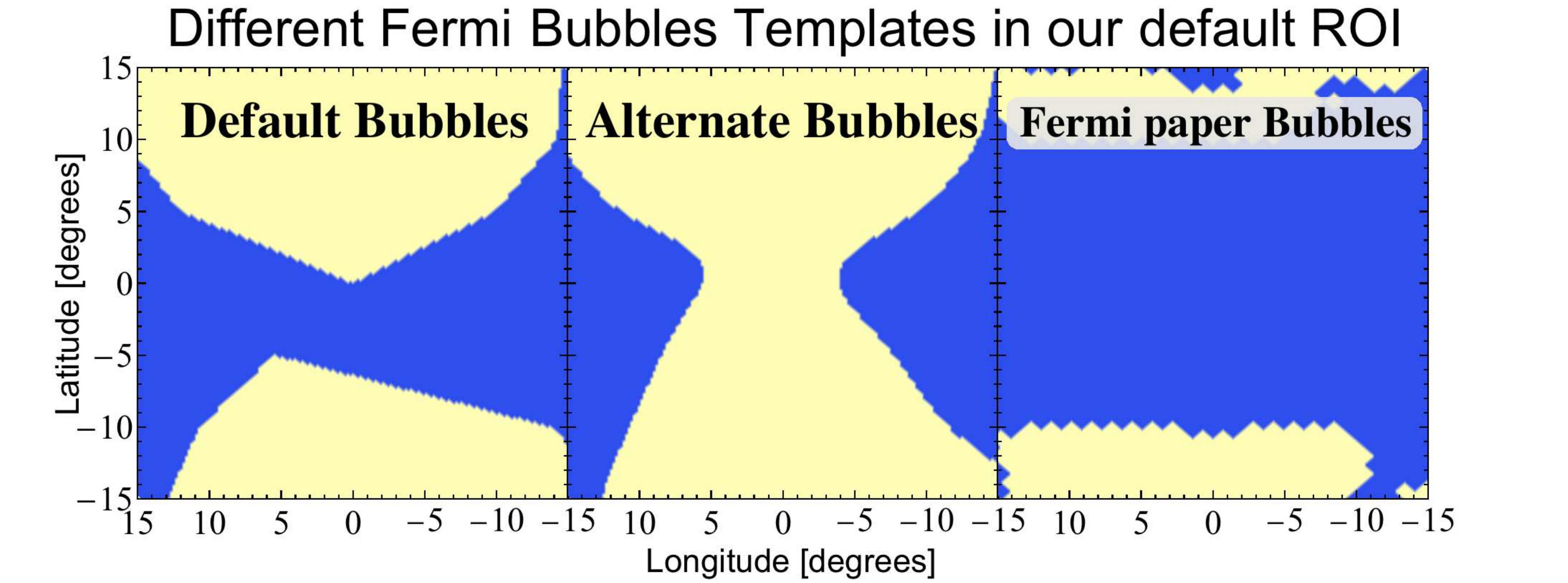}
\end{tabular}
\caption{\footnotesize{Spatial morphology of our default template for the {\it Fermi} Bubbles and two alternatives, within our region of interest for the IG analysis. The cream region corresponds to the interior of the Bubbles, and the blue region to the exterior. The templates are taken from~\citep{Su:2010qj} (``Default Bubbles''),~\citep{mengprivate} (``Alternate Bubbles''), and~\citep{Fermi-LAT:2014sfa} (``Fermi paper Bubbles'').}}
\label{fig:DiffBubTemp}
\end{figure*}

\begin{figure}[t!]
\centering
\begin{tabular}{c}
\includegraphics[scale=0.45]{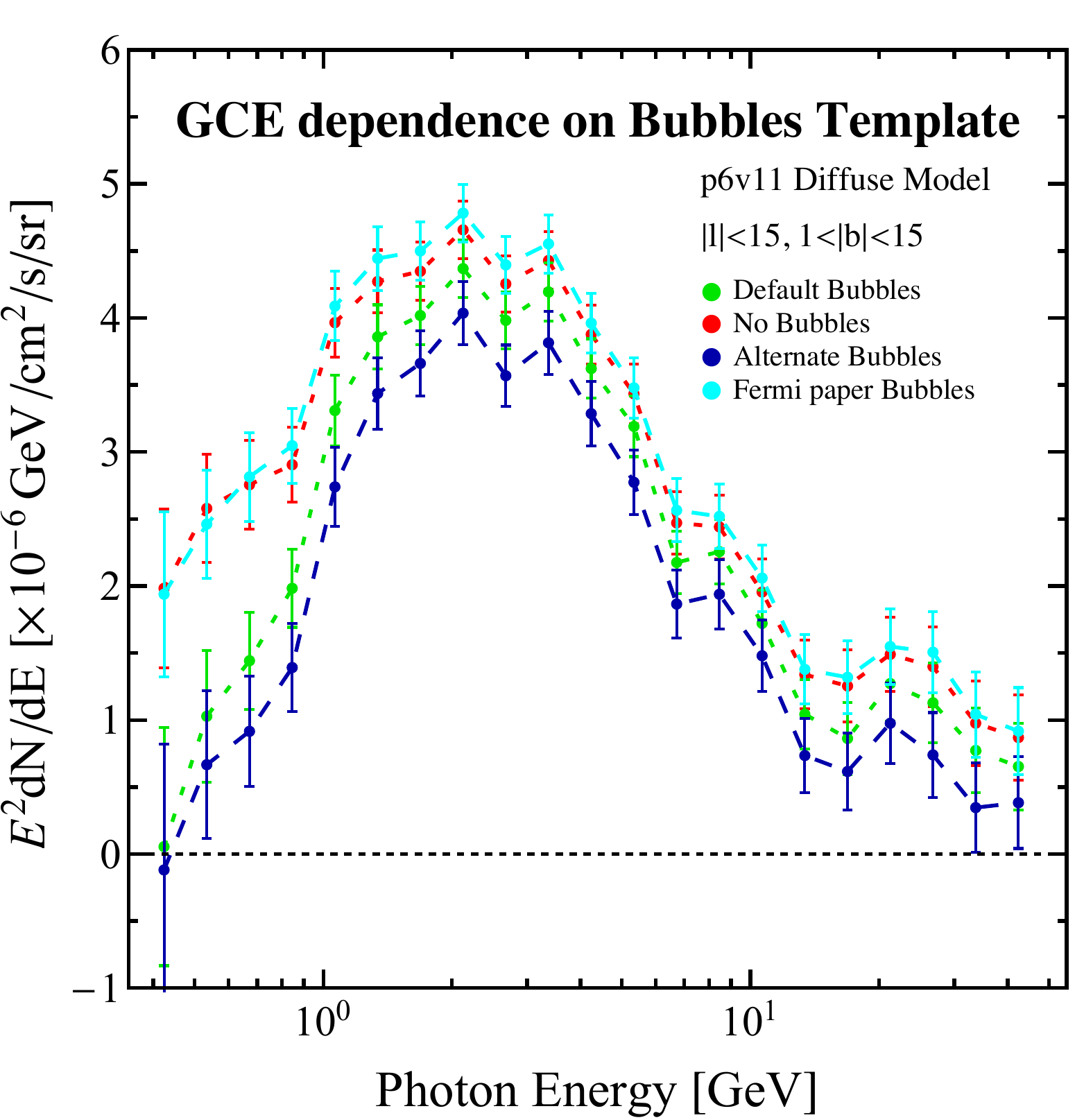}
\end{tabular}
\caption{\footnotesize{The spectrum of the GCE template in the default case (green points), when the {\it Fermi} Bubbles template is omitted entirely (red points), and when the default {\it Fermi} Bubbles template is replaced by an alternative prescription (dark and light blue points).}}
\label{fig:DiffBubFit}
\end{figure}

As a cross-check of our results, we test the impact in the IG analysis of replacing our default Bubbles template with each of two alternative templates, one derived by the {\it Fermi} Collaboration~\citep{Fermi-LAT:2014sfa}, and the other one based on modeling by Su~\citep{mengprivate}. We label these the ``Fermi paper Bubbles'' and ``Alternate Bubbles'' respectively. We also test the impact on the GCE spectrum if the Bubbles template is omitted completely. All templates correspond to constant emission per sr within the Bubbles; the default template and the two alternatives are shown in Fig.~\ref{fig:DiffBubTemp}. The results of the corresponding fits are shown in Fig.~\ref{fig:DiffBubFit}.

We see that the impact of changing this template is modest in the IG, and our default template gives results lying between those obtained with the {\it Fermi} Collaboration template and the ``Alternate Bubbles''. The {\it Fermi} Collaboration template has very little support in our ROI, and so is almost indistinguishable from including no template for the Bubbles at all; this slightly increases the flux associated with the GCE, with the impact being largest at sub-GeV energies. The ``Alternate Bubbles'' template has more support close to the GC, and so including it slightly reduces the flux associated with the GCE. However, the general features of the high-energy spectrum are similar in all cases. We have also tested the variation in the preferred GCE axis ratio as we vary the Bubbles templates in the IG and find results consistent with those shown in the main text.

\begin{figure*}[t!]
\centering
\begin{tabular}{c}
\includegraphics[scale=0.22]{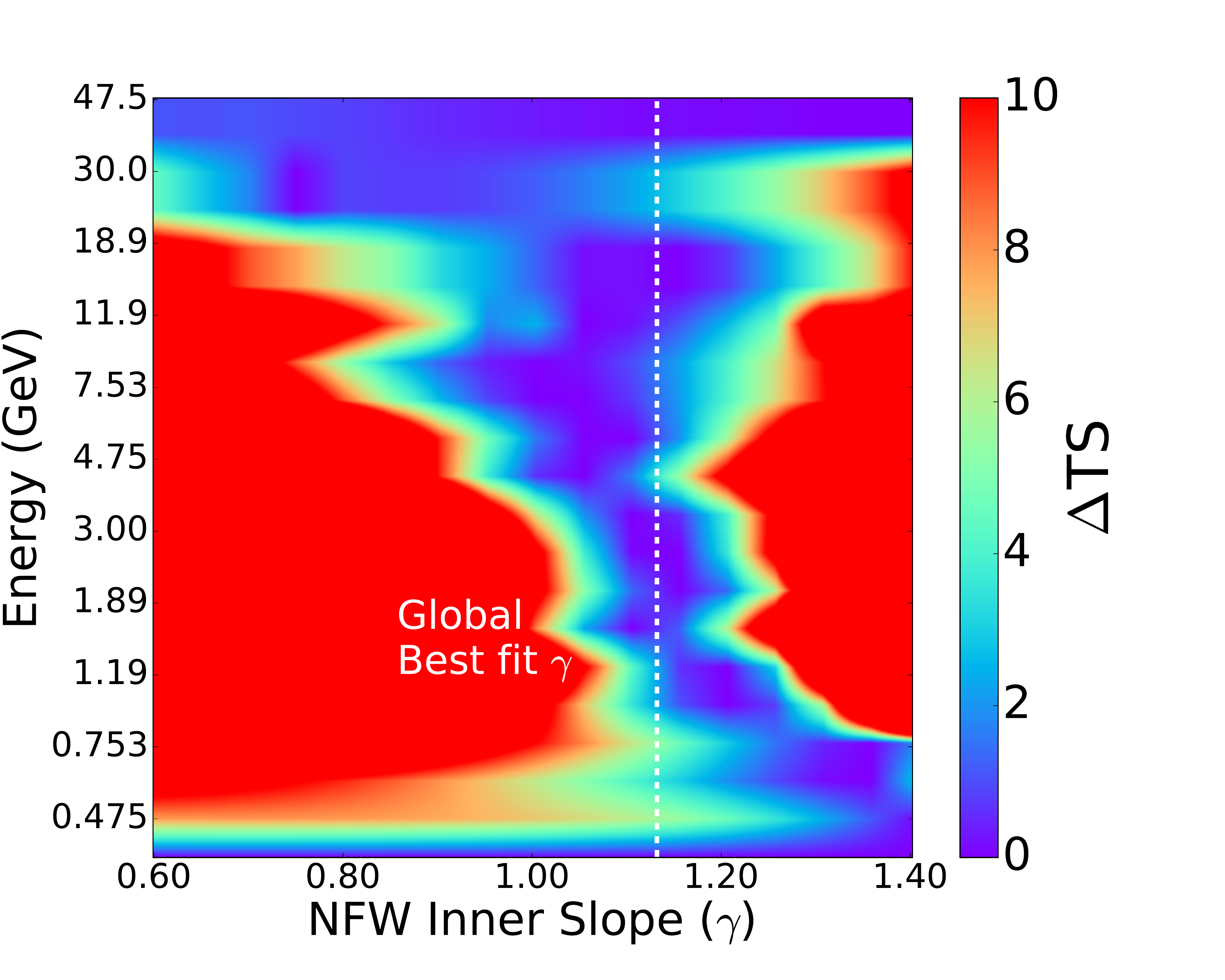} \hspace{0.15in}
\includegraphics[scale=0.22]{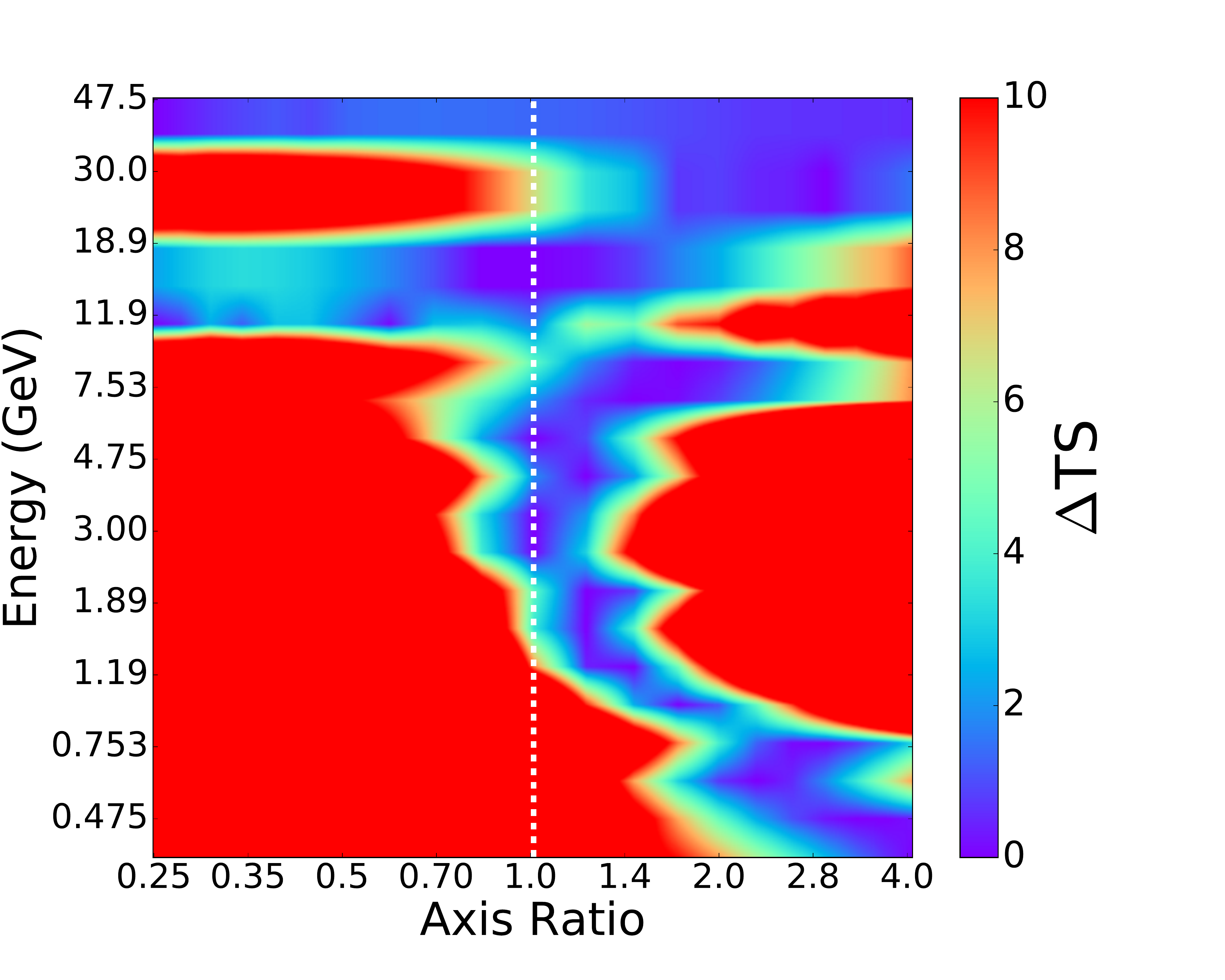}
\end{tabular}
\caption{\footnotesize{Same as Fig.~\ref{fig:gc_modA_fits} for a model where the default Bubbles template is replaced by the Alternative Bubbles template shown in Fig.~\ref{fig:DiffBubTemp}. We find almost no difference between these models in terms of the parameters of the GCE fit, indicating that the Bubbles template and the GCE template are not in any way degenerate, and that the Bubbles only contribute marginally to the $\gamma$-ray data within the GC ROI. The best fit $\gamma$ in this case is $\gamma=1.12$, and the best fitting axis ratio is $1.44$.}}
\label{fig:gc_Bubbles}
\end{figure*}

The difference between the {\it Fermi} Bubbles templates shown in Fig.~\ref{fig:DiffBubTemp} is especially pronounced near the GC, where the default Bubbles template is set to identically 0, while the Alternative Bubbles template has an equivalent surface brightness to the entirety of the Bubbles region. In Fig.~\ref{fig:gc_Bubbles} we show the resulting fits to the morphology and ellipticity of the GCE in the GC ROI when utilizing the Alternative Bubbles template. We find almost no difference in the fits of the GCE components, demonstrating that this version of the Bubbles template also provides only a marginal flux to the current $\gamma$-ray data in the GC ROI. Additionally, we find that the extracted spectrum of the GCE component varies by $<1\%$ when using the ``Alternative Bubbles" template. We note two possible explanations for the minimal contribution of the Bubbles in the GC analysis. First, the Bubbles may not provide any $\gamma$-ray emission near the GC, as might be the case if the Bubbles are powered by old leptonic jets that are no longer active. Second, the morphology of the Bubbles template may vary considerably from the ``equal surface brightness'' model that fits the high-latitude data. This second option seems reasonable given that the {\it Fermi} Bubbles appear to be generated near the GC, where they may assume a more complex morphology. 

\subsection{Varying the Smoothing Procedure for the Diffuse Model}

As outlined in the main text, in order to smooth our templates to the data we use the {\it Fermi} Science Tools for the Galactic diffuse model, whilst we employ much simpler Gaussian smoothing for the remaining templates. The motivation here is that as the diffuse model is significantly brighter than the other contributions in our ROI, even if smoothing it incorrectly only introduces errors at the percent level, this could still noticeably impact our results. On the other hand, we found that our GCE results are essentially unchanged even if we do not smooth the GCE template at all. The only caveat is that for large values of $\gamma$, where the GCE template becomes more sharply peaked, there can be a slight impact.

Although smoothing the diffuse model through the {\it Fermi} Science Tools is the correct procedure, we can compare how much our results would change if we instead used Gaussian smoothing. The results are shown in Table~\ref{table:SmoothGam}; note this check was done early on, so these results were determined using the Pass 7 Reprocessed dataset, which we will discuss in detail in App.~\ref{app:SelCr}. We see that uniformly moving to Gaussian smoothing increases the preferred $\gamma$ for the GCE template. Nonetheless in all cases $\Delta \gamma < 0.1$ and so the impact does not appear significant. Finally note that moving to Gaussian smoothing did not change the relative quality of fit (in the IG ROI) of the various background models discussed above.

\begin{table}[h]
\begin{center}
\begin{tabular}{| c || c | c |}
    \hline
    Diffuse Model & {\it Fermi} Smoothing & Gaussian Smoothing \\ \hline \hline
    \texttt{p6v11} & $1.12$ & $1.16$ \\ \hline
    \texttt{p7v6} & $1.16$ & $1.18$ \\ \hline
    Model F & $1.01$ & $1.10$ \\
    \hline
\end{tabular}
\end{center}
\caption{Variation in the preferred value of $\gamma$ over the full energy range in our default ROI when we use Gaussian smoothing instead of smoothing with the {\it Fermi} Science Tools, considered for three different diffuse models. This cross check was performed using Pass 7 data. The 1$\sigma$ statistical uncertainties are $\sim 0.01$ or less and are omitted.}
\label{table:SmoothGam}
\end{table}

\subsection{Changing the Point-Source Model}

\begin{figure*}[t!]
\centering
\begin{tabular}{c}
\includegraphics[scale=0.22]{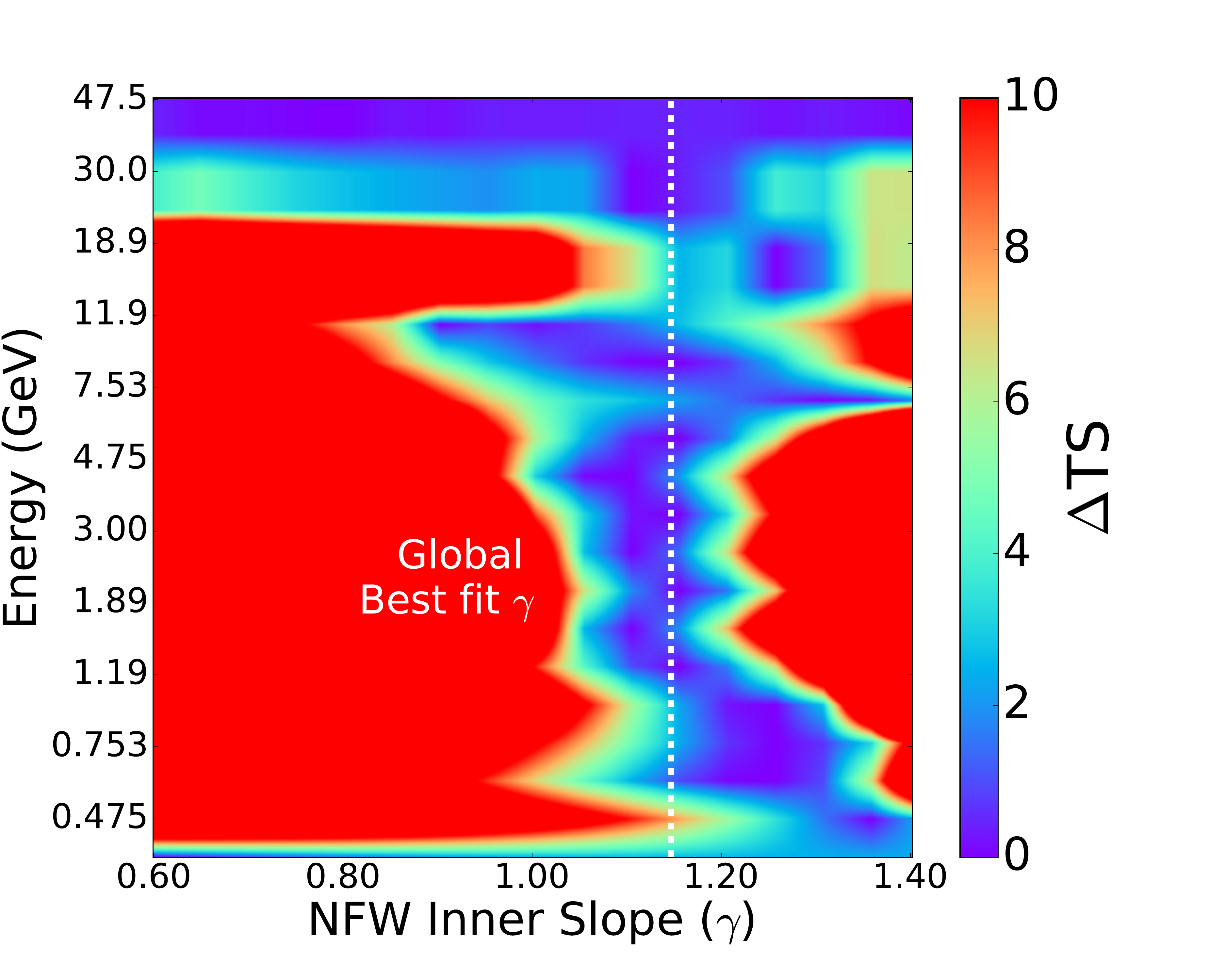} \hspace{0.15in}
\includegraphics[scale=0.22]{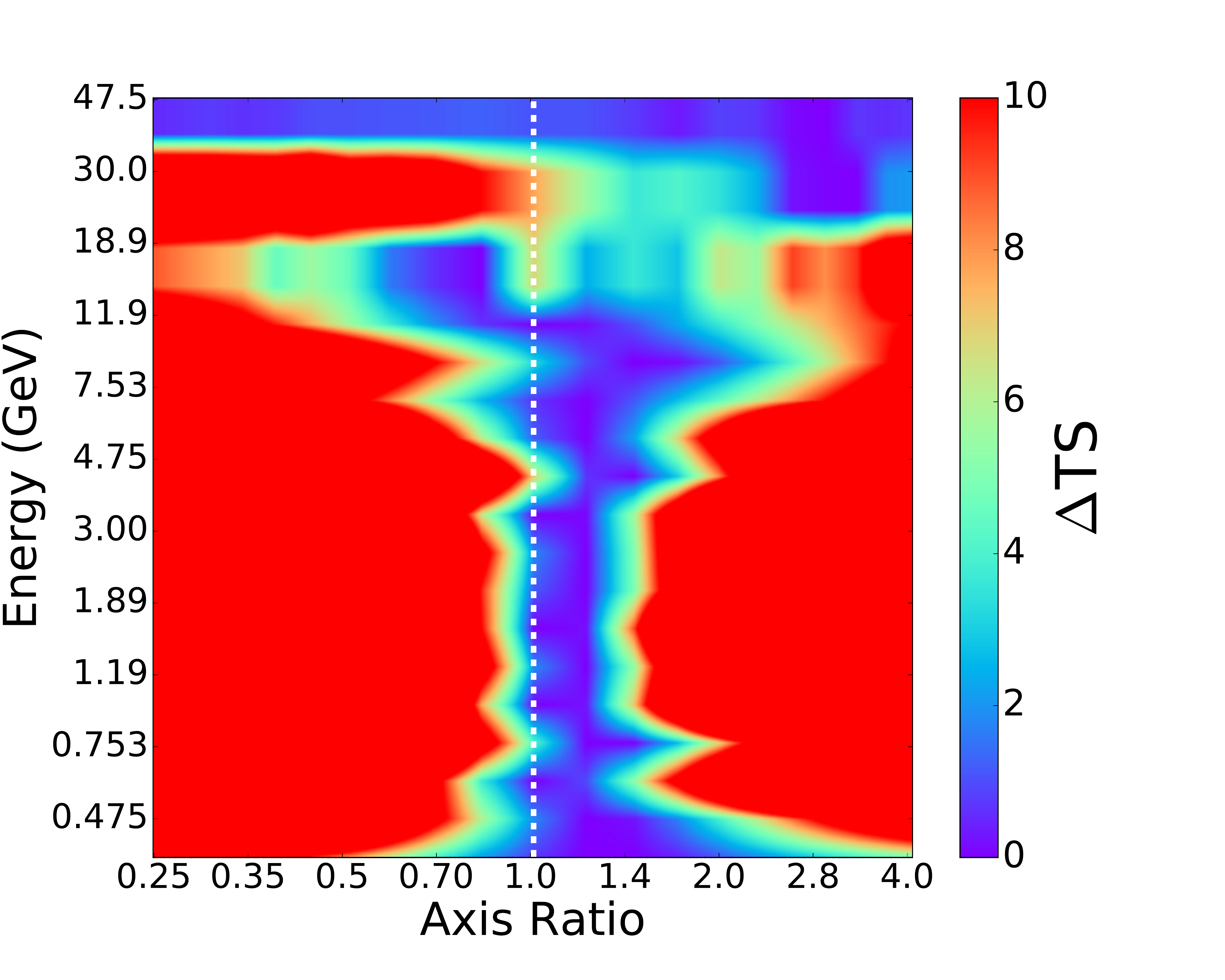}
\end{tabular}
\caption{\footnotesize{Preferred value for $\gamma$ and the axis ratio for a GC analysis that utilizes the 2FGL PS catalog, instead of the 3FGL catalog used in our default analysis. This provides an easier comparison to previous works, such as~\citep{Daylan:2014rsa}. We note that, unlike our default analysis, which utilized $\gamma=1.00$ for the ellipticity analysis, owing to the more peaked profile preferred in the 2FGL analysis, we calculate the ellipticity assuming a value $\gamma=1.20$.} The globally best fitting value of $\gamma$ for our 2FGL analysis is $\gamma=1.14$.}
\label{fig:2FGL_fits}
\end{figure*}

In the GC analysis we are unable to mask out even the brightest PSs without significantly affecting our signal region. Thus, we model all PSs, making our results possibly sensitive to the employed PS model. In our default analysis we utilize the 3FGL catalog, which corresponds to an update from previous analyses (e.g.~\citep{Daylan:2014rsa}) which utilized the 2FGL catalog. Here we recalculate our results utilizing the 2FGL catalog to provide an easier comparison to those works. As shown recently in~\citep{Carlson:2016iis}, the PS degrees of freedom are generally not degenerate with the smooth emission profile of the GCE. However, degeneracies may exist within the inner $\sim 1^\circ$ of the GC, where the GCE template is extremely peaked, and potentially degenerate with several PSs located within the region. Within this small region, the 2FGL and 3FGL templates differ considerably. The 2FGL template includes 4 PSs within $1^\circ$ of Sgr A*, only one of which lies within $0.1^\circ$. The 3FGL includes 7 PSs within $1^\circ$ of Sgr A*, two of which lie within $0.1^\circ$ of the GC. 

In Fig.~\ref{fig:2FGL_fits} we show the best fitting value of $\gamma$ and the ellipticity for a model that utilizes the 2FGL PS catalog. We note one interesting change: the preferred value of $\gamma$ increases significantly when the 2FGL PS catalog is used, preferring a best fit value of $\gamma=1.14$ over the full energy range of the analysis. This is in significantly better agreement with the best results for the GC analysis performed in~\cite{Daylan:2014rsa}, which preferred a value $\gamma=1.17$. Secondly, we note that, compared to our default analysis, models utilizing the 2FGL template continue to prefer $\gamma$-ray emission from the GCE template in the energy range of $18.9-30.0$~GeV. We note that the fits in this energy range additionally prefer relatively large values of $\gamma$, compared to our default analysis, indicating that the residual emission here may be degenerate with PSs that exist in the 3FGL catalog, but not in 2FGL. 

Given the preference for a more steeply sloped $\gamma$-ray emission profile, we assume a value $\gamma=1.20$ to perform the ellipticity analysis, compared to the value $\gamma=1.00$ adopted for our default results. We find only mild differences between our results utilizing the 2FGL and 3FGL PS templates. We note that the high-energy emission is more consistent with sphericity than found in the 3FGL case; while there is still a marginal preference for ellipticity perpendicular to the Galactic plane, its significance is reduced. Since, as argued above, PSs are only likely to be degenerate with the GCE template very close to the GC, this result may be related to a strong preference for spherically symmetric and extended emission very close to the GC, that can not be absorbed by the 2FGL catalog.

\section{Impact of Varying the Data Selection}
\label{app:SelCr}

\subsection{Move from Pass 7 to Pass 8 data}
\label{app:P7P8}

During the course of this project the {\it Fermi} Pass 8 dataset was released, and so we updated our analysis framework to make use of the improvements Pass 8 brings. As part of this process we completed a number of sanity checks on the variation of the GCE between Pass 7 (Reprocessed) and Pass 8; we reproduce some of these results in this appendix, showing only results for the IG analysis. As will be seen, whilst there can be small changes in the specifics, our qualitative conclusions are largely unchanged by the move to the updated dataset. Given this general consistency, several cross checks discussed in the appendices are only performed in either Pass 7 or 8 (simply determined by when they were done), but they will be noted as such where appropriate. Unless otherwise stated results throughout this work were produced using Pass 8 data.

In Sec.~\ref{sec:models} we outlined the details of the Pass 8 dataset used throughout. The equivalent details for the Pass 7 (V15) Reprocessed dataset are as follows. We used data collected between August 4, 2008 and March 8, 2015, and applied the recommended selection criteria for that dataset: zenith angle $<100^{\circ}$, instrumental rocking angle $<52^{\circ}$, \texttt{DATA\_QUAL} = 1, \texttt{LAT\_CONFIG}=1. Although this is a slightly shorter time period than used for the Pass 8 data, note that it also included data taken during the period where the {\it Fermi} satellite adjusted its search to enhance exposure of the GC.

The division of the data into event classes is also somewhat different in Pass 7. There is no UCV class; the highest-quality event class is ``Ultraclean'' (UC). While events are separated into front-converting and back-converting as in Pass 8, there is no additional separation into quartiles as in Pass 8. However, a photon sample similar to the top PSF quartile in Pass 8 can be obtained by selecting front-converting photons with high values of the CTBCORE parameter, corresponding to good directional reconstruction~\citep{Portillo:2014ena}. Following ~\citep{Daylan:2014rsa}, we denote the top 50\% of photons ranked by CTBCORE as Q2 (top two quartiles), and the whole dataset as Q4 (all four quartiles). The Q2 front-converting sample (Q2F) is approximately the top 25\% of the overall sample sorted by PSF, so can be broadly compared to Pass 8 BestPSF data.

\subsubsection{Consistency of Results for the Full GCE}

The first thing to note is that in both Pass 7 and Pass 8 data, the GCE is preferred in the data with a large TS. For example, the statistical preference for the GCE as a whole claimed in~\citep{Daylan:2014rsa} was ${\rm TS} \sim 1100$. This analysis used Q2F Pass 7 Reprocessed data collected up to December 2013, in a slightly larger ROI than the one we use in our default IG analysis ($40^\circ \times 40^\circ$ rather than $30^\circ \times 30^\circ$), and employed the $\texttt{p6v11}$ Galactic diffuse model. Repeating the same analysis with Pass 8 UCV BestPSF data, we find a TS of $1175$ in the $40^\circ \times 40^\circ$ ROI, and $1542$ in our smaller $30^\circ \times 30^\circ$ ROI.

We can also perform a direct comparison between our default IG analysis using All Source Pass 8 data and the same analysis using the All Source Pass 7 dataset described above.  In the Pass 8 data, we find a TS of 2859 for the GCE, whereas in Pass 7 data we find a TS of 2706. If we employ the \texttt{p7v6} Galactic diffuse model rather than \texttt{p6v11}, the normalization and significance of the excess falls markedly in both cases (as noted above, we suspect this is due to the inclusion of spatially and spectrally fixed templates for extended diffuse excesses in the \texttt{p7v6} model), to a TS of 1205 for Pass 8 and 1197 for Pass 7. Thus in terms of raw significance, the GCE seems very similar in the Pass 7 and Pass 8 datasets, independent of the event selection.

\begin{figure}[t!]
\centering
\includegraphics[scale=0.53]{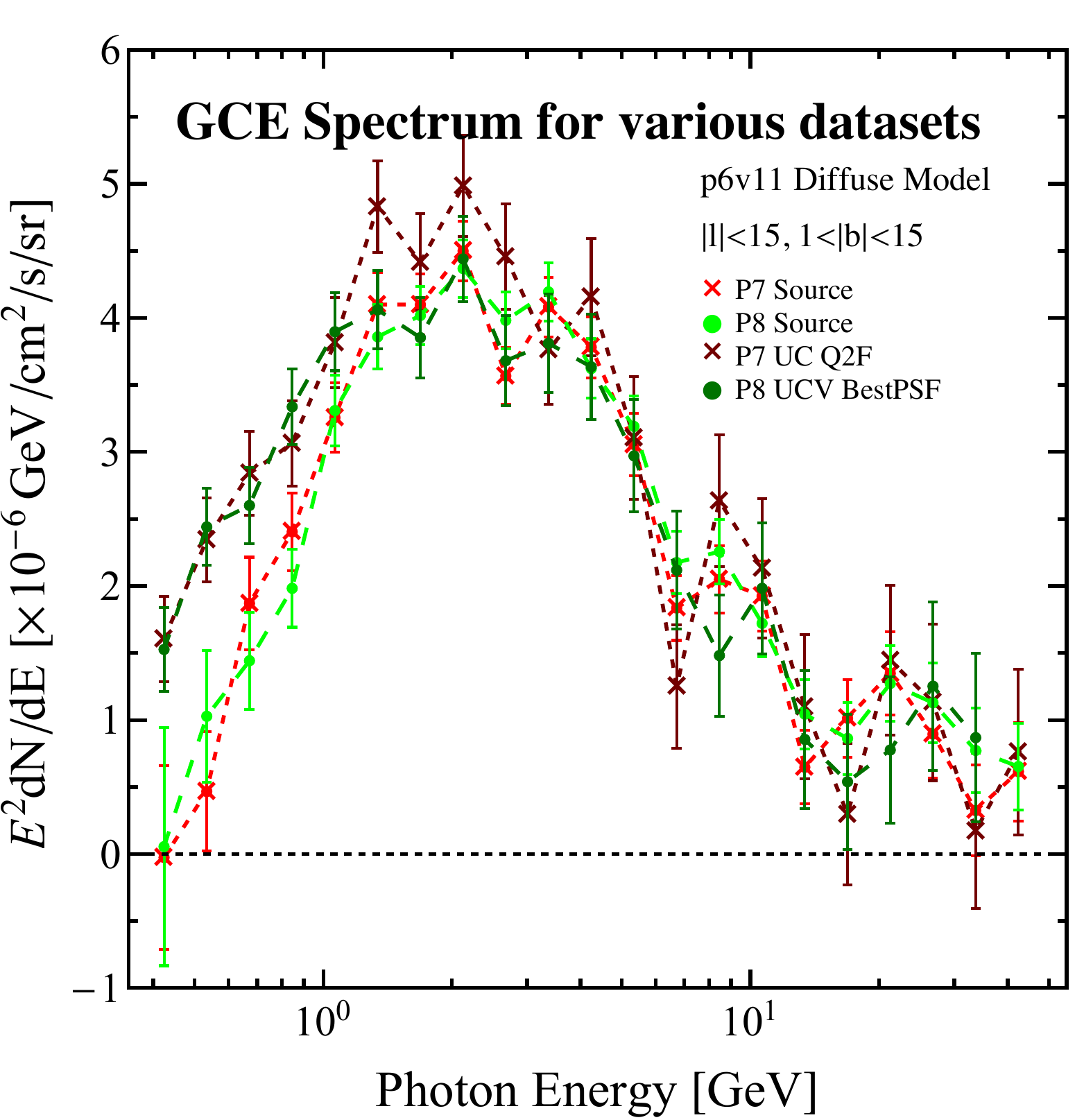}
\caption{\footnotesize{The spectrum extracted in the IG analysis by a $\gamma=1.1$ GCE template for four different datasets: Pass 7 All Source (P7 Source), Pass 8 All Source (P8 Source), Pass 7 Ultraclean Q2 Front-Converting (P7 UC Q2F), and Pass 8 UCV BestPSF (P8 UCV BestPSF). All other options are set to the IG defaults. We see the difference between Pass 7 and 8 is far less important than the differences between varying event selections within Pass 7/8.}}
\label{fig:SpecVsDataSet}
\end{figure}

Another basic check is how much the extracted spectrum for the GCE varies as we shift between different {\it Fermi} data reconstructions. In Fig.~\ref{fig:SpecVsDataSet} we show the spectrum extracted by a $\gamma=1.1$ GCE template for four different datasets: Pass 7 All Source, Pass 8 All Source, Pass 7 Ultraclean Q2F (as used in~\citep{Daylan:2014rsa}), and Pass 8 UCV BestPSF. We see that very little changes between Pass 7 and Pass 8 -- there is a much more pronounced variation between different choices of event quality class and angular resolution quartile(s). This difference is most appreciable at low energies, and can in fact be traced to the difference in size between the PS mask in the various datasets (which has its origin in the different PSFs), as we will explore App.~\ref{app:BkgROI}.

\begin{figure}[t!]
\centering
\includegraphics[scale=0.55]{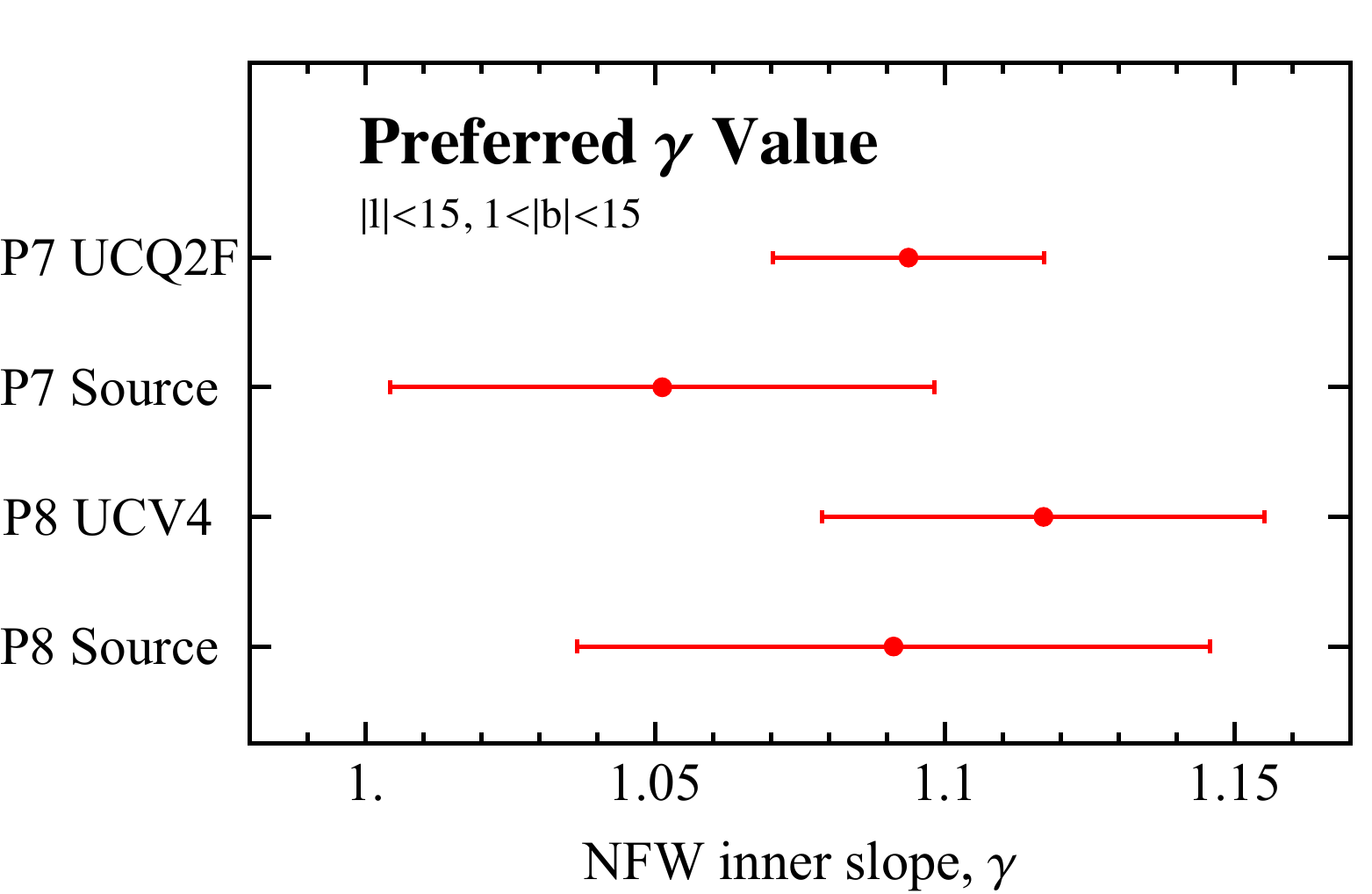}
\caption{\footnotesize{Plotted are the mean and one standard deviation on the best fit $\gamma$ value for the GCE template determined using 16 (17) diffuse models for Pass 7 (Pass 8) data. This is done using our default IG analysis for the same four datasets described in Fig.~\ref{fig:SpecVsDataSet}, where UCV4 is shorthand for UCV BestPSF.}}
\label{fig:GamVsDataSet}
\end{figure}

Finally we can test the stability of the spatial morphology between datasets, by seeing how much the preferred $\gamma$ value varies, for the same four datasets we just mentioned. In Fig.~\ref{fig:GamVsDataSet} we show the mean and one standard deviation on the extracted preferred $\gamma$ value for 16 (17) diffuse models for Pass 7 (Pass 8), where the models used are described in App.~\ref{app:Bkg}; note we do not include the \texttt{p8v6} model for an analysis of Pass 7 data. The figure makes it clear that the spatial morphology is consistent between the datasets, although there can clearly be differences in the finer details.

\subsubsection{Consistency of High-Energy Results}

We can already see in Fig.~\ref{fig:SpecVsDataSet} that the extracted high-energy spectrum is roughly independent of the dataset and event selection. In Table~\ref{table:TSP7P8} we show the TS for a GCE template on a bin-by-bin basis for Pass 7 and Pass 8. Note this calculation uses the default energy binning laid out in the main text, which combines several of the logarithmic spaced bins together; further details can be found in App.~\ref{app:Binning}. From the table we see that whilst generally the move to Pass 8 increases the TS, it is not always clear cut with the TS decreasing in the first bin. In both Pass 7 and Pass 8 there is a dip in the spectrum at energies $\sim 12-20$~GeV; we scrutinize this apparent feature in more depth in App.~\ref{app:15G}.

\begin{table}[h]
\begin{center}
\begin{tabular}{| c || c | c | c | c |}
\hline
& \multicolumn{4}{ |c| }{TS for GCE} \\ \cline{2-5}
DataSet & Bin 1 & Bin 2 & Bin 3 & Bin 4 \\ \hline \hline
Pass 7 & 59.2 & 18.4 & 27.8 & 3.8 \\ \hline
Pass 8 & 51.3 & 27.6 & 37.5 & 10.4 \\ \hline
\end{tabular}
\end{center}
\caption{TS for a GCE template for our four high-energy bins (using the default binning described in the main text) for Pass 7 and 8 data using our default IG analysis.}
\label{table:TSP7P8}
\end{table}

We can also examine the consistency of our results for the radial variation and sphericity, shown in Sec.~\ref{sec:SpPr}, between Pass 7 and Pass 8. For the radial variation, the preferred $\gamma$ at high energies is somewhat less consistent with the low-energy value in Pass 7 (compared to our main Pass 8 results); a stretch perpendicular to the plane, however, is more restricted in Pass 7 than in Pass 8. In both cases any extension along the Galactic plane is disfavored at high energies.

Overall, however, the qualitative conclusions reached in the main text are essentially unchanged by reverting to the older Pass 7 {\it Fermi} data set.

\subsection{Dependence of the Full GCE on Event Quality Cuts}

We have already presented a number of results for two extreme cases of the data selection: All Source data (all source-converting events, with no further cut on quality) and UCV BestPSF (for Pass 8) or Ultraclean Q2F events (for Pass 7). To study the impact of the data selection in more detail, we study the preferred slope of the NFW profile for several intermediate cases. Since we are interested in systematics here we use data at all energies to reduce statistical uncertainties. In this section we use Pass 7 data, since in the previous subsection we have shown that the results are very consistent between Pass 7 and Pass 8, and it facilitates comparison to previous work. Our results are shown in Table~\ref{table:DataSet} for three different ROIs and two different Galactic diffuse models: we consider both our default ROI and the two ROIs considered in~\citep{Daylan:2014rsa}, for ease of comparison.

\begin{table}[h]
\begin{center}
\begin{tabular}{| c || c | c | c | c |}
    \hline
    Diffuse Model & S FB & C FB & UC FB & UC Q2F \\ \hline \hline
    \multicolumn{5}{ |c| }{$|l|<15^{\circ}$ and $1^{\circ}<|b|<15^{\circ}$} \\ \hline 
    \texttt{p6v11} & $1.12$ & $1.12$ & $1.12$ & $1.07$ \\
    \texttt{p7v6} & $1.16$ & $1.16$ & $1.16$ & $1.07$ \\ \hline \hline
    \multicolumn{5}{ |c| }{$|l|<20^{\circ}$ and $1^{\circ}<|b|<20^{\circ}$} \\ \hline 
    \texttt{p6v11} & $1.21$ & $1.21$ & $1.21$ & $1.19$ \\
    \texttt{p7v6} & $1.25$ & $1.25$ & $1.25$ & $1.19$ \\ \hline \hline
    \multicolumn{5}{ |c| }{Full Sky except $|b|<1^{\circ}$} \\ \hline 
    \texttt{p6v11} & $1.14$ & $1.17$ & $1.17$ & $1.22$ \\
    \texttt{p7v6} & $1.12$ & $1.19$ & $1.19$ & $1.22$ \\
    \hline
\end{tabular}
\end{center}
\caption{Impact on the preferred value of $\gamma$ in the IG analysis over the full energy range of changing our data selection criteria for three different ROIs and two diffuse models. The four different datasets are: 1. all Source class front- and back-converting photons (S FB); 2. all Clean front and back events (C FB); 3. all Ultraclean front and back events (UC FB); and 4. Ultraclean front Q2 events (UC Q2F). All these results are for Pass 7 data. The 1$\sigma$ statistical uncertainties are $\sim 0.01$ or less and are omitted.}
\label{table:DataSet}
\end{table}

In smaller ROIs such as the one we consider, the distinction between Source, Clean and Ultraclean data is minimal, motivating the choice of Source class in order to maximize statistics. There is a more noticeable difference when moving from all UC data to UC Q2F data, and consistency between the  \texttt{p6v11} and \texttt{p7v6} Galactic diffuse models is improved in this case; however, this corresponds to a roughly four-fold reduction in statistics, making study of the high-energy regime difficult.

We also checked the case of Source front-converting photons, to determine whether simply removing the back-converting events was enough to yield a change in $\gamma$. The impact was very minimal, for both the \texttt{p6v11} ($\Delta \gamma = 0.02$) and \texttt{p7v6} (no change) Galactic diffuse models.

We can now compare our results directly to those of Daylan et al~\citep{Daylan:2014rsa}; that analysis found a preferred value of $\gamma=1.18$ for their GCE template fit over the full energy range, using a $40^\circ \times 40^\circ$ ROI, and UC Q2F Pass 7 data. This is very consistent with our results for UC Q2F data (best-fit $\gamma=1.19$). In our Pass 7 All Source analysis in our default ROI we find a best-fit $\gamma=1.12$, compared to $\gamma=1.14$ for our default Pass 8 analysis (over the full energy range). Thus to the extent that there is any small discrepancy between our results and those presented previously in~\citep{Daylan:2014rsa}, it appears to be fully explained by the different data selection and ROI. 

The comparison with~\citep{Calore:2014xka} is less straightforward. This is because their analysis employs a fundamentally different technique for determining the best inner slope. Whilst our approach is to repeat our template analysis using GCE models with various values of $\gamma$ and compare the quality of fit, the authors of that reference broke their ROI into ten different regions and used the weighting of the GCE template in the various regions to determine the slope. When we repeat our analysis in their ROI ($|l|<20^{\circ}$ and $2^{\circ}<|b|<20^{\circ}$), data selection (Clean All photons) and background model (Model F), we find a preferred $\gamma$ of $1.05$, significantly lower than their quoted best-fit value of $1.28$. We suspect this difference is mainly the result of the different method employed to determine $\gamma$.

\subsection{Dependence of GCE Center on Point-Spread Function Cuts}
\label{subsec:GCEcenter}
In Fig.~\ref{fig:gc_centering}, we showed the best-fit location for the center of the NFW profile in our Galactic Center analysis. We did not repeat this exercise in the Inner Galaxy analysis, as it masks the region $|b|<2^\circ$, and is thus insensitive to the exact center of the NFW profile. However, the fact that the NFW profile center is most sensitive to the $\gamma$-ray emission closest to the position of Sgr A* is potentially worrying for the GC analysis as well, since this region may be highly affected by systematic issues. 

In~\citep{Daylan:2014rsa}, the center of the GCE emission profile was found to lie within $\approx 0.05^\circ$ of Sgr A*, with a best fit location that fell $0.025^\circ$ from Sgr A* and was preferred by a factor of $\Delta{\rm TS} \sim 9$ to a profile coincident with the dynamical center of the Milky Way. However, in Fig.~\ref{fig:gc_centering} we found that a center located $0.1^\circ$ from Sgr A* was preferred by a factor $\Delta{\rm TS} \sim 37$ over a location coincident with the dynamical center of the Milky Way. We note that the statistical change between the two datasets is not unexpected (in the case that the offset is real), as the GC analysis in this paper uses a dataset that is more than a factor of 5 larger than that employed in \citep{Daylan:2014rsa} (due to the fact that this analysis uses front and back events, places no cuts on CTBCORE, and utilizes an additional year of data.). Even though the additional events have less statistical strength than the best-event classes used in  \citep{Daylan:2014rsa}, they should still moderately contribute to the location of the GCE center.   

We note three important changes relative to the analysis of \citep{Daylan:2014rsa} that may also contribute to the movement of the GCE template center farther from Sgr A*. First, we have updated our study to utilize the 3FGL, rather than 2FGL, point source catalog. While the 2FGL catalog had 1 point source within $0.1^\circ$ of Sgr A*, and 4 point sources within $1^\circ$ of Sgr A*, the 3FGL catalog has 2 and 7 point sources, respectively. All of these point sources are labeled with the `c' tag in each catalog, indicating that their attributes (and sometimes their very existence) depends sensitively on the diffuse background model of the GC region. If these sources are spurious, they may affect the centering of the GCE template. While these point sources are allowed to have 0 flux in our fit, and thus can be eliminated by the GCE template, they add degrees of freedom into particular sky locations, which might affect the best fitting location of the GCE residual. Second, in the current paper, we have allowed the normalization of each point source to float independently in each energy bin, rather than fixing the point sources to a single spectral template --- this significantly increases the capability of point sources to soak up excesses at given positions in our ROI, potentially affecting the morphology of the floating GCE component.

Finally, we note that in this analysis we have used all P8R2\_Source events, including those that were observed only in the ``back" of the Fermi-LAT instrument. This procedure is well-motivated for our main scientific inquiry, the nature of the GCE $\gamma$-ray emission at high energies. However, this dataset is not optimal for careful evaluations of the GCE center at low energies, since it includes a large number of photons with poor angular reconstructions. While this would be immaterial in the case that systematic errors in the astrophysical $\gamma$-ray foregrounds were well constrained, this is not true near the GC. In particular, in Fig.~\ref{fig:GCResidual}, we note a large region of over-subtraction centered at a latitude $1^\circ$ from the GC, exactly opposite the location of the best-fit GCE center. This over-subtraction region is within the 68\% containment radius of the majority of back-converting events at energies near our spectral peak of $\sim 1$~GeV.

\begin{figure*}[t!]
\centering
\includegraphics[scale=0.45]{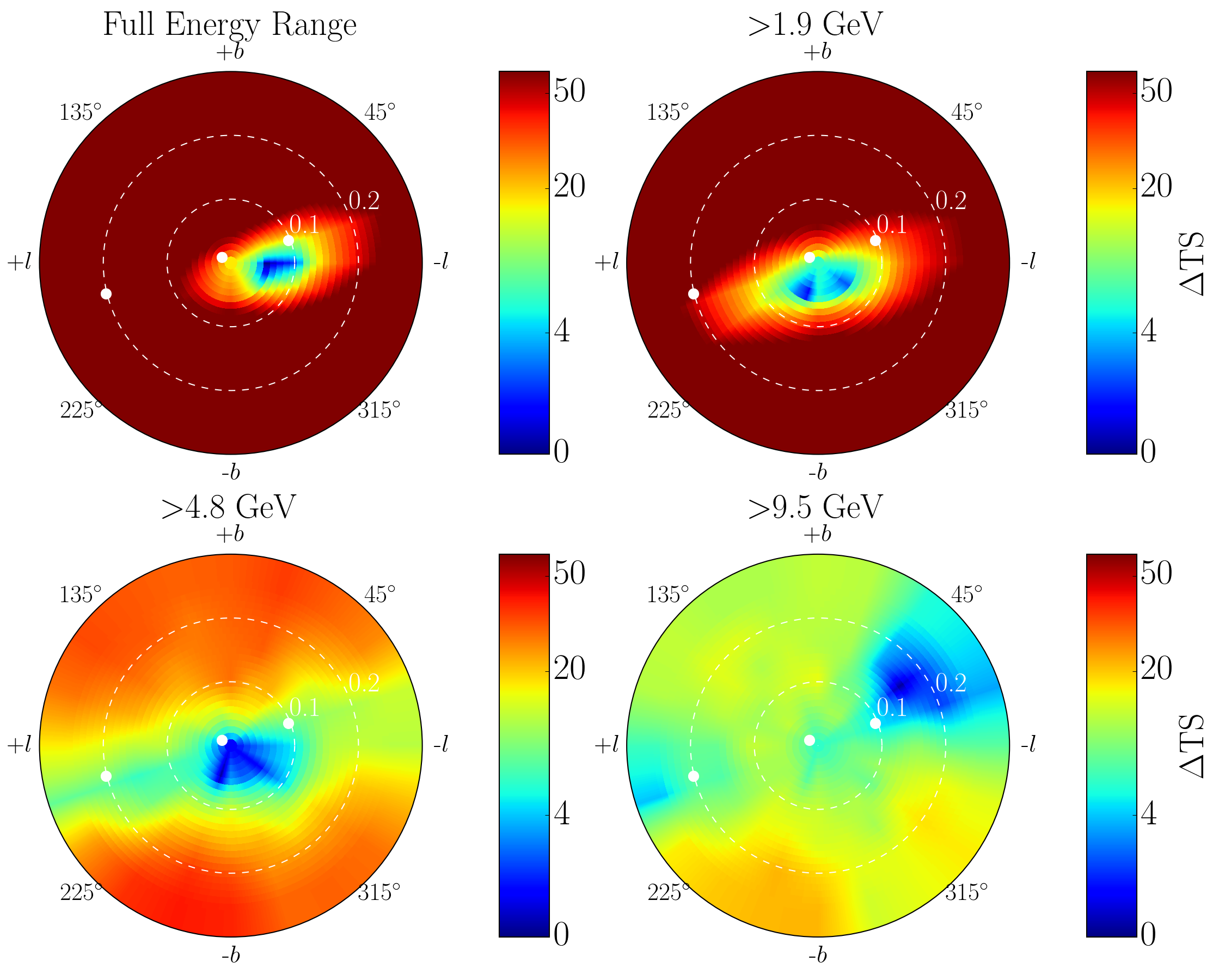} 
\caption{\footnotesize{Same as Fig.~\ref{fig:gc_centering} for an analysis of the Galactic Center ROI that uses only the 25\% of P8R2\_Source class events with the best angular reconstruction (PSF3). Compared to an analysis utilizing the full dataset, the low-energy analysis prefers a center that is closer to Sgr A*, though we still do note a statistically significant offset. The three white dots denote the positions of nearby 3FGL point sources (from left to right 3FGL J1746.3-2851c, 3FGL J1745.6-2859c, 3FGL J1745.3-2903c).}}
\label{fig:gc_centering_psf3}
\end{figure*}

In Fig.~\ref{fig:gc_centering_psf3} we show an analysis where we have utilized only the top quartile P8R2\_Source class photons with the best angular reconstruction (PSF3). We find that in this case, the profile becomes somewhat more centered, preferring a position only $0.05^\circ$ from the GC, with a significance that has dropped to $\Delta {\rm TS} = 18$. This is not only due to the smaller photon statistics, as the PSF3 events provide nearly 2/3 of the total statistical significance in our fit to the data. We do note a deviation in this model in the very high energy range ($>9.5$~GeV), where an offset of nearly $0.2^\circ$ is preferred compared to emission from the GC. However, this statistical preference is at $\Delta{\rm TS} \sim 4$, and it is unclear whether this is due to a statistical fluctuation, or perhaps the presence of an unmodeled point source at high energies. 

\section{Impact of Changing The Masked Regions}
\label{app:BkgROI}

\subsection{Changing the Region of Interest}

In this section we explore the impact of varying the ROI on the IG analysis. It was noted by the authors of~\citep{Daylan:2014rsa} that the spectrum of the GCE, particularly along the plane, was sensitive to the choice of ROI. That work hypothesized that the effect was due to the outer Galaxy preferring a higher normalization for the Galactic diffuse model, causing the inner Galaxy to suffer from systematic oversubtraction, especially along the Galactic plane. (A recent study of the {\it Fermi} Bubbles using similar background models found that the spectrum of the Bubbles in the $|b| = 10-20^\circ$ latitude range was also sensitive to the choice of ROI~\citep{Narayanan:2016nzy}.) For this study we consider four ROIs: 1. our default ROI, referred to as ``30''; 2. our default ROI but reducing the mask of the plane to $|b|<0.3^{\circ}$, referred to as ``0.3''; 3. the smaller of the two ROI considered in~\citep{Daylan:2014rsa} of $|l|<20^{\circ}$ and $1^{\circ}<|b|<20^{\circ}$, referred to as ``40''; and 4. the Full Sky masking $|b|<1^{\circ}$, referred to as ``FS''. 

As a first test, we examine the overall significance of the GCE as a function of the choice of ROI, using Pass 8 data and three different diffuse models; we show results for All Source data in Table~\ref{table:TSSource}, and for UCV BestPSF data in Table~\ref{table:TSUCV}. The TS is generally broadly comparable between the different ROIs, and there is no consistent trend in which ROI has a larger significance for the excess.

\begin{table}[h]
\begin{center}
\begin{tabular}{| c || c | c | c |}
\hline
& \multicolumn{3}{ |c| }{TS for GCE} \\ \cline{2-4}
Model & 30 & 40 & FS \\ \hline \hline
\texttt{p6v11} & 2858.9 & 2289.7 & 3333.6 \\ \hline
\texttt{p7v6} & 1204.7 & 720.9 & 982.9 \\ \hline
\texttt{p8v6} & 266.0 & 239.9 & 798.6 \\ \hline
\end{tabular}
\end{center}
\caption{TS for the GCE template in the IG analysis using All Source data, which includes models for the Galactic diffuse emission (either \texttt{p6v11}, \texttt{p7v6} or \texttt{p8v6}), isotropic emission and the {\it Fermi} Bubbles. See text for details.}
\label{table:TSSource}
\end{table}

\begin{table}[h]
\begin{center}
\begin{tabular}{| c || c | c | c |}
\hline
& \multicolumn{3}{ |c| }{TS for GCE} \\ \cline{2-4}
Model & 30 & 40 & FS \\ \hline \hline
\texttt{p6v11} & 1541.8 & 1175.4 & 1268.5 \\ \hline
\texttt{p7v6} & 779.4 & 477.6 & 439.5 \\ \hline
\texttt{p8v6} & 521.7 & 405.8 & 560.4 \\ \hline
\end{tabular}
\end{center}
\caption{Same as Table~\ref{table:TSSource}, but using UCV BestPSF data.}
\label{table:TSUCV}
\end{table}

More striking differences are apparent when we examine the preferred value of the NFW slope $\gamma$. Results are shown for three background models in Table~\ref{table:BkgChoice}, using Pass 7 data.

\begin{table}[h]
\begin{center}
\begin{tabular}{| c || c | c | c | c |}
    \hline
    Diffuse Model & 30 & 0.3 & 40 & FS \\ \hline \hline
    \texttt{p6v11} & $1.12$ & $1.11$ & $1.21$ & $1.14$ \\ \hline
    \texttt{p7v6} & $1.16$ & $1.12$ & $1.25$ & $1.12$ \\ \hline
    Model F & $1.01$ & $0.87$ & $1.05$ & $\ll 1$ \\
    \hline
\end{tabular}
\end{center}
\caption{The best fit value of $\gamma$ as we change the ROI of the IG analysis for three diffuse models. Note Model F is a poor description of the Full Sky and so when a GCE template is added it wants to be as flat as possible to fix modeling issues over a wide area, which drives the preferred $\gamma$ to very small values. All these results are for Pass 7 data. The 1$\sigma$ statistical uncertainties are $\sim 0.01$ or less and are omitted. See text for details.}
\label{table:BkgChoice}
\end{table}

The results of Table~\ref{table:BkgChoice} make it clear that the choice of ROI has far more impact on the inferred radial variation than the choice of dataset or event quality cuts (App.~\ref{app:SelCr}). Note that changing the mask of the plane from $1^{\circ}$ to $0.3^{\circ}$ does not have a strong impact for \texttt{p6v11} or \texttt{p7v11}, but has a large impact for Model F.

Model F was found to provide a better fit to the data than the \texttt{p6v11} model in~\citep{Calore:2014xka}, but only within their $40^\circ \times 40^\circ$ ROI (with the plane masked at $|b| = 2^\circ$). It appears to perform poorly generally very close to the Galactic plane, which may explain the preference for a very flat ``excess'' (to absorb unmodeled emission over a broad area) in the case where the plane mask is reduced to $0.3^\circ$. This preference for a flat GCE template persists across a wide energy range, suggesting the problems with Model F near the plane are not isolated to just a few energy bins. As such it is not surprising that Model F also performs poorly in our GC analysis, which is why we have chosen to use Model A instead in that region.

To understand in greater depth the dependence of the GCE morphology on the ROI, and its relevance for our current dataset (Pass 8 All Source Data), we follow the analysis of~\citep{Daylan:2014rsa} and divide the GCE template into halves, with one half covering the north and south regions ($|b|>|l|$) and the other the east and west regions ($|l|>|b|$). Comparing the spectrum of these templates to the full GCE template provides a rough metric for sphericity; a perfectly spherical template in isolation of any background would give an identical spectrum for all three. For this analysis we set $\gamma=1.2$ in all cases, although the best-fit $\gamma$ will vary somewhat between ROIs. We show results with the \texttt{p6v11} and \texttt{p7v6} Galactic diffuse models; we do not consider the \texttt{GALPROP}-based models here as they tend to systematically underperform the \texttt{p6v11} model over the whole sky. We do not show results for the \texttt{p8v6} model, given its unsuitability for analysis of extended sources; however we have confirmed that results obtained using it are also strongly ROI dependent (and qualitatively similar to \texttt{p7v6}).

\begin{figure*}[t!]
\centering
\begin{tabular}{c}
\includegraphics[scale=0.4]{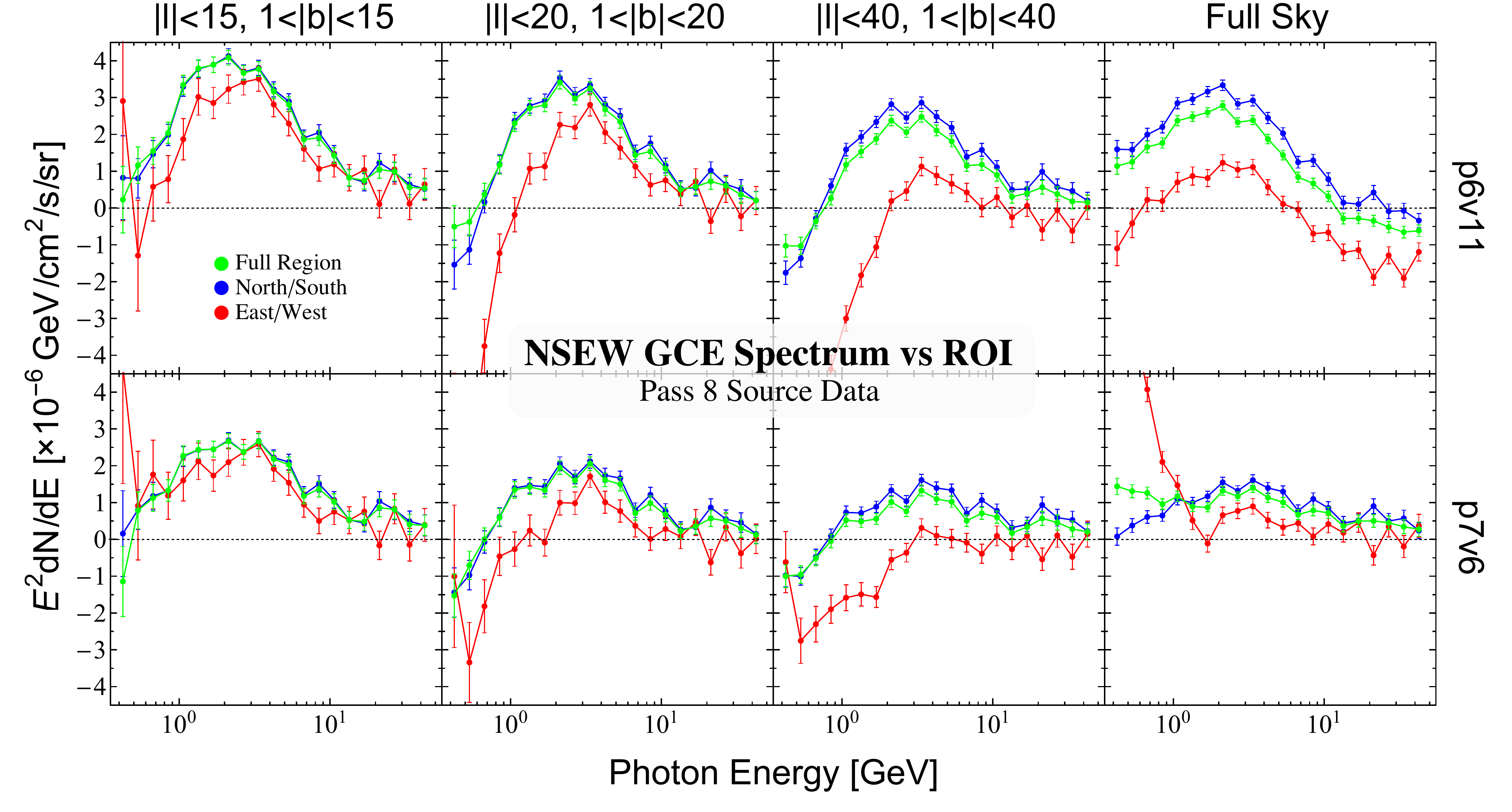} 
\end{tabular}
\caption{\footnotesize{Here we show the IG analysis spectra obtained by three different GCE templates with $\gamma=1.2$, in green that associated with a full template, in blue the results for the north and south regions of the template (where $|b|>|l|$) and in red the east and west contribution (where $|l|>|b|$). We show this using the \texttt{p6v11} diffuse model (top) and \texttt{p7v6} model (bottom) for four different regions: $|b|,|l|<15^{\circ}$ (left), $|b|,|l|<20^{\circ}$ (second from left), $|b|,|l|<40^{\circ}$ (second from right), and the full sky (right). All ROIs mask the plane at 1 degree. See text for details.}}
\label{fig:ROIDep}
\end{figure*}

In Fig.~\ref{fig:ROIDep} we show the result of this analysis for progressively larger sky regions: $30\times 30$, $40\times 40$, $80\times 80$ and then the full sky, in all cases masking the plane at $1^{\circ}$. Given the large PS mask needed for All Source data, it is difficult to make the ROI smaller than $30\times 30$, as much of the region will be masked. (An alternative approach would be to remove the PS mask and carefully model PSs, as we do in our GC analysis; however, this becomes more difficult as the size of the ROI and the number of PSs is increased.) 

As in~\citep{Daylan:2014rsa} we find that in the smallest ROI there is relatively good agreement between the three spectra above about $1$~GeV, whereas for larger regions the disagreement is more pronounced. Unlike~\citep{Daylan:2014rsa}, even in our smallest ROI we find noticeable disagreement at the lowest energies. This is likely the result of our use of lower quality photons with a larger PSF (since we use All Source data as opposed to Ultraclean Q2F). As the {\it Fermi} PSF increases with decreasing energy, at lower energies the GCE template will inevitably begin to absorb emission associated with the Galactic disk, which is very non-spherical. 

Our conclusions are similar for the two Galactic diffuse models. Very similar behavior is also seen when we revert to Pass 7 data, make modest changes to our mask of the plane, or use UCV BestPSF data instead of All Source data (although in that case the discrepancies are somewhat less acute); this appears to be a fairly stable feature of the models we have tested.

One interesting difference between the two Galactic diffuse models occurs in the Full Sky region where we see the East/West contribution becomes very large at low energies when the \texttt{p7v6} model is used. This tends to suggest that whilst the \texttt{p6v11} data has a tendency to oversubtract the plane when fit over large ROIs, the \texttt{p7v6} model appears to be undersubtracting the data. 

Thus we see that especially when morphology is at issue, the choice of ROI can potentially have a larger impact on the properties and spectrum extracted for the GCE than any of the other systematics discussed already, at least for the IG analysis. Our relatively small ROI should help mitigate the severity of the problem. Note that whilst we have focused on the entire excess in this section, this issue is particularly acute for the high-energy analysis. For example the top right panel of Fig.~\ref{fig:ROIDep} highlights that were we to perform our analysis in the full sky ROI, we would conclude that there is no emission at high energies.

\subsection{Changing the Point-Source Mask}

As discussed previously, Fig.~\ref{fig:SpecVsDataSet} demonstrates that the spectrum extracted for the GCE can vary somewhat depending on the dataset used, especially at low energies. In this section we point out that this difference is most likely due to the different size of the PS masks used for the two datasets. Recall by default we mask the 300 brightest sources in the 3FGL catalogue, with a mask radius determined by the 95\% containment radius of the PSF at that energy. As the PSF varies with dataset, so does the size of our mask, and it will be much larger for the All Source data than for UCV BestPSF.

\begin{figure*}[t!]
\centering
\begin{tabular}{c}
\includegraphics[scale=0.45]{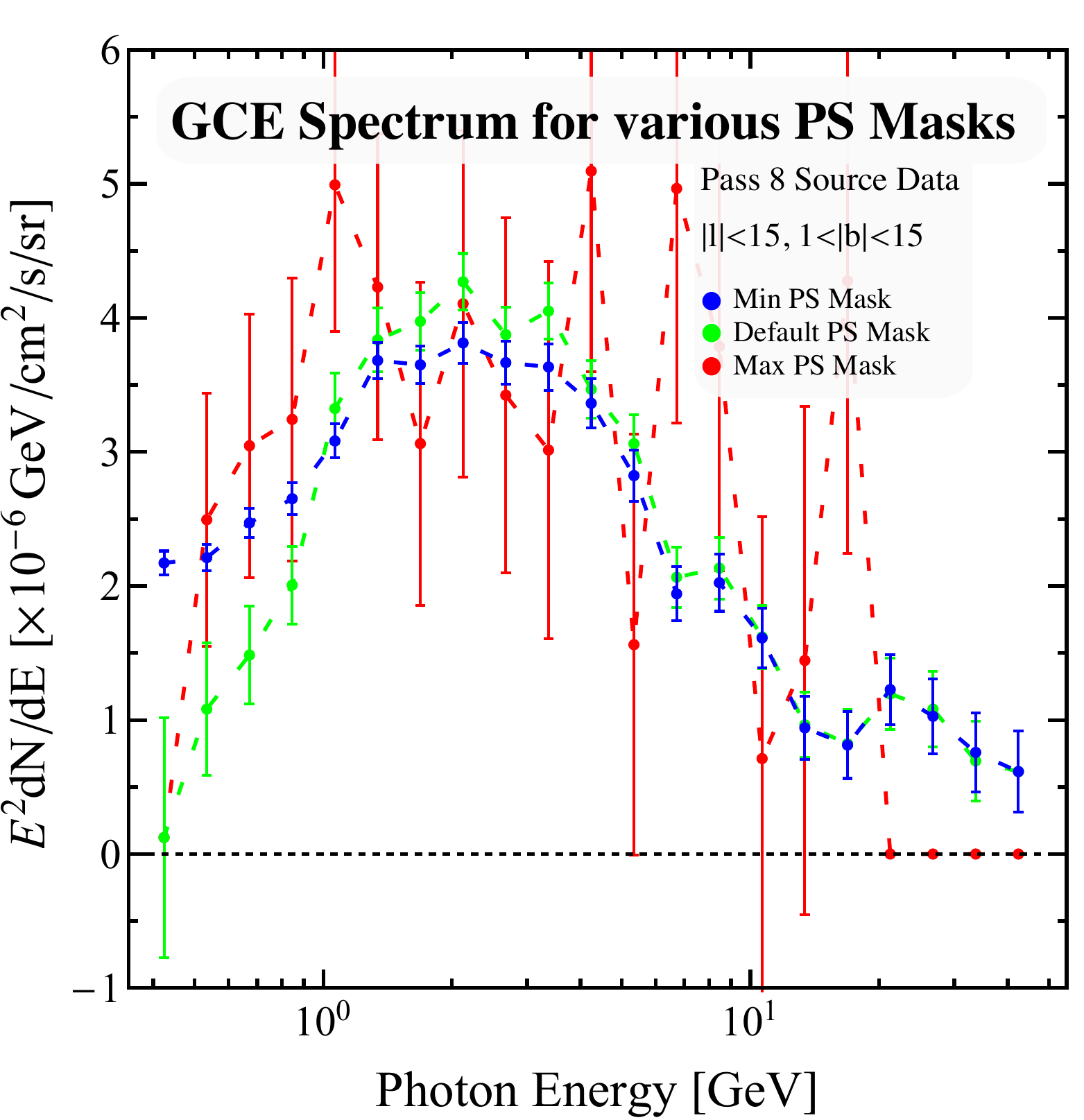} \hspace{0.15in}
\includegraphics[scale=0.45]{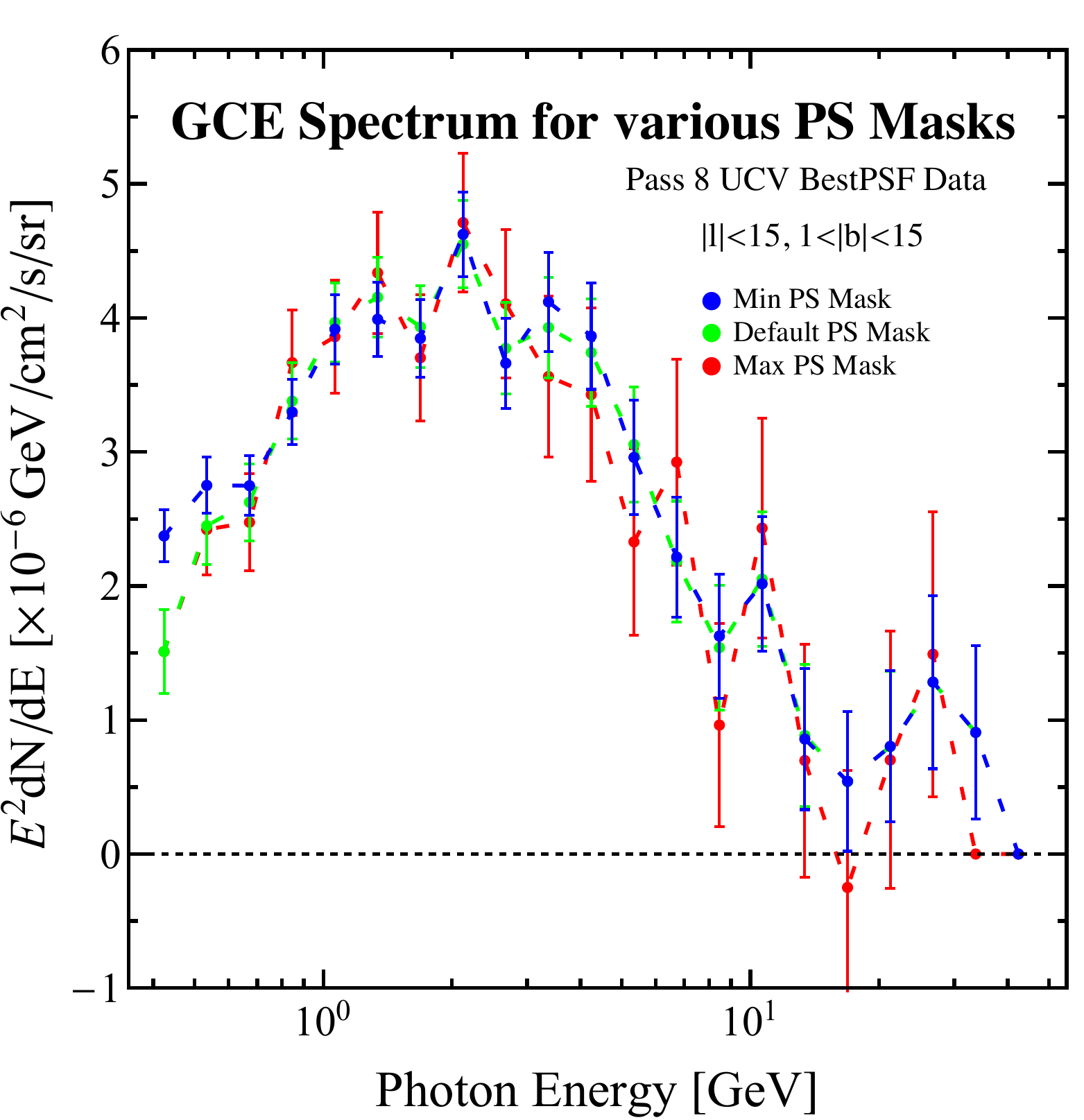}
\end{tabular}
\caption{\footnotesize{Here we show the impact of changing the size of the PS mask on the spectrum extracted for the best fit GCE template in the IG analysis. In green we show the default mask, which varies with energy according to the PSF of {\it Fermi}, in blue use a mask size fixed at the highest energy considered, and in red the lowest energy. We show this for All Source data on the left and UCV BestPSF data on the right, from which it is clearly a much bigger issue in source data. Note that in both cases for the largest mask, in red, where the high-energy points go to zero correspond to a failure of the fit to converge given the very low statistics.}}
\label{fig:PSMask}
\end{figure*}

In order to check this, in Fig.~\ref{fig:PSMask} we show how the extracted spectrum in the IG analysis changes as we vary this masking procedure for two additional cases. In green we show our default masking procedure, whilst in blue (red) we use a fixed mask radius set to the 95\% containment radius for the highest (lowest) energy considered. Thus the blue points represent a much smaller PS mask than the red points. 

In the case of UCV BestPSF, there is very little difference in the extracted spectrum between these cases. This is in line with a similar cross check performed in~\citep{Daylan:2014rsa}, which used Pass 7 Ultraclean Q2F events and thus also photons with high quality angular reconstruction.

However, in the case of All Source data the difference is dramatic, particularly in the case where we use the largest mask. The larger error bars on the spectrum in this case are due to the PS mask removing a large part of the ROI. The results with the minimal PS mask are similar to those for our default analysis except at low energies; perhaps more interestingly, they are also much closer to the results with UCV BestPSF data. This suggests that much of the apparent difference between the All Source and UCV BestPSF results at low energies may be due less to the photon quality and more to the different PS masks.

However, since our analysis focuses primarily on high energies, the PS mask appears to be small enough that the results are converged within reasonable choices for the mask.

\section{Comparison of Results in the Galactic Center and Inner Galaxy}
\label{app:gc_spec_appendix}

\begin{figure*}[t!]
\centering
\includegraphics[scale=0.45]{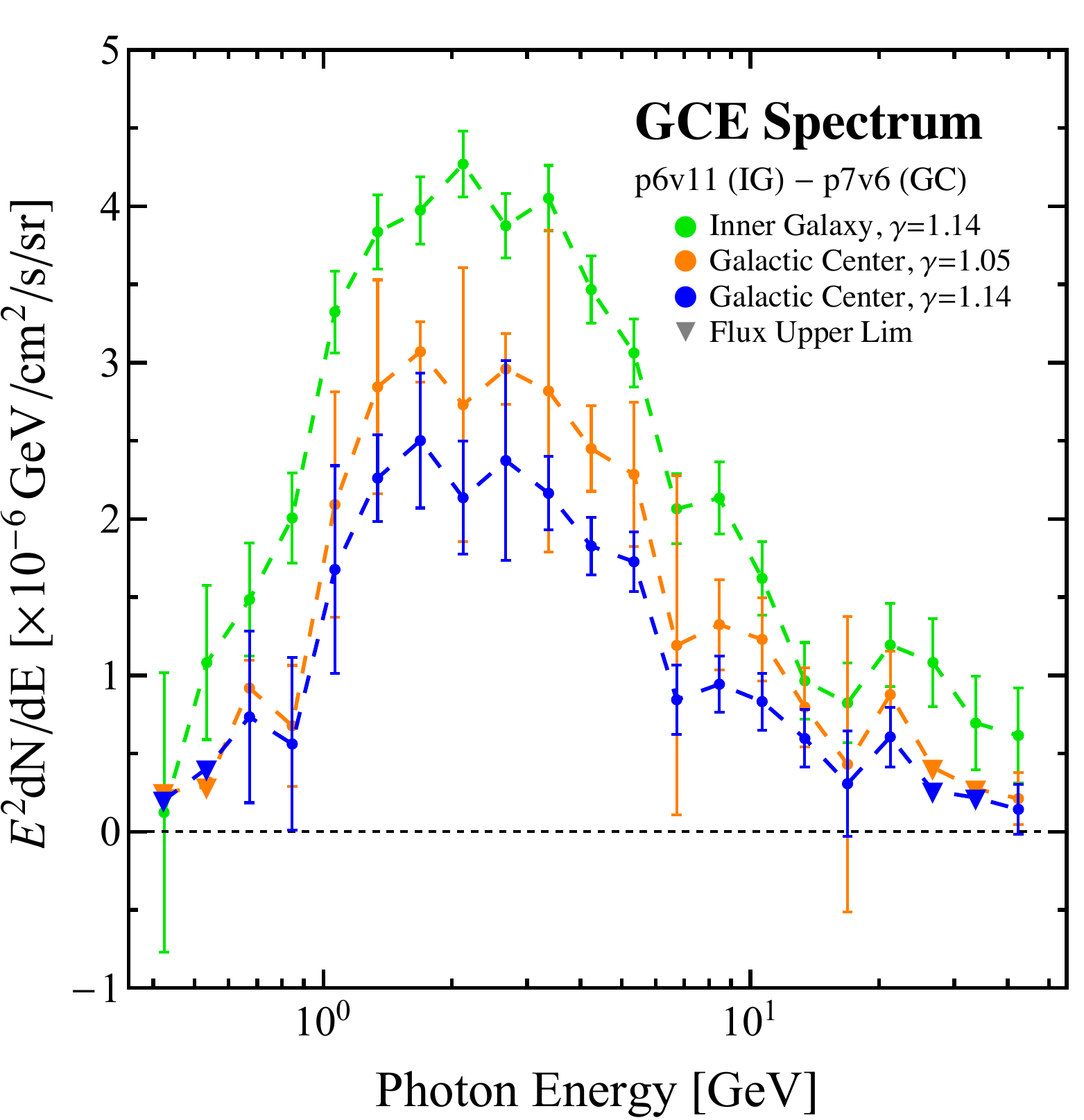} \hspace{0.15in}
\includegraphics[scale=0.45]{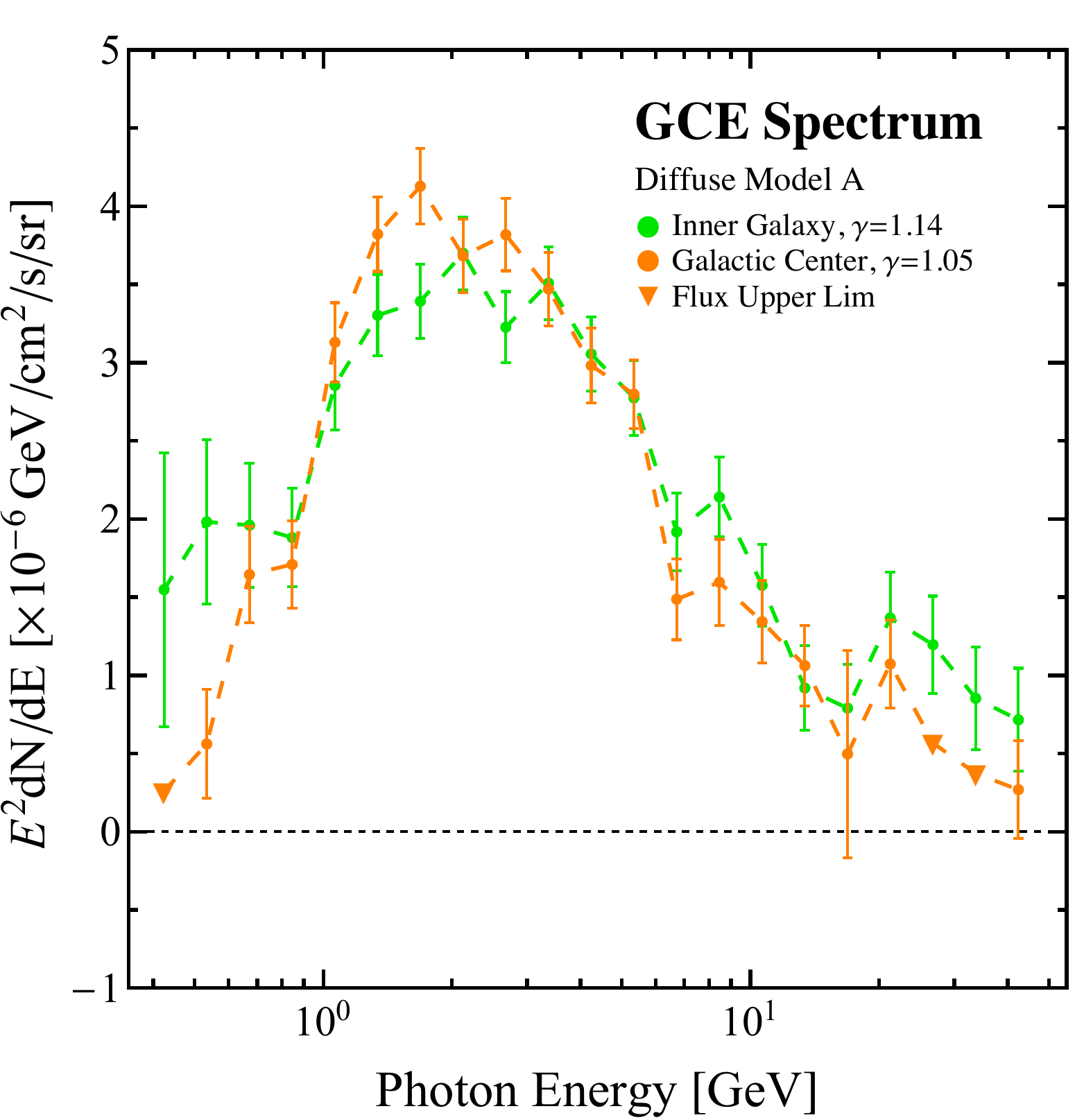}
\caption{\footnotesize{On the left we show the spectrum and normalization of the GCE in the IG analysis for the best fit profile slope of $\gamma=1.14$, compared to the GC analysis for an identical best-fitting profile slope. We note that the GCE in the GC ROI contributes a smaller normalization than in the IG ROI, a result which hints at a potential steepening of the inner-profile slope as a function of radius. On the right, the GCE spectrum for the default IG and GC analyses is shown with both using the Model A diffuse emission model. Here we use the best-fit profile values from the default analyses, $\gamma=1.14$ for the IG and 1.05 for the GC. Here we see that when using the same diffuse emission models the GCE templates can produce similar fluxes at $5^\circ$ from the GC, but note this is not true for all diffuse models. See text for details.}}
\label{fig:gcSpec_compare}
\end{figure*}

Another worthwhile test of the consistency of the GCE in different analyses concerns the best-fit normalization of the GCE component in the GC and IG ROIs. Since the normalizations of the GCE are allowed to float independently in each analysis, it is conceivable that the GCE excess could float to very different normalizations, suggesting a breakdown of our simple power-law/NFW-like model for the radial variation. In Fig.~\ref{fig:BaseSpec} we compared the spectrum and normalization of our best fitting models at a position $5^\circ$ from the GC, and found that the GC analysis prefers a normalization that is approximately 30\% smaller than the IG analysis. Additionally, the two analyses prefer slightly different inner profile slopes of $\gamma=1.05$ ($\gamma=1.14$) for the GC (IG) analysis. 

On the left of Fig.~\ref{fig:gcSpec_compare} we show the resulting spectrum and normalization of the GCE component when the NFW inner profile slope is fixed to be the same value ($\gamma=1.14$) in each analysis. We note two immediate conclusions. First, we find that the spectral features of the GCE as obtained from the GC analysis are almost entirely unchanged by the new value of $\gamma$, including the bin-to-bin variation in the spectral features. However, the total intensity of the GCE (normalized to the emission intensity at $5^\circ$ from the GC) decreases by approximately 20\%. This is not surprising, as this region lies near the edge of our GC analysis ROI, and thus changes that make the excess more peaked will tend to decrease the intensity of the excess in regions far from the GC. However, this increases the already existing offset between the best fit intensity of the excess at $5^\circ$ from the GC between the GC and IG analyses (Fig.~\ref{fig:BaseSpec}), bringing the mismatch up to approximately a factor of 2. This provides some indication that the emission mechanism producing the GCE does not represent a pure power-law throughout the full ROI of the GC and IG analysis. This result, while shown clearly via the comparison between Fig.~\ref{fig:BaseSpec} and Fig.~\ref{fig:gcSpec_compare}, has been hinted at through many analyses performed over the last several years~\citep{Hooper:2013rwa, Abazajian:2014fta, Daylan:2014rsa, Calore:2014xka}. Notably, when the ROI of the $\gamma$-ray analysis is increased, the preferred value of $\gamma$ increases as well (c.f. Table~\ref{table:BkgChoice}). The significance of this is, as of yet, unclear. Radially dependent changes in $\gamma$ can be naturally accommodated in leptonic outburst, or Millisecond Pulsar models, but can also be accommodated in dark matter models if baryonic effects alter the dark matter density profile very near the GC, as is predicted in many structure formation models. An alternative explanation involves the many degrees of freedom in the GC analysis, compared to that in the IG, which may allow PSs near the GC to soak up much of the $\gamma$-ray excess, artificially dimming the GCE component. Using the 2FGL PS catalog rather than the more recent 3FGL increases the apparent amplitude of the excess by $\sim 30\%$.

The GC and IG analyses also use different default Galactic diffuse models (as explained in the main text); one might wonder whether the observed differences arise from this choice. Using the \texttt{p7v6} model in the IG does not decrease the preferred value of $\gamma$, but does reduce the overall normalization of the GCE spectrum, making the result consistent with the GC analysis. On the other hand, the converse is not true, utilizing the \texttt{p6v11} diffuse model in the GC analysis actually further decreases the intensity of the GCE by 20\%, increasing the mismatch with the default IG results. This second scenario is well understood, however, as the \texttt{p6v11} diffuse model has an angular scale of $0.5^\circ$ and has no latitude bin corresponding to b=0.0, making it unsuitable for studies of the dense GC ROI. On the right side of Fig.~\ref{fig:gcSpec_compare} we show the comparison when both analyses use the Model A diffuse emission model, which is applicable in both ROIs, and there we see agreement between the spectra extracted in both regions. Nevertheless we emphasize that in general the spectral differences between the two analyses depend on a number of systematics as discussed, and should not be thought of as just originating from the differences in diffuse modeling.

\section{Cross-checks of the non-Poissonian Template Fit Analysis}
\label{app:NPTFCheck}

In this appendix we consider variations of the NPTF analyses presented in Sec.~\ref{sec:NPTF} and shown in Fig.~\ref{fig:Bayes}.

By default we let all of the parameters in the source-count function, shown in Eq.~\ref{SourceCount}, float independently. However, one might worry that this gives the fit too much freedom, especially as there is some a priori expectation for the low-luminosity slopes of the GCE and disk PS source-count functions. If the spherical PS population arises from a population of millisecond pulsars, then it may be natural to expect $n_2 \approx 1.5$~\cite{Petrovic:2014xra,Cholis:2014noa,Strong:2006hf,Venter:2014zea}.  The disk PS population was found in~\cite{Lee:2015fea} to have an index $n_2 \approx 1.4$.  

\begin{figure*}[t!]
\centering
\begin{tabular}{c}
\includegraphics[scale=0.5]{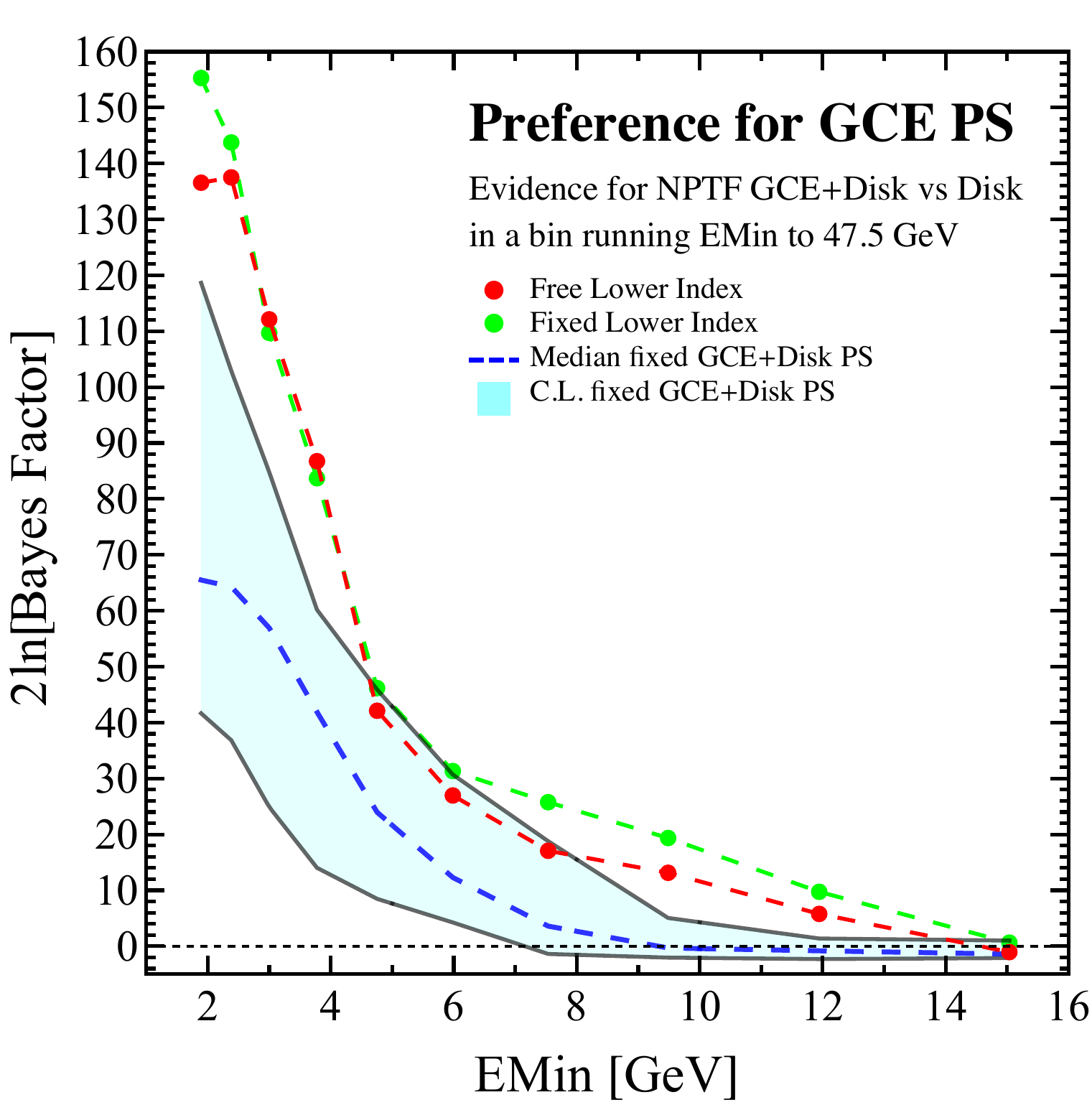} \hspace{0.15in}
\includegraphics[scale=0.5]{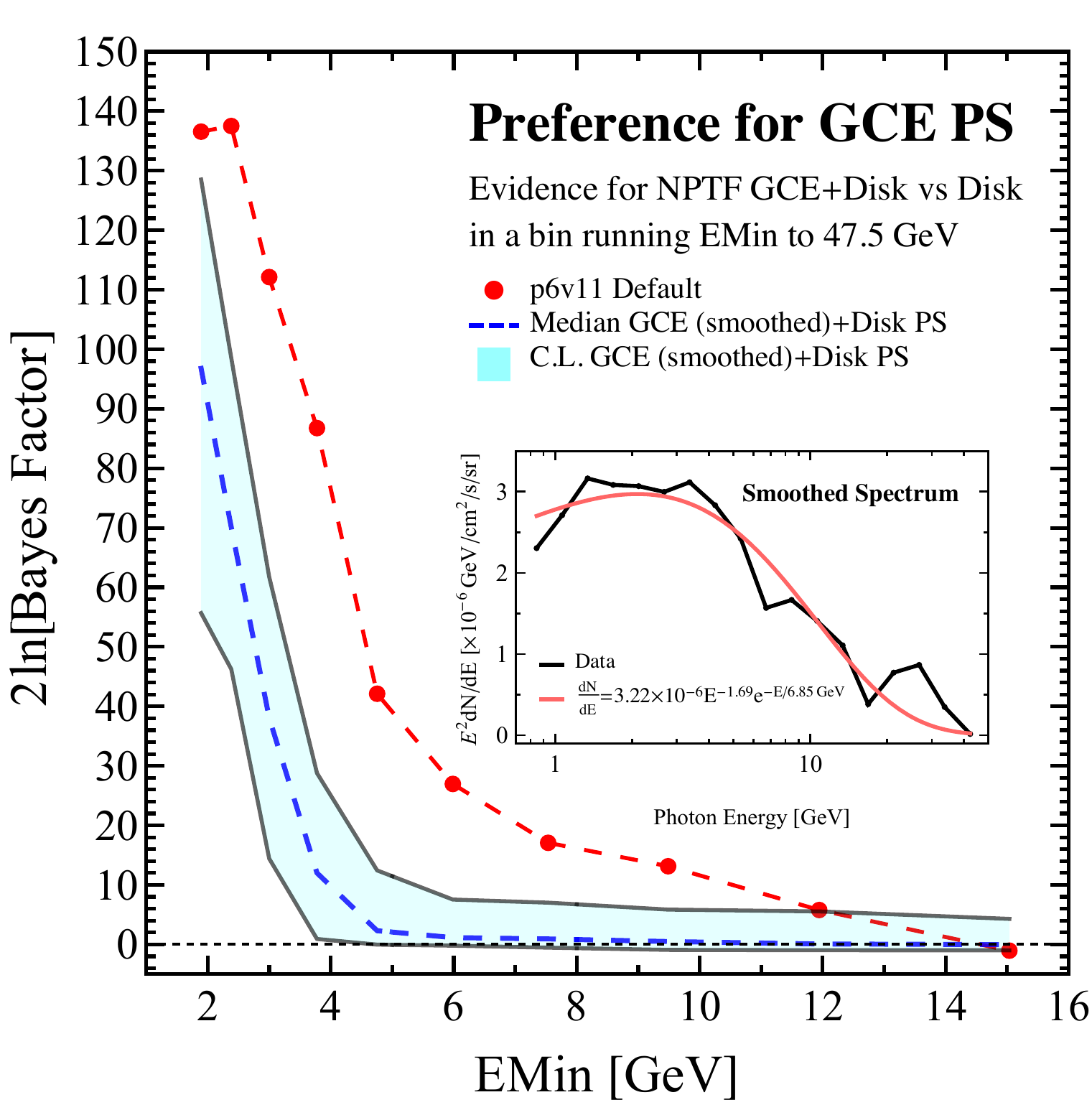}
\end{tabular}
\caption{\footnotesize{Here we show two variations on the results shown in the left-hand panel of Fig.~\ref{fig:Bayes}. On the left we show the default results in red, but in green show what happens if we fix the lower index ($n_2$ in Eq.~\ref{SourceCount}) to be $1.5$ for the GCE PS template and $1.4$ for the disk PS template. The simulated data 90\% confidence interval (blue) shown also assume a fixed value of $n_2$ both in the production of the simulated data and the analyses. On the right, we again show the default analysis in red, but this time the simulated data is generated assuming a smoothed spectrum for the GCE shown inset, rather than the spectrum extracted for the Poissonian GCE template shown in black. The smoothed spectrum is determined by fitting the data to a power law with an exponential cutoff. See text for details.}}
\label{fig:BayesCheck}
\end{figure*}

We have studied the effect of fixing these two indices to these values, while letting all other parameters float.  The resulting energy-dependent Bayes factor in preference for the model with spherical PSs is shown in green in the left panel of Fig.~\ref{fig:BayesCheck}.  We have verified explicitly that small variations in the values of the fixed indices do not affect the results.  For reference, we also show the Bayes factor for our default NPTF analysis (shown in red), where all indices are allowed to float. Medians and 90\% confidence limits determined from simulated data, produced and analyzed assuming a fixed lower index, are shown in blue. The similarity between the cases with fixed and floating indices suggests that our choice to float all parameters has not biased our conclusions.

In producing the simulated data, we assume the GCE PS template has the energy spectrum extracted for the Poissonian GCE in a fit performed with no non-Poissonian templates. We do this because the approach used in this work does not allow the spectrum associated with a given non-Poissonian template to be carefully extracted. One might wonder how uncertainties in this data-driven (and hence fluctuating) spectrum could propagate and impact our results. 

To test this question, we remade the simulated data assuming a smoothed spectrum for the GCE PS template, derived by fitting a power law with an exponential cutoff to the GCE spectrum; specifically, the smoothed spectrum was given by $dN/dE \propto E^{-1.69} e^{-E/6.85{\rm~GeV}}$.\footnote{Such a spectrum might be expected if the GCE originates from millisecond pulsars, although the presence of the high-energy tail leads us to derive a larger value for where the exponential cutoff enters than usually associated with millisecond pulsars. See, e.g.,~\citep{Bartels:2015aea}.} Our new energy-dependent medians and 90\% confidence limits for the Bayes factor, derived from this new simulated data, are shown on the right of Fig.~\ref{fig:BayesCheck}, with the smoothed spectrum shown in the inset. This variation on the assumed GCE PS spectrum has a non-negligible impact on the derived confidence limits from multiple Monte Carlo simulations, especially at high-energy, where we see that even if the GCE PS population is present, it becomes difficult to confirm this presence above $4$~GeV (as compared to $6$~GeV in our default analysis). This cross-check further emphasizes our main result for the NPTF analysis, i.e. that there is simply insufficient statistical power at present to determine if the high-energy tail of the excess has a PS origin, at least via this method.

\begin{figure}[t!]
\centering
\includegraphics[scale=0.53]{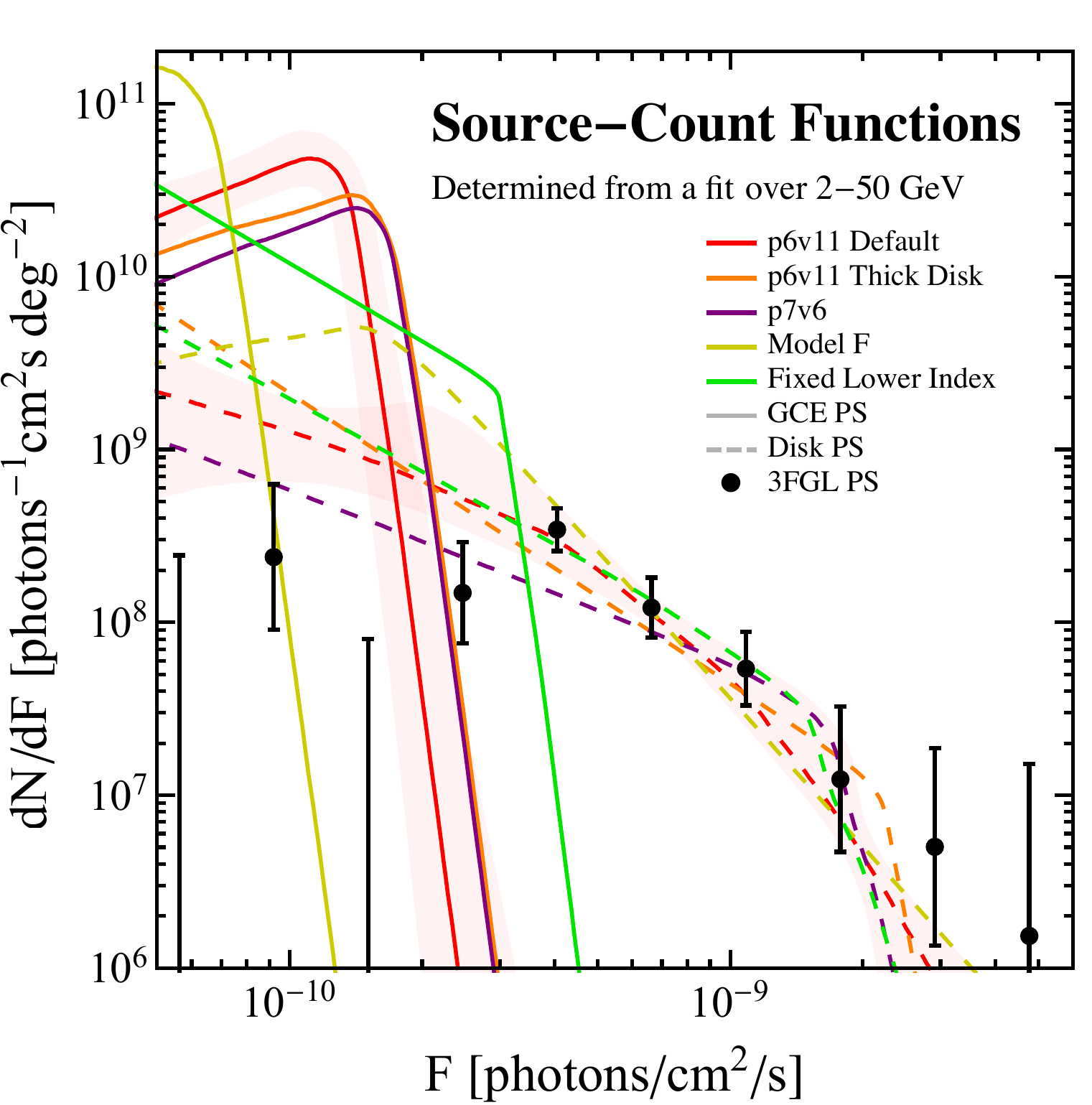}
\caption{\footnotesize{Best fit source-count functions derived from a fit over $2-50$~GeV for a GCE PS (solid curve) and disk PS (dashed curve). The various colors correspond to five variations on the modeling: default analysis with \texttt{p6v11} (red), use of a thick disk (orange), move to \texttt{p7v6} (purple), Model F (khaki), and default analysis but fixing the lower index ($n_2$ in Eq.~\ref{SourceCount}) of the GCE and disk PS to be $1.5$ and $1.4$ respectively (green). For the default analysis only we show the 68\% confidence limits also, to give some indication of the statistical uncertainty. In black dots we show a histogram of the observed flux from 3FGL sources in that energy range range, with 68\% Poissonian error bars simply coming from counting statistics.}}
\label{fig:SourceCount}
\end{figure}

In Fig.~\ref{fig:SourceCount} we show the extracted source-count function for both GCE and disk PS populations in our default NPTF analysis over the $2-50$~GeV range,\footnote{For the default analysis, we checked that the two source-count functions extracted in the $5-50$~GeV range are consistent with those shown. This means that values of $n_1$ and $n_2$ are consistent, as the overall curve is moved to lower fluxes, as expected if the point sources emit fewer photons at higher energies.} and also the variations considered in this appendix. As previously we study a large energy range in order to probe the impact of systematic effects in a regime where statistics are large. For our default analysis only, in shaded red, we also illustrate the 68\% confidence interval, constructed point-wise from the posterior distribution, for the source-count function.  The black points come from histogramming the 3FGL PS catalog within this energy range and our ROI.  In constructing this histogram, which is for illustrative purposes only, we do not exclude PSs that are explicitly known to be of extragalactic origin. It is clear that the identified 3FGL sources are well-described by the disk PS population, while the spherical PS population is predominantly found to be below the current PS detection threshold in all of our analyses (this conclusion agrees with the analysis of~\citep{Lee:2015fea}, which used a larger ROI and different data selection cuts).

\section{Scrutinizing the Behaviour in the Bin at 11.9--18.9~GeV}
\label{app:15G}

One notable feature in the high-energy spectrum, as shown in the right hand panel of Fig.~\ref{fig:BaseSpec}, is the dip at $11.9-18.9$~GeV. If this were a genuine feature of the excess it could potentially be a useful handle to help understand the origin of the GCE. However, in this section we argue that the dip appears to be associated with background mismodeling, suggesting that this feature may not be a robust property of the GCE. 

To begin with, however, we note that in the IG analysis, this structure in the high-energy spectrum is not peculiar to the default choice of diffuse model, ROI or dataset. It appears in all 17 diffuse models considered in this work, the details of which are given in App.~\ref{app:Bkg}; it persists if the ROI is increased, if we apply quality cuts on the photons being used, or if we revert to the Pass 7 Reprocessed data.

A similar spectral feature appears to be present in the spectrum calculated from the GC analysis, albeit at a lower statistical significance. This lends some credence to the interpretation that the dip is physical. However, it should be remembered that the GC and IG datasets are not entirely independent, as the ROIs of the two analyses overlap. Additionally, systematic errors in the diffuse emission templates may propagate through both analyses (although the IG template utilizes the {\tt p6v11} diffuse model and the GC analysis utilizes {\tt p7v6}, these models are based on similar physics, multi-wavelength data, and modeling techniques).

\begin{figure*}[t!]
\centering
\begin{tabular}{c}
\includegraphics[scale=0.45]{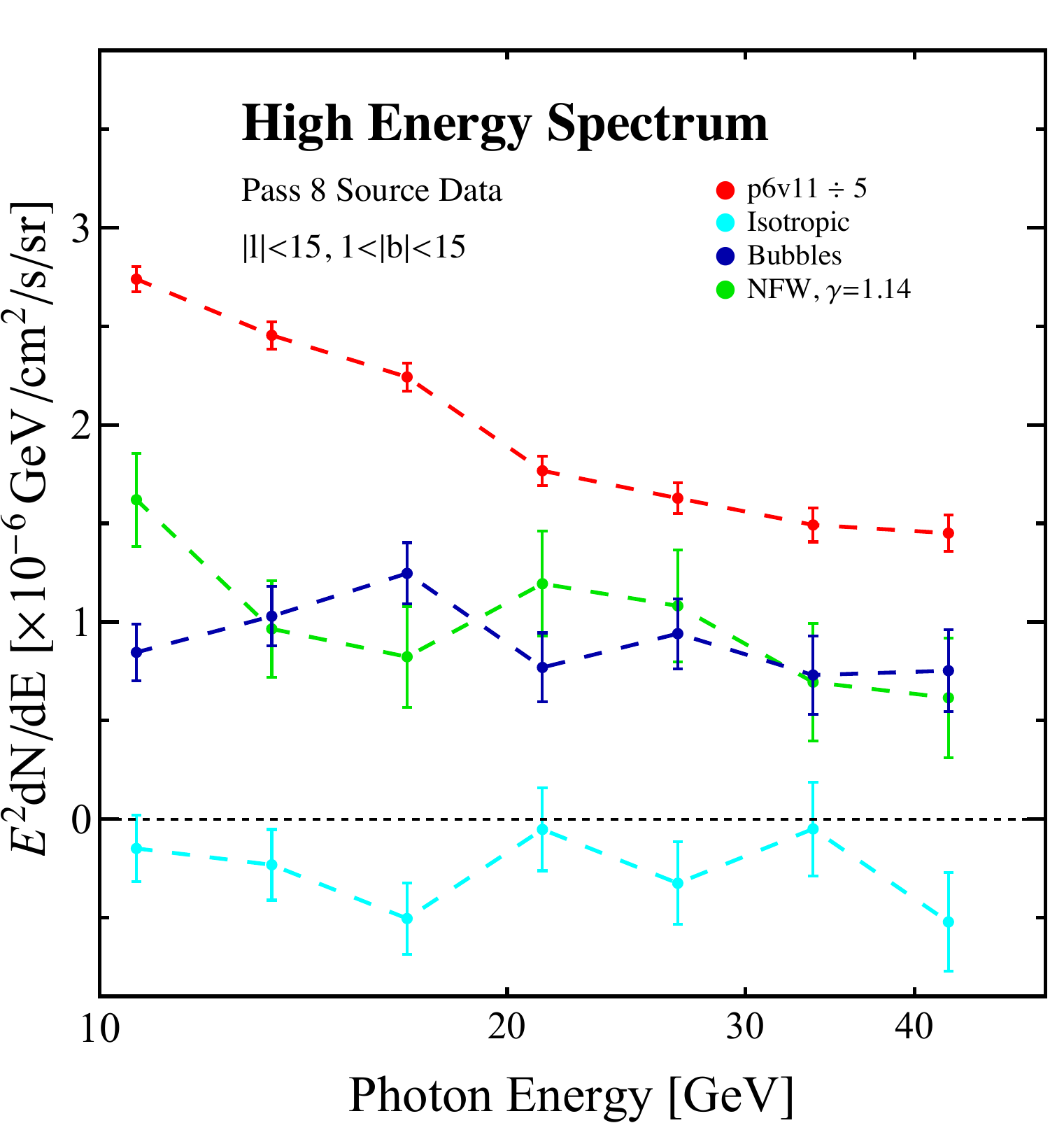} \hspace{0.15in}
\includegraphics[scale=0.45]{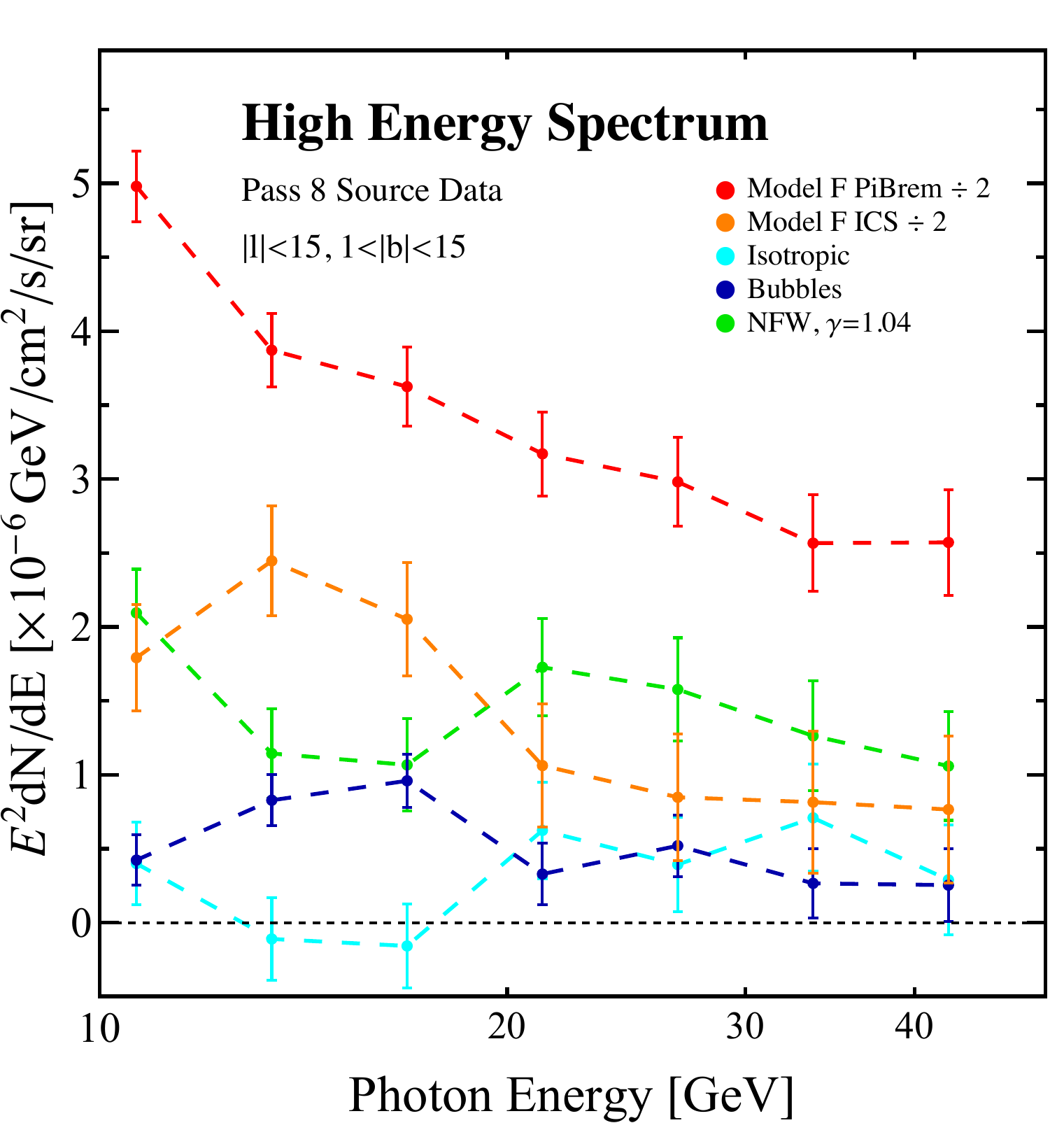}
\end{tabular}
\caption{\footnotesize{Here we show the spectrum associated with all templates at high energies in the IG analysis. We show this for \texttt{p6v11} (left) and Model F (right) diffuse models. Note in both the coefficients of the diffuse components have been rescaled to make the comparison between templates more straightforward. See text for details.}}
\label{fig:Coeff15G}
\end{figure*}

Fig.~\ref{fig:BaseSpec} only shows the spectrum for the GCE template, so in Fig.~\ref{fig:Coeff15G} we show the spectra for all the templates that were included in the fit. For the left panel we used the \texttt{p6v11} diffuse model for the Galactic diffuse emission, whereas for the right panel we replaced it with Model F, which recall has the $\pi^0$ and the bremsstrahlung template floated independently of the ICS component. In both panels we show the spectra for the GCE based on the best-fit slope for the generalized NFW profile, and we have also rescaled the non-GCE spectra to aid the comparison.

Looking firstly at the \texttt{p6v11} case, we can see that where the GCE spectrum dips, the spectrum for the {\it Fermi} Bubbles rises, whilst the diffuse model falls less steeply than in the subsequent (higher-energy) bin. If this dip was a real feature of the GCE it need not be accompanied by any features in the spectra of the other templates; the presence of these apparently correlated features raises the concern that the dip may be related to mismodeling of the other components.

To follow up on this, we can also consider the Model F background, recalling that this model has an additional degree of freedom in the modeling of the diffuse emission, due to the free ICS template. In this case we see a clear bump in the reconstructed ICS spectrum, in the same energy bin as the dip in the GCE spectrum, suggesting a mismodeling of the ICS (or a mismodeling that spatially overlaps the ICS template) might be responsible for the dip.  As an additional check we tried rerunning the Model F case, but fixing the ICS component to have a value in the bins between $11.9-18.9$~GeV that follows the slope on either side of these bins. When doing this we find the $\pi^0$ plus bremsstrahlung template now fluctuates upwards and the GCE still dips. If we scan for the preferred $\gamma$ of the GCE in this case, the fit prefers a very small value of $\gamma$, suggesting that the GCE template is absorbing some spatially broad discrepancy between the model and the data.

\begin{figure*}[t!]
\centering
\begin{tabular}{c}
\includegraphics[scale=0.65]{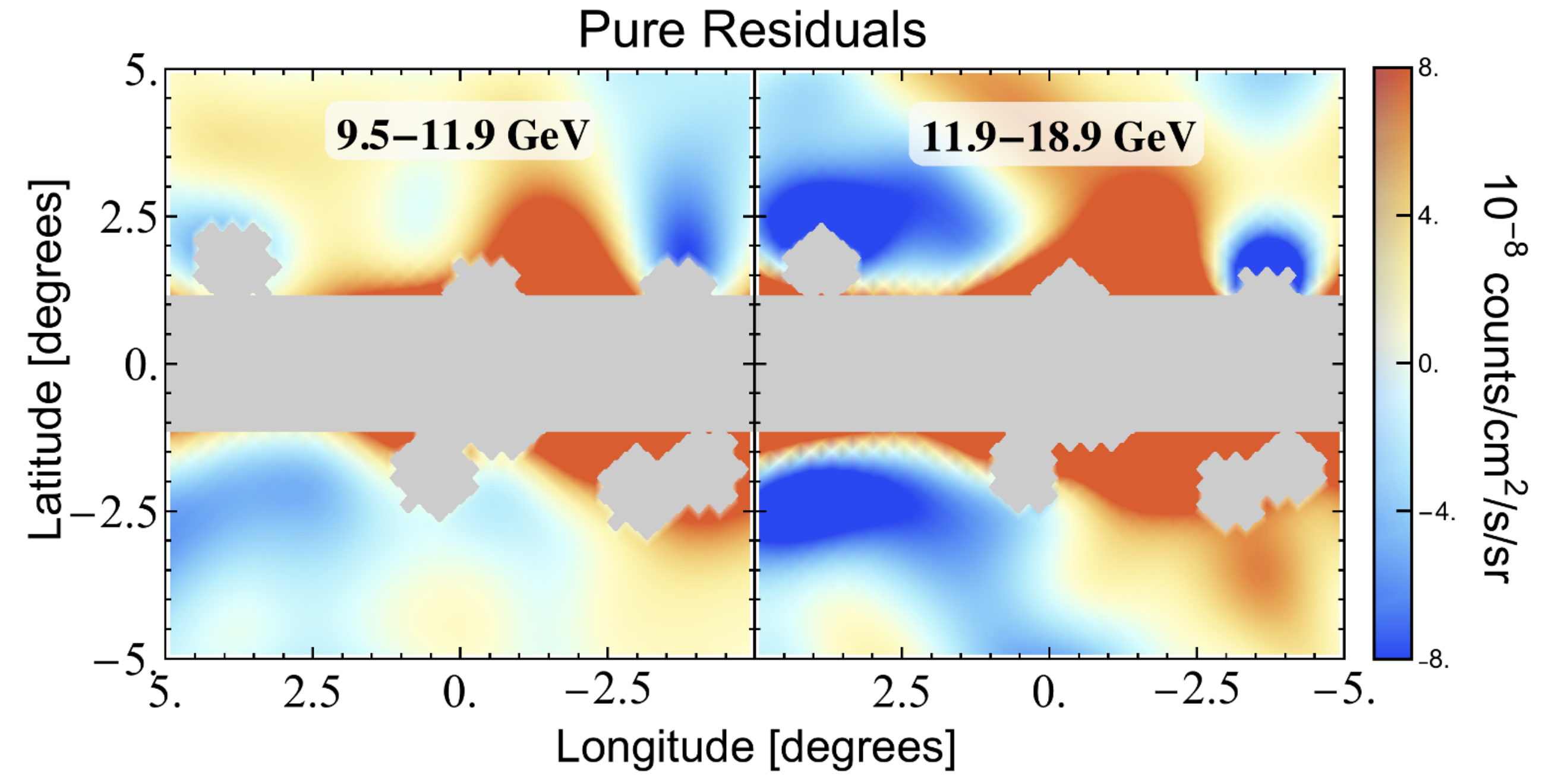}
\end{tabular}
\caption{\footnotesize{Here we show the analogue of the right hand side of Fig.~\ref{fig:FullResidual}, but for the first two high-energy bins separately. The second energy bin, shown on the right, is the focus of App.~\ref{app:15G}. It is clear that in this bin oversubtraction is more pronounced than in the lowest energy bin. The residuals for the two higher energy bins not shown are much closer to the left hand plot than the right.}}
\label{fig:ResBin1and2}
\end{figure*}

We can gain more insight into this mismodeling by examining the residual maps for individual energy bins. For the first and second energy bin we show the residual maps in Fig.~\ref{fig:ResBin1and2}, with a different scale to highlight oversubtraction. From these we see that while there is under and over subtraction in both energy bins, it is particularly bad around the center in the second bin. This could lead to the GCE template preferring an artificially small coefficient, as its presence (with a positive coefficient) would worsen any oversubtraction near the center, worsening the fit. These conclusions were found to be consistent between the various Galactic diffuse models, ROIs and datasets we have tested. The residual plots for the two higher energies not shown also show some oversubtraction, but appear closer to the map for the first bin, rather than for the second (where the dip is present).

To quantitatively determine whether the issue was restricted to a particular area of the sky, within the region where the GCE is bright, we imposed an exponential radial cutoff on the GCE, at a cutoff radius that was allowed to vary (to a minimum of 2 degrees). This did not remove the spectral feature. We also repeated the analysis examining only the north, south, left or right halves of the sky, and found that the dip was consistently present in all regions.

While the regions of large oversubtraction present in the residual maps, and the unexpected behavior of the diffuse background   components in this energy bin, do not definitively rule out the possibility that the dip is a physical feature of the GCE spectrum, any such interpretation should be treated with great caution. A better understanding of this feature will likely require a better understanding of the diffuse gamma-ray backgrounds around the GC region.

\section{Energy Binning Considerations}
\label{app:Binning}

\subsection{Choice of High-Energy Binning}

\begin{table}[h]
\begin{center}
\begin{tabular}{| c || c | c | c | c | c | c | c |}
\hline
Bin & 1 & 2 & 3 & 4 & 5 & 6 & 7 \\ \hline \hline
TS & 59.2 & 16.8 & 10.8 & 22.0 & 15.5 & 6.0 & 4.4 \\ \hline
\end{tabular}
\end{center}
\caption{Analogue of Table~\ref{table:TSP7P8} for the energy bins before combining them, for our default analysis using Pass 8 data.}
\label{table:TSAllBins}
\end{table}

\begin{figure*}[t!]
\centering
\begin{tabular}{c}
\includegraphics[scale=0.45]{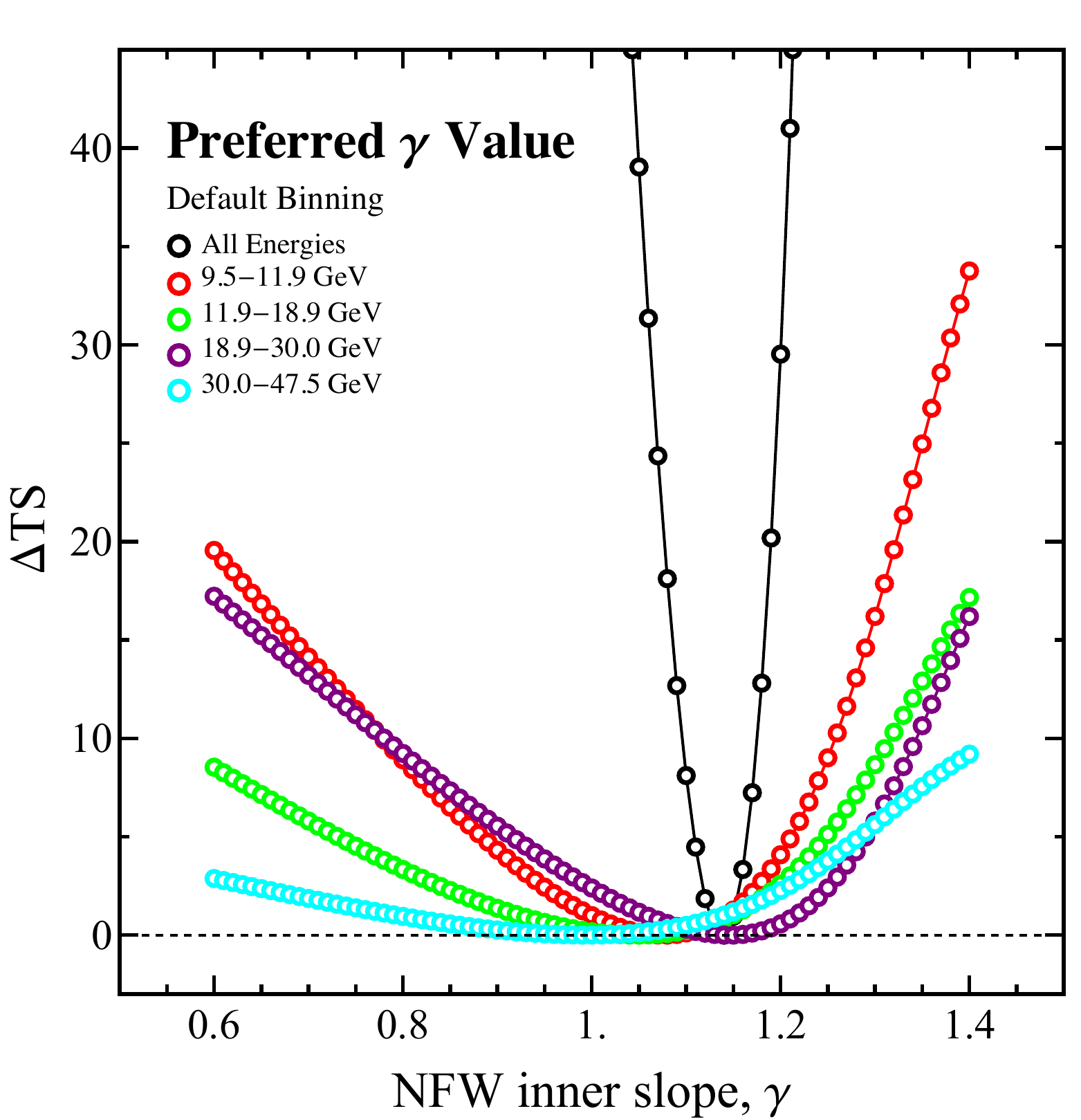} \hspace{0.15in}
\includegraphics[scale=0.45]{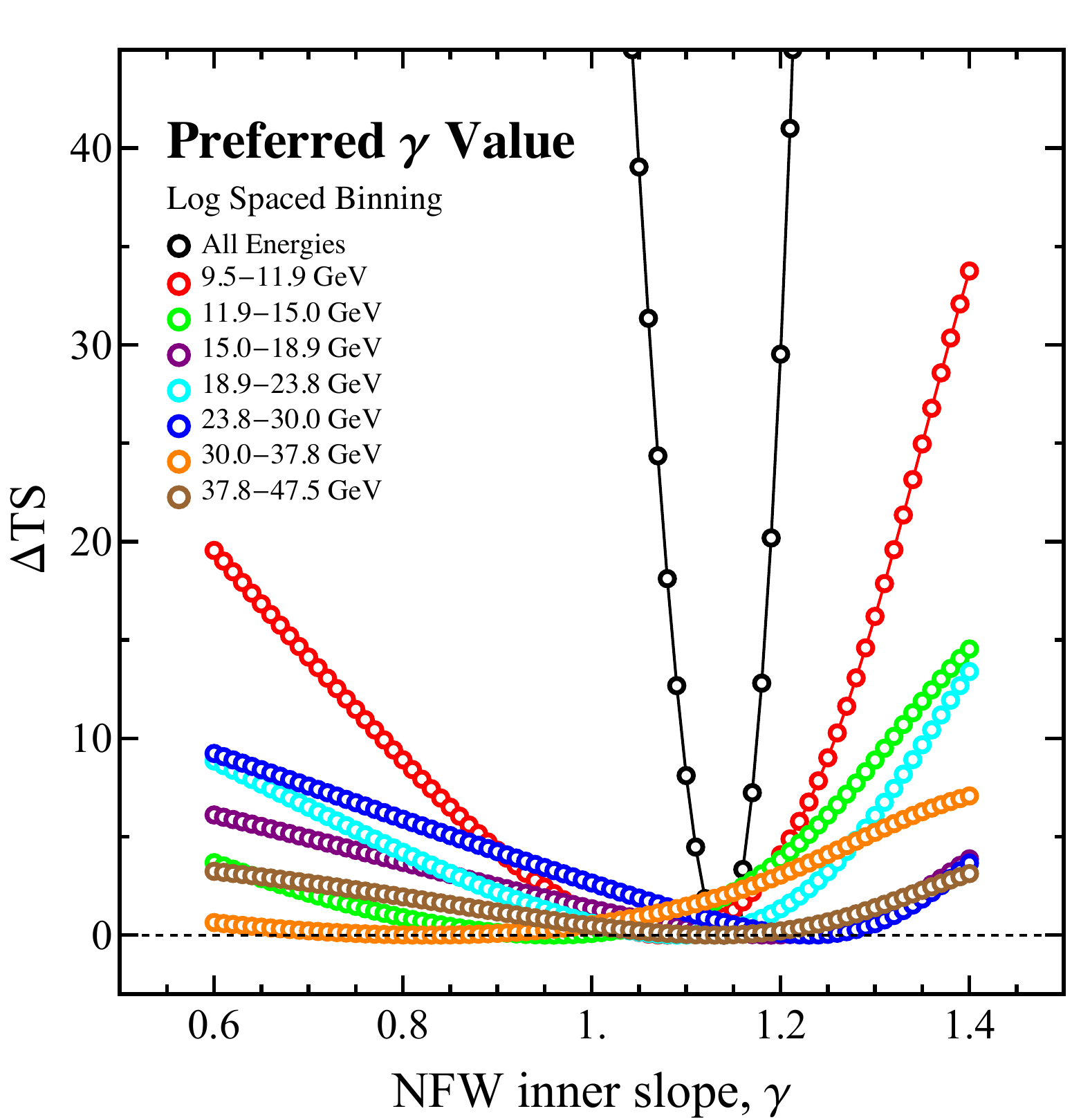}
\end{tabular}
\caption{\footnotesize{Preferred $\gamma$ for the default analysis for two different energy binnings: the binning employed in the main text for statistical analyses (left) and equal log spaced bins (right).}}
\label{fig:Rebin}
\end{figure*}

In addition to choosing our dataset to maximize statistics, as discussed in Sec.~\ref{sec:SpPr}, we combined several of the high-energy bins together to increase statistics. By default we have 7 equally log spaced bins between $9.5$ and $47.5$~GeV, whilst our rebinning combines the top 6 of these into adjacent pairs reducing us to the 4 bins described in the main text. Results without this rebinning are largely similar to what has already been shown, although understandably with lower statistical significance. To estimate the difference in statistics per bin, the TS for the GCE shown for the combined bins in Table~\ref{table:TSP7P8} can be compared to the uncombined bins in Table~\ref{table:TSAllBins}. As another example, in Fig.~\ref{fig:Rebin} we show the bin-by-bin TS as a function of $\gamma$ for the two different binnings just discussed. As claimed, the qualitative behavior is independent of the binning, but of course the statistical significance of results in any one bin depends on the bin size.

In a Bayesian analysis, instead of the TS one would instead examine the $2 \ln \left[ {\rm Bayes~factor} \right]$ for a GCE in each energy bin. Note that unlike for TS, $2 \ln \left[ {\rm Bayes~factor} \right]$ can be less than 0, as the model with the GCE is penalized for having an additional degree of freedom. If this were to occur it would indicate the model without the GCE should be considered a better fit to the data. With this in mind we repeated our default analysis, but examined the $2 \ln \left[ {\rm Bayes~factor} \right]$ for adding the GCE in each of the log spaced energy bins. In the first energy bin there is a clear preference for the GCE template, but for higher energy bins the evidence was often marginal and sometimes negative. Nonetheless when we repeated this with the combined energy binning, there was clear evidence ($2 \ln \left[ {\rm Bayes~factor} \right] > 9$) for the GCE in all bins but the highest, which gives an additional reason to make use of the combined binning.

\subsection{Cumulative Analysis of Spatial Properties}

In the main text we showed how the $\Delta$TS behaves in each energy bin as a function of the radial variation parameter $\gamma$ and the GCE sphericity. This is useful for seeing how the preferred spatial properties vary between bins, but it is difficult to determine the statistical preference for a particular value of $\gamma$ or axis ratio, accounting for \emph{all} the data above a specified energy threshold. This quantity is less noisy than the bin-by-bin preferences and may be preferred for that reason.

\begin{figure*}[t!]
\centering
\includegraphics[scale=0.55]{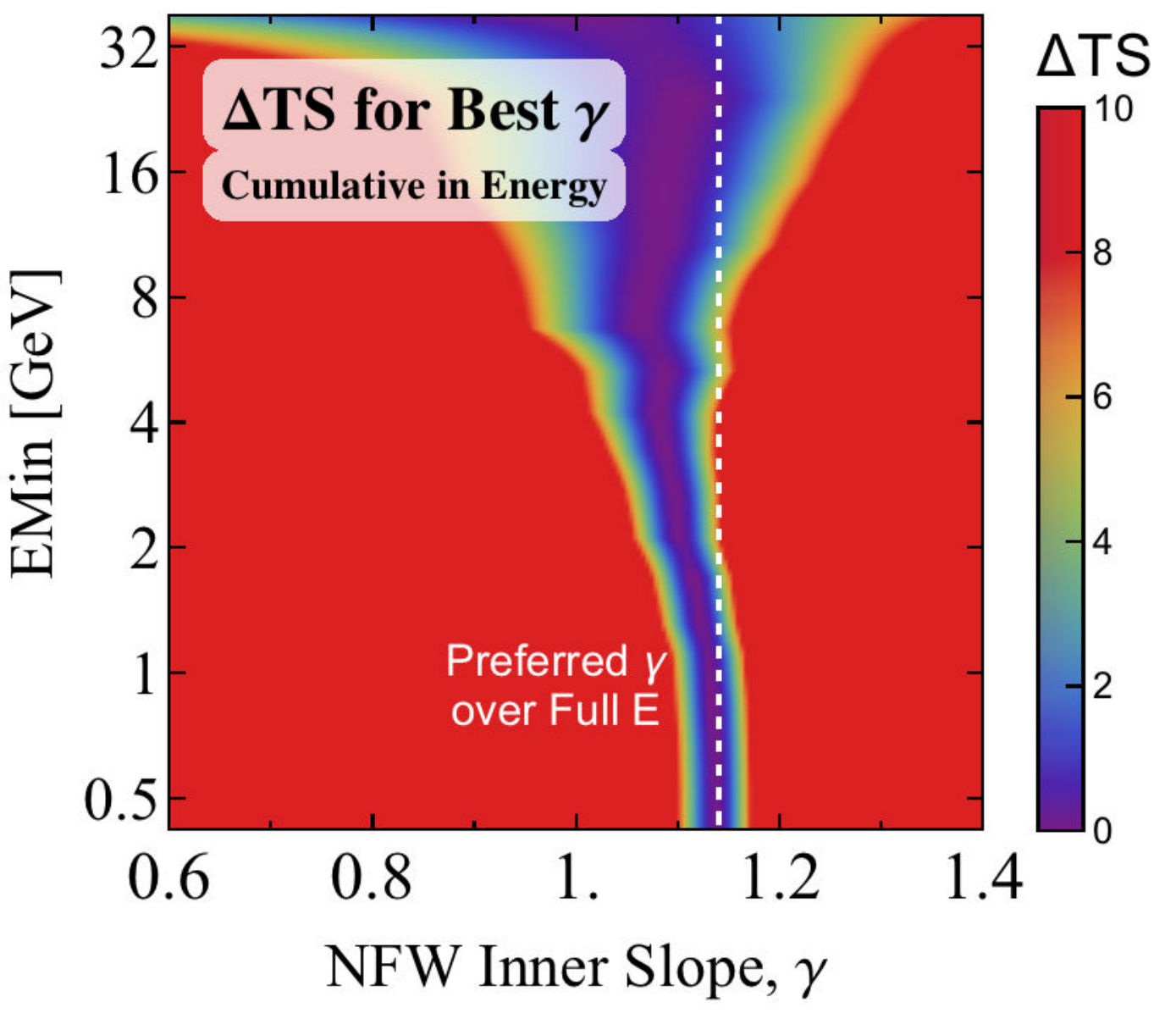} \hspace{0.15in}
\includegraphics[scale=0.24]{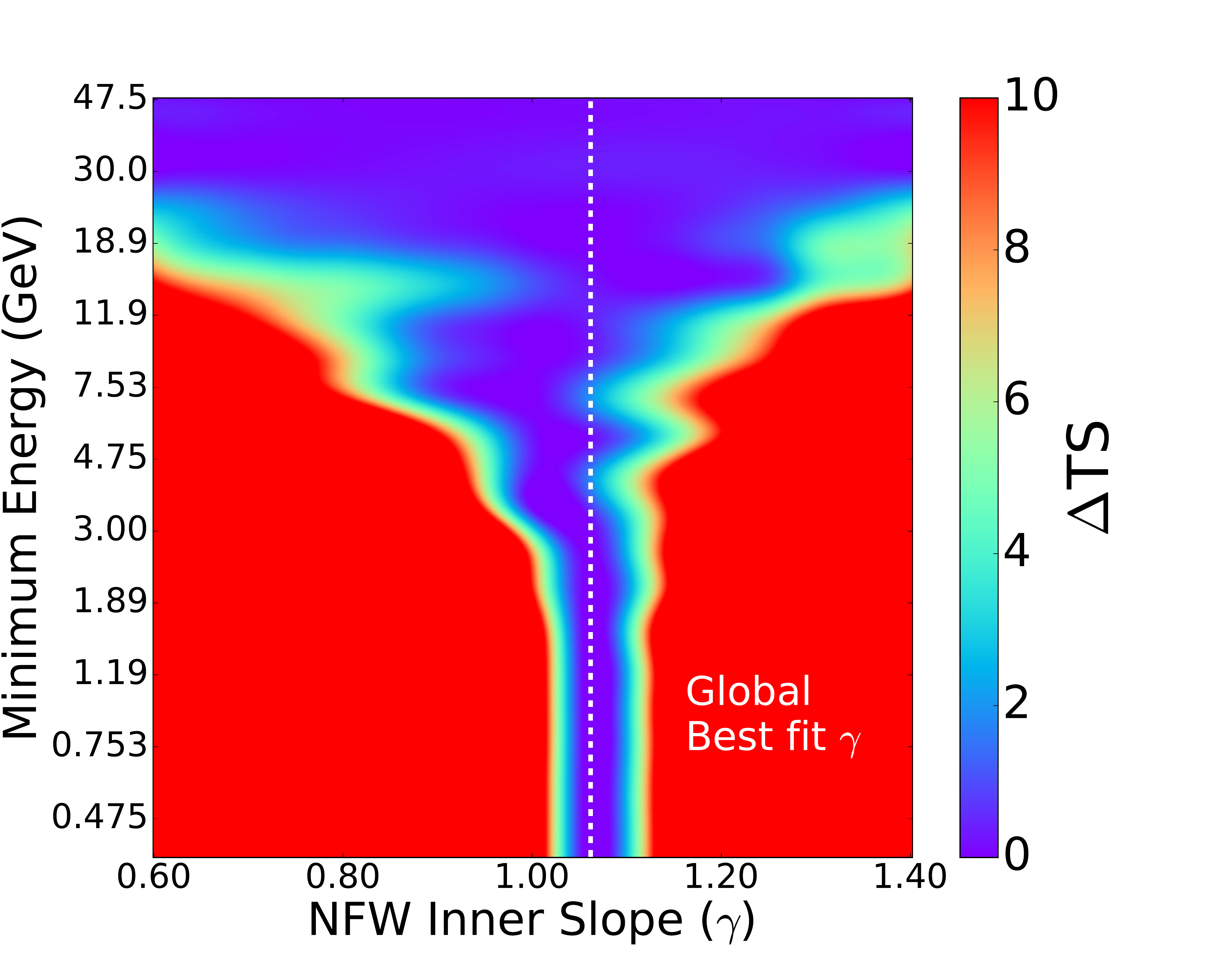}
\caption{\footnotesize{Cumulative version of Fig.~\ref{fig:IGsphericity1} and \ref{fig:gc_sphericity}, for the IG (left) and GC (right).}}
\label{fig:BestGammaCumulative}
\end{figure*}

\begin{figure*}[t!]
\centering
\includegraphics[scale=0.55]{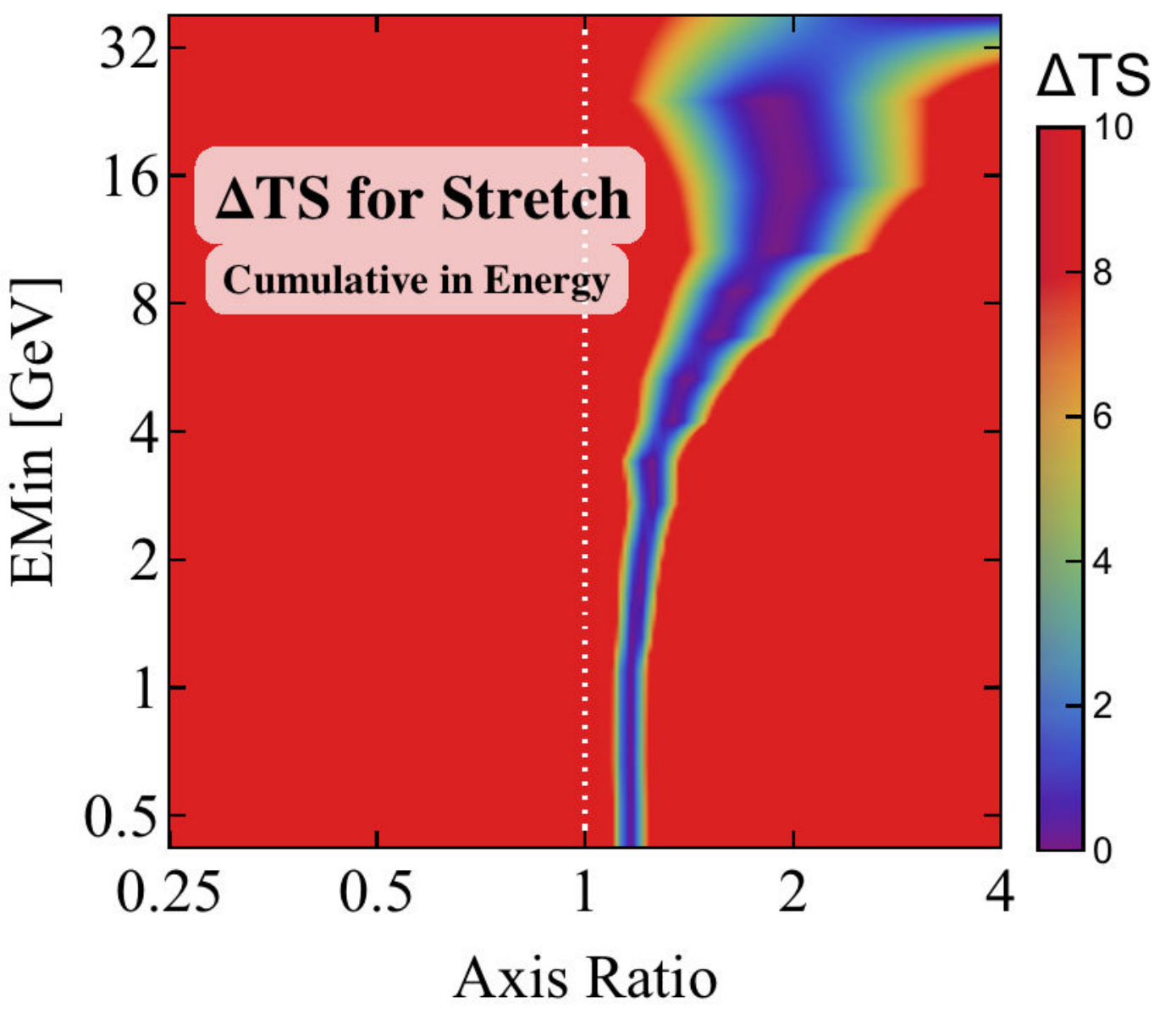} \hspace{0.15in}
\includegraphics[scale=0.24]{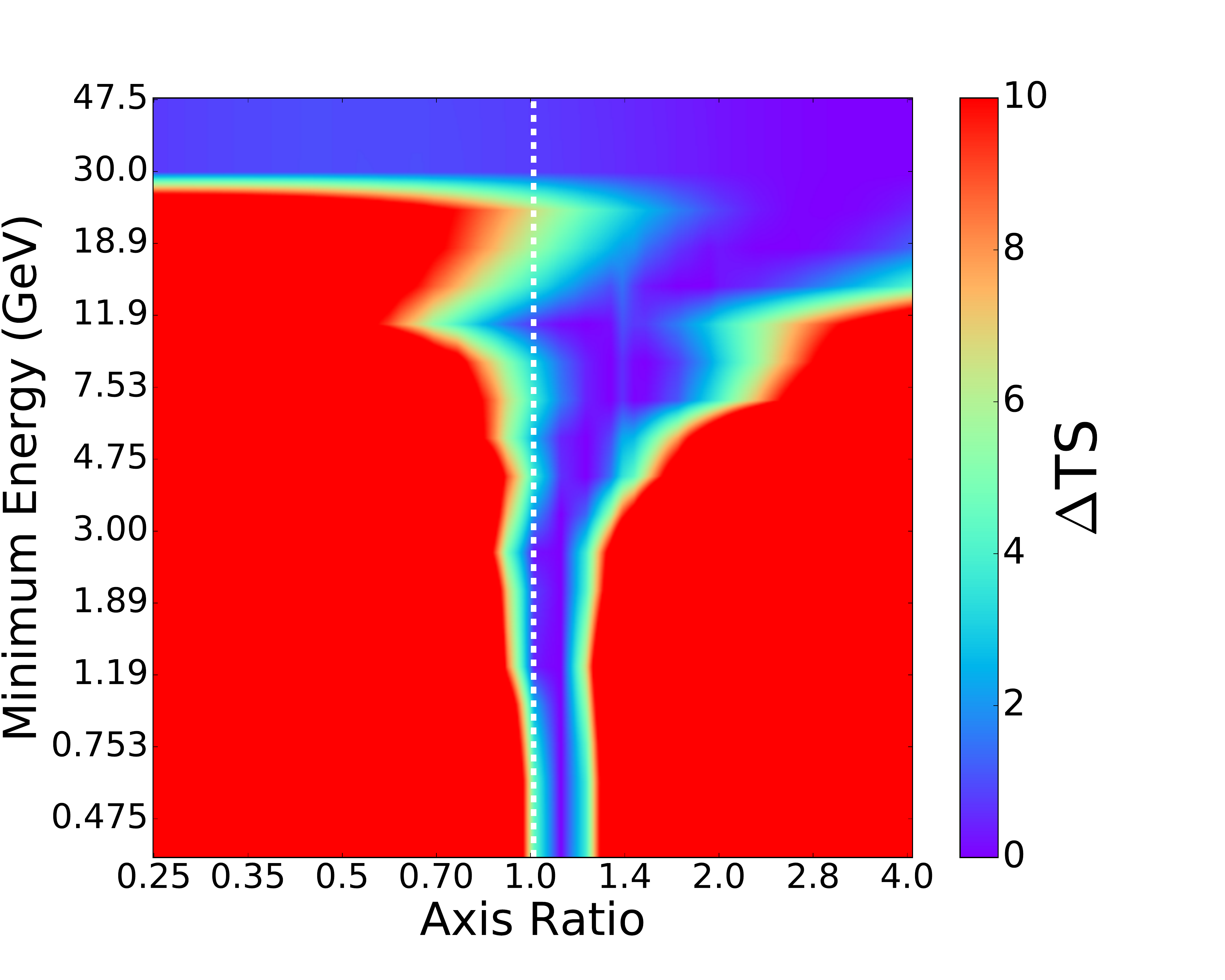}
\caption{\footnotesize{Cumulative version of Fig.~\ref{fig:IGsphericity1} and \ref{fig:gc_sphericity}, for the IG (left) and GC (right).}}
\label{fig:BestStretchCumulative}
\end{figure*}

In this subsection we combine the results for the preferred inner slope (shown in Fig.~\ref{fig:BestGamma} and \ref{fig:gc_morphology_default}) and sphericity (shown in Fig.~\ref{fig:IGsphericity1} and \ref{fig:gc_sphericity}) into a single energy bin and then vary the minimum energy of that bin. The resulting cumulative plots are shown in Fig.~\ref{fig:BestGammaCumulative} and \ref{fig:BestStretchCumulative}. The preferred values and the statistical errors on these are shown in Table~\ref{table:CumulativeVals}.

\begin{table}[h]
\begin{center}
\begin{tabular}{| c || c | c | c | c |}
\hline
EMin & \multicolumn{2}{ |c| }{Preferred $\gamma$} & \multicolumn{2}{ |c| }{Preferred axis ratio} \\ \cline{2-5}
(GeV) & IG & GC & IG & GC \\ \hline \hline
0.4 & $1.14\pm 0.01$ & $1.06 \pm 0.02$ & $1.17^{+0.01}_{-0.02}$ & $1.10\pm0.03$ \\ \hline
2 & $1.10^{+0.02}_{-0.01}$ & $1.06 \pm 0.02$ & $1.21^{+0.03}_{-0.02}$ & $1.09^{+0.04}_{-0.09}$ \\ \hline
5 & $1.09^{+0.02}_{-0.03}$ & $1.02^{+0.05}_{-0.04}$ & $1.39^{+0.09}_{-0.06}$ & $1.20^{+0.13}_{-0.14}$\\ \hline
10 & $1.08\pm 0.05$ & $1.01^{+0.08}_{-0.11}$ & $1.9 \pm 0.2$ & $1.21^{+0.33}_{-0.25}$ \\ \hline
\end{tabular}
\end{center}
\caption{Preferred $\gamma$ and axis ratio in the IG and GC analyses as we vary the minimum energy. Unlike in previous tables we here include the statistical errors as they can become large as we move to high energies and thereby lower statistics.}
\label{table:CumulativeVals}
\end{table}

\subsection{Impact of the {\it Fermi} Energy Dispersion}

The energy dispersion of the {\it Fermi} telescope is small ($\sim 6-9$\% for the dataset and energy range we consider), but not negligible.\footnote{See e.g. the discussion at \texttt{http://fermi.gsfc.nasa.gov/ssc/} \texttt{data/analysis/documentation/Pass8\_edisp\_usage.html}} We have neglected the energy dispersion in our analysis, treating the reconstructed energy as the true energy of the photons, since the energy dispersion is small compared to the width of our smallest energy bins. However, because the spectrum of the GCE falls steeply at higher energies, one might worry that even a relatively small fraction of low-energy photons leaking into high-energy bins might substantially bias the high-energy spectrum, and in the worst case, even fake a high-energy signal that is not actually present.

As a simple estimate of the possible impact of the energy dispersion, we take our \emph{extracted} spectrum for the GCE, and ask what the impact of energy dispersion would be if this were the \emph{true} spectrum of the GCE, by convolving the photon spectrum $dN/dE$ with the energy dispersion function. The resulting smoothed spectrum is not physically meaningful, but the difference between the original and smoothed spectra gives some estimate of the impact of neglecting the energy dispersion for a spectrum similar to that of the GCE. A more careful analysis would require convolving all the model maps by the energy dispersion function, with some prescription for the spectrum of each component, before comparing the template model to the data; this is beyond the scope of this work.

\begin{figure}[t!]
\centering
\begin{tabular}{c}
\includegraphics[scale=0.45]{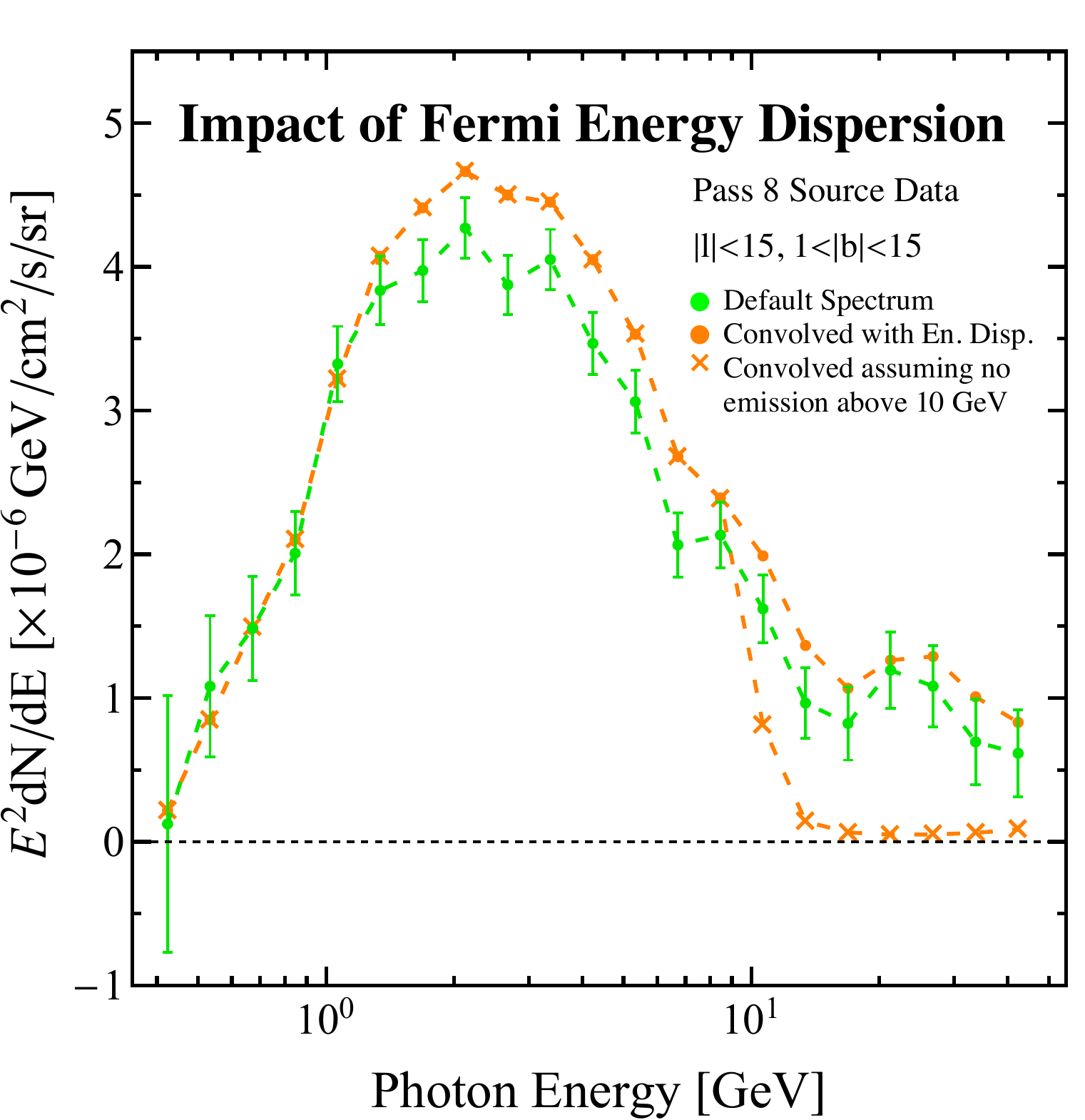}
\end{tabular}
\caption{\footnotesize{The effect of convolving the extracted spectrum for the GCE with the energy dispersion of Fermi. Green dots show the original spectrum, orange dots the spectrum after convolution with the energy dispersion function (note that this is \emph{not} an estimate of the ``true'' spectrum), and orange crosses the spectrum after convolution if the original spectrum is truncated above $10$~GeV.}}
\label{fig:EnDisp}
\end{figure}

The results of this simple analysis are shown in Fig.~\ref{fig:EnDisp}. We see that indeed, at and above the peak of the excess where the spectrum begins to fall with increasing energy, convolution with the energy dispersion can noticeably change the spectrum. However, at high energies the resulting shift appears to be comparable to or smaller than the statistical errors (as well as other systematic uncertainties we have discussed, e.g. due to changing the diffuse model or the {\it Fermi} Bubbles template).

To test the plausibility of the worst-case scenario where \emph{all} the emission we see above $10$~GeV is due to leakage from lower-energy bins, we also test the effect of truncating the GCE spectrum at $10$~GeV before convolving with the energy dispersion function. The result is also shown in Fig.~\ref{fig:EnDisp}; as expected, since the width of the energy dispersion function is much smaller than the bin width, in this case we would expect essentially no emission in the higher-energy bins.

The relatively small sizes of the energy dispersion effects are consistent with expectations that the bias to photon number above $10$~GeV should be at the $10\%$ level or less, even for quite steeply falling spectra.

\bibliography{GCEHighE}

\end{document}